\documentclass[reprint,onecolumn,english,nofootinbib,letterpaper,nobalancelastpage]{revtex4-1}

\usepackage[utf8x]{inputenc}
\usepackage{amsmath}
\usepackage{amssymb}
\usepackage{bbm}
\usepackage{babel}
\usepackage{graphicx}
\usepackage{verbatim}
\usepackage{latexsym}
\usepackage{mathptm}
\usepackage{color}

\newcommand{\f}{\begin{equation}}
\newcommand{\ff}{\end{equation}}

\begin{document}
\title{Relative-locality distant observers and the phenomenology
of momentum-space geometry}

\author{$~$\\
{\bf Giovanni Amelino-Camelia}$^{1,2}$,
  {\bf Michele Arzano}$^{3}$, {\bf Jerzy Kowalski-Glikman}$^{4}$, {\bf Giacomo Rosati}$^{1,2}$, {\bf Gabriele Trevisan}$^{1}$
\\
$^1${\footnotesize Dipartimento di Fisica, Universit\`a ``La Sapienza" P.le A. Moro 2, 00185 Roma, Italy }\\
$^2${\footnotesize Sez.~Roma1 INFN, P.le A. Moro 2, 00185 Roma, Italy }\\
$^3${\footnotesize Institute for Theoretical Physics,
Utrecht University,
Leuvenlaan 4, Utrecht 3584 CE, The Netherlands}\\
$^4${\footnotesize Institute for Theoretical Physics, University of Wroclaw,  Pl. Maxa Borna 9, 50-204 Wroclaw, Poland}}

\begin{abstract}
We study the translational invariance of the relative-locality
framework proposed in arXiv:1101.0931, which had been previously  established only for
the case of a single interaction. We provide an explicit example
of boundary conditions at endpoints of worldlines, which indeed
ensures the desired translational invariance
for processes involving several interactions,
even when some of the interactions are causally connected (particle exchange).
We illustrate the properties of the associated
relativistic description of distant observers
within the example of
a $\kappa$-Poincar\'e-inspired
 momentum-space geometry, with de Sitter metric and parallel transport governed
by a non-metric and torsionful connection. We find that in such a theory
simultaneously-emitted massless particles do not reach simultaneously a distant detector,
 as expected in light of the findings
of arXiv:1103.5626 on the implications of non-metric connections.
We also show that the theory admits a free-particle limit, where the relative-locality
results of arXiv:1102.4637 are reproduced. We establish that the torsion
of the $\kappa$-Poincar\'e connection introduces a small (but observably-large)
 dependence of the time of
detection, for simultaneously-emitted particles, on some properties of the
interactions producing the particles at the source.
\end{abstract}

\maketitle

\newpage

\tableofcontents

\newpage

\section{Introduction and summary}
The relative-locality framework of Refs.~\cite{prl,grf2nd} is centered on
the possibility of a non-trivial geometry for momentum space,
and links to those geometric properties some effects of relativity
of spacetime locality, such that events established to be coincident
by nearby observers are not described as coincident in the coordinatization
of spacetime by distant observers. Interestingly,
just like the relativity of simultaneity
implies that there is no observer-independent projection from spacetime
to separately space and time, relative locality
implies that there is no observer-independent projection from a one-particle
phase space to a description of the particle separately in spacetime and in
momentum space.
Besides its intrinsic interest from the point of view of relativity research,
this relative-locality framework appears to be also relevant~\cite{prl,grf2nd}
for the understanding
of several issues which have emerged in the recent quantum-gravity literature.
Indeed several approaches to the study of the quantum-gravity problem
have led to speculations about nonlinearities in momentum space that may admit
geometric description (see, {\it e.g.}, Refs.~\cite{majidCurvedMomentum,kpoinap,gacdsr1,leedsrPRD,jurekLONGPAPER,florianCurvedMomentum}
and references therein), and some related studies had hinted at possibly
striking implications of such nonlinearities for the fate of locality
at the Planck scale~\cite{gacIJMPdsrREV,Schutzhold:2003yp,Arzano:2003da,grilloSTDSR,dedeo,Hossenfelder:2010tm}.

Evidently a pivotal role for the success (or failure) of this relative-locality
proposal will be played by investigations of the relativistic description
of distant observers, which is the main focus of the study
we are here reporting.
This is in itself an aspect of relative locality which is rather intriguing
from a conceptual perspective. In previous evolutions of our relativistic theories
the most subtle issues always concerned boost transformations, and therefore
the relativistic description of pairs of observers with a relative boost.
Translational invariance, and therefore the relativistic description
of distant observers, always admitted an elementary and fully intuitive
description.
This is particularly clear when looking at
the transition from Galilean relativity, and its description of relative rest,
to special relativity, with Einstein's description of relative simultaneity and rest:
grasping the physical content of special relativity proved challenging
because of properties of special-relativistic boosts, which force us
to abandon a ``common sense understanding". But special relativistic translations
are no less trivial than Galilean translations.
This is because the special-relativistic notion of relative
simultaneity is already fully characterized when focusing
on pairs of observers connected
by a pure boost transformation.
But the relative-locality notion that an interaction established to be local by
nearby observers
may be described as nonlocal by distant observers evidently implies
a crucial role for the identification of a corresponding formalization
of translational symmetries, and we must therefore expect that ensuring
a relativistic description of distant observers should be one of
the main challenges for the formalization of relative locality.

\begin{figure}[hb!]
\begin{center}
\includegraphics[width=0.3 \textwidth]{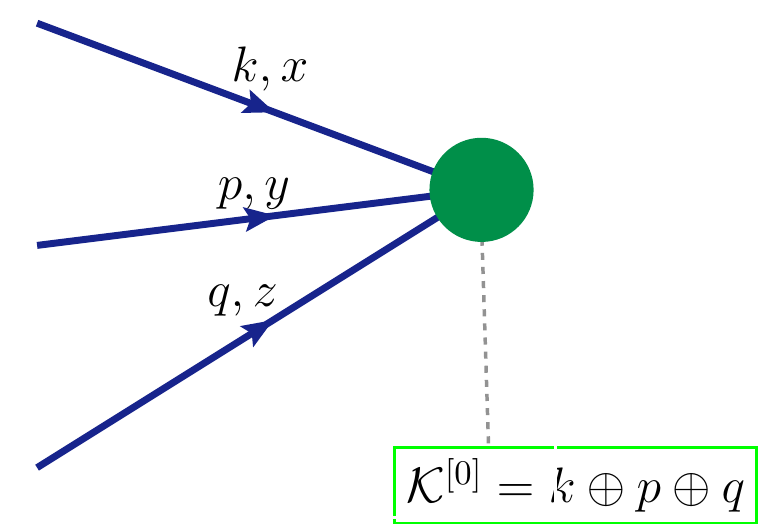}
\caption{We show here a simple example of process
for which the observations reported in Ref.~\cite{prl}
suffice for establishing translational
invariance (with relative locality).
Specifically the graph intends to describe the idealized case
of a process
with 3 particles of energy-momenta $k_\mu$, $p_\mu$, $q_\mu$,
all incoming. Here and elsewhere in this manuscript we describe
pictorially some processes using a graphical scheme which is mainly
of evocative valence. The lines in the graph are not intended
as representatives of worldlines of particles or other fine aspects of
the evolution of variables in terms of the affine parameter.
They should rather be looked at, going from left to right,
as a schematic portrait of the discrete steps in the redistribution
of momentum among particles, changing
at every subsequent interaction (but the case in this figure is a single-interaction process). A similar graphical characterization
of processes is often adopted for quantum-field-theory Feynman diagrams
(but our entire analysis
is confined to the context of classical particles).}
\label{oksofarfig}
\end{center}
\end{figure}

A crucial step toward the understanding of these issues here of interest,
concerning the interplay between relative locality and translation symmetries,
is already found in Refs.~\cite{prl,grf2nd}.
For example, according to Ref.~\cite{prl},
one could describe the process in Figure~\ref{oksofarfig},
which is the idealized case
with 3 particles of energy-momenta $k_\mu$, $p_\mu$, $q_\mu$
all incoming,
on the basis of the following action:
\begin{equation}
\begin{split}
\mathcal{S}^{example}= & \int_{-\infty}^{s_{0}}ds\left(x^{\mu}\dot{k}_{\mu}+y^{\mu}\dot{p}_{\mu}
+z^{\mu}\dot{q}_{\mu}+\mathcal{N}_k\mathcal{C}\left[k\right]
+\mathcal{N}_p\mathcal{C}\left[p\right]+\mathcal{N}_q\mathcal{C}\left[q\right]\right) -\xi^{\mu}\mathcal{K}_\mu(s_{0}) \ .
\end{split}
\label{actionFIRST}
\end{equation}

As we shall here discuss in greater detail in Section~\ref{setup},
the bulk part of $\mathcal{S}^{example}$ ends up characterizing~\cite{prl} the
propagation of the 3 particles, with the Lagrange multiplier
$\mathcal{N}_k$ (and similarly $\mathcal{N}_p$ and $\mathcal{N}_q$)
enforcing the on-shell relation ${\cal C}[k]= D^2(k)-m^2$,
with $D^2(k)$ in turn derived from the metric on momentum space
as the distance of $k_\mu$ from the origin
of momentum space.
And the form of the boundary term $\xi^{\mu}\mathcal{K}_\mu(s_{0})$
is such that~\cite{prl} the Lagrange multipliers $\xi^{\mu}$
enforce the condition $\mathcal{K}_\mu(s_{0}) = 0$,
so that by taking for $\mathcal{K}_\mu$ a suitable composition
of the momenta $k_\mu$, $p_\mu$, $q_\mu$ the boundary terms enforces
a law of conservation of momentum at the interaction.
The form of the law of composition of momenta used
for the conservation law $\mathcal{K}_\mu(s_{0}) = 0$ is governed
by the affine connection on momentum space~\cite{prl}, and may involve
nonlinear terms which are ultimately responsible for the relativity
of spacetime locality. This is indeed seen by studying the invariance
of the action $\mathcal{S}^{example}$ under translations of the
coordinates of worldline points $x^{\mu}(s)$, $y^{\mu}(s)$, $z^{\mu}(s)$,
which each observer introduces as variables that are
canonically conjugate to the coordinates
on momentum space $k_\mu(s)$, $p_\mu(s)$, $q_\mu(s)$.

The observations used in Ref.~\cite{prl} in the derivations that
established the presence of this translational invariance
appeared to rely crucially
on some simplifications afforded only by the case of a single-interaction process.
As here stressed in Section~\ref{setup}, for a single interaction several alternative
choices of boundary terms enforcing the same conservation law are consistent
with the presence of relativistic translation symmetries within the relative-locality
framework of Ref.~\cite{prl}. However, the demands of translational invariance
become much more constraining when 2 or more interactions are
causally connected, {\it i.e.} when a particle outgoing from one interaction is incoming
into another interaction.
In Section~\ref{main} we provide an explicit example of formulation
of the boundary terms which ensures translational invariance when several
causally-linked interactions
are analyzed.
And the logical structure of our proposal is easily described:
translation transformations are generated by the total-momentum charge
(obtained from individual particle momenta via the connection-induced
composition law) and the boundary terms are written as differences between the total
momentum before the interaction and after the interaction (so that the
associated constraints automatically
ensure conservation of the total momentum).

Before getting to that main part of our analysis it will be useful to
do some preparatory
work. In the next section we motivate our focus on results that
are obtained only at leading order in the deformation scale, by observing
that, if indeed the deformation scale is roughly given by the Planck scale,
the experimental sensitivities foreseeable at least for the near future
will not afford us investigations going
much beyond the leading-order structure of the
geometry of momentum space. And we also show that working at leading order not only
simplifies matters in the way that is commonly encountered in physics, by shortening
some computations, but in this case also provides some qualitative simplifications,
including most notably the fact that at leading order the momentum-composition
laws of the relative-locality framework are automatically associative.

Then in Sections~\ref{kappaMtotherightDIR}
and~\ref{geometry} we characterize the specific
example of relative-locality momentum space on which we shall test our
proposal for relativistic translational symmetries. This is based on results
obtained in the $\kappa$-Minkowski/$\kappa$-Poincar\'e framework, where indeed
evidence of a deformed on-shell relation and of a nonlinear law of composition
of momenta has been discussed for more than a decade, but without appreciating
the implications for how distant observers would characterize events that
are found to be coincident by nearby observers.
We find that this $\kappa$-Poincar\'e-inspired momentum space has
 de Sitter metric and parallel transport governed
by a non-metric and torsionful connection.

Equipped with these preliminary observations we then discuss, in Section~\ref{setup},
the challenges that must
be dealt with in seeking a translationally-invariant description of chains
of causally connected interactions within the relative-locality framework
proposed in Ref.~\cite{prl}.
The main insight gained from the analysis reported in  Section~\ref{setup}
is that in order to achieve translational invariance
 it is not sufficient to ensure that the boundary terms at endpoints of worldlines
 enforce some suitable momentum-conservation laws, since in general two such choices
 of boundary terms at the endpoints of a finite worldline (a worldline going from
 one interaction to another) will spoil translational invariance.

Section~\ref{kbobreview} is a short aside on some characterizations
of relative locality confined to a Hamiltonian description
of free particles~\cite{whataboutbob,leeINERTIALlimit,arzkowaRelLoc,kappabob}
which serves two purposes: it prepares the rest of our analysis
by reviewing some concepts about the
type of symplectic structure that is traditionally used in
the $\kappa$-Minkowski/$\kappa$-Poincar\'e literature,
and its characterization of relative locality for free particles
is then of reference for our description of a free-particle limit,
which is an important corollary result of
 our proposal for interacting particles.

It is in Section~\ref{main} that the reader finds
our main results concerning
the proposal and analysis of a relativistic formulation of  processes
involving several interactions,
within the general framework of Ref.~\cite{prl},
with translational invariance assured by a corresponding specification
of the boundary conditions that implement momentum conservation.
We test the robustness of our proposal mainly by applying it to
the illustrative example of
the $\kappa$-Poincar\'e-inspired momentum space.

Section~\ref{secgrbs} contains our results that are
of particular significance from the perspective of phenomenology.
We show that, within our setup for $\kappa$-Poincar\'e interacting
 particles, simultaneously-emitted massless particles
do not necessarily reach the same detector at the same time.
Since our $\kappa$-Poincar\'e momentum space has nonmetricity,
this is consistent with the thesis put forward in Ref.~\cite{leelaurentGRB},
according to which these time delays at detection for simultaneously-emitted
massless particles are to be expected when nonmetricity is present.
In addition we also investigate how the torsion of
our $\kappa$-Poincar\'e momentum space
affects these time-of-detection delays,
an issue for which no previous result is applicable.
And we find that the torsion does affect the time delays, by essentially
rendering the effect non-systematic: the time-of-detection difference for
two simultaneously emitted massless particles depends non only on the momenta
of the two particles involved but also on some properties of the
events that emitted the two particles. We also discuss the first
elements of a phenomenology that could exploit this striking feature.

In deriving the results reported in
Sections~\ref{main} and~\ref{secgrbs}
we use $\kappa$-Poincar\'e illustrative example non only
in the sense that we adopt the metric and connection
of the ``$\kappa$-momentum space" (of Sections~\ref{kappaMtotherightDIR}
and~\ref{geometry}) but also by imposing upon us the use of the
nontrivial symplectic
structure that is preferred in the $\kappa$-Poincar\'e literature.
However, in
Section~\ref{itsgeometry} we keep the $\kappa$-Poincar\'e momentum space
while switching to a trivial symplectic structure, and we reproduce
again the results of Section~\ref{secgrbs}. This allows us to establish
that the predictions derived in Section~\ref{secgrbs} are purely
manifestations of momentum space geometry.

Section~\ref{closingsec} contains some closing remarks,
mostly focusing on the outlook of the relative-locality research program.

\bigskip

{ We adopt units such that the speed-of-light scale
(speed of massless particles in the infrared limit)
and the reduced Planck constant
are $1$ ($c=1=\hbar$).
And we denote by $\ell$ the momentum-space-deformation scale.
Of course, $\ell$ carries dimensions of inverse momentum, and
a natural quantum-gravity-inspired estimate
would be to have $|\ell^{-1}|$ roughly of the order of the Planck scale.
The issues studied in this manuscript are of exactly the same nature
in the case of a 4D momentum space and in  the case of a 2D momentum space,
and we shall often (but not always)
focus for definiteness and simplicity on the 2D case. When not otherwise specified
we shall switch between 4D and 2D formulas by simply denoting with $p_0,p_j$
the momentum in the 4D case and with $p_0,p_1$ the momentum in the 2D case.}

\section{Leading-order anatomy of relative-locality momentum spaces}\label{anatomy2D}

Refs.~\cite{prl,grf2nd} (also see Refs.~\cite{leelaurentGRB,soccerball})
raised the issue of determining experimentally
the geometry of momentum space, much like it is traditional in physics
to study experimentally the geometry of spacetime.\\
It is however important to notice a crucial difference: while we do have
experimental access to distance scales larger than the scales of
curvature of spacetime, it is very unlikely that in the foreseeable future
we could have
experimental access to momentum
scales even just comparable to the Planck scale, which is the natural
candidate for the scale of curvature of
the relative-locality momentum space~\cite{prl}.\\
It should be appreciated that
this disappointing limitation of our horizons on the geometry of
momentum space can also be turned in some sense into a powerful weapon
for the phenomenology of momentum-space geometry: evidently
all we need is a characterization of the geometry
of momentum space near the origin, where  $|p| \ll |\ell|^{-1} \simeq M_p$.
And at least at first this will essentially be focused on the
search of leading-order evidence of a nontrivial geometry of momentum space.

For what concerns the affine connection on momentum space,
responsible for the nontrivial properties of the law of composition
of momenta~\cite{prl}, all we need for the purposes of
this phenomenology are the
($\ell$-rescaled) connection coefficients on momentum space
 evaluated at $p_\mu = 0$, which we denote
 by $\Gamma_\mu^{\alpha\beta}$:
$$\left(p \oplus q\right)_\mu \simeq p_\mu + q_\mu -
\ell \Gamma_\mu^{\alpha\beta}\, p_\alpha\,
q_\beta + \cdots$$
And evidently the fact that the phenomenology only needs
leading-order results implies (also considering that we already rescaled
the connection coefficients by the Planck scale)
that we can treat the $\Gamma_{\mu}^{\alpha\beta}$
as pure numbers.

Analogous considerations lead us to focus on momentum-space metrics
that are at most linear in the momenta:
\begin{equation}
g^{\mu \nu} = \eta^{\mu \nu} +
\ell h^{\mu \nu \rho} p_\rho
\label{leadingmetric}
\end{equation}
and, just like the $\Gamma_\mu^{\alpha\beta}$, we should handle
the coefficients $h^{\mu \nu \rho}$ as pure numbers
in our leading-order phenomenology.

In this manuscript we shall mainly work only at leading order in the deformation
scale, and it will be evident that this provides with significant advantages.
In particular, at leading order in the deformation scale
the momentum-composition law is always associative. This can be established by
writing a general leading-order composition law as follows:
\begin{equation}
(k \oplus p)_\mu = k_\mu + p_\mu - \ell \, \Gamma^{\alpha \beta}_\mu k_\alpha p_\beta  ~,
\label{connectiontorsyNOK}
\end{equation}
and then noticing that  indeed (of course to leading-order accuracy)
\begin{equation}
[(k \oplus p) \oplus q]_\mu =
 k_\mu + p_\mu + q_\mu
-  \ell \, \Gamma^{\alpha \beta}_\mu \left( k_\alpha p_\beta +
k_\alpha q_\beta +p_\alpha q_\beta \right) = [k \oplus (p \oplus q)]_\mu ~.
\label{tribodytorsyNOK}
\end{equation}
Beyond leading order the composition law could be nonassociative, and in that
case one could appreciate the curvature of the momentum-space connection,
with interesting but technically challenging consequences which we shall
not encounter in this manuscript, and will never be encountered when working
at leading order in the deformation scale.

The fact that our horizons on the geometry of momentum space
probably are confined to leading order may be viewed as an
unpleasant philosophical limitation, but pragmatically can be turned
 into a powerful asset for phenomenology work
 on relative-locality momentum spaces, since the task of phenomenologists
 then is very clearly and simply specified: the target should be
 to determine experimentally (as accurately as possible)
  a few dimensionless numbers for the leading-order (and possibly
  the next-to-leading order) geometry of momentum space.

  To make this point fully explicit let us for simplicity
  imagine a
by 2D relative-locality momentum space.
In the 2D case a full leading-order characterization of the momentum-space
 geometry requires establishing experimentally (in hypothetical 2D experiments)
 the 8 dimensionless parameters of the affine connection
  on momentum space,
\begin{gather*}
\Gamma_{0}^{00}, \quad \Gamma_{0}^{01}, \quad \Gamma_{0}^{10}, \quad \Gamma_{0}^{11}, \\
\Gamma_{1}^{00}, \quad \Gamma_{1}^{01}, \quad \Gamma_{1}^{10}, \quad \Gamma_{1}^{11}
\label{eightparameters}
\end{gather*}
and the 6 dimensionless parameters of the leading-order description
of the metric, which one can conveniently encode\footnote{We are here implicitly using
the fact that our ``leading-order momentum-space metrics" can be parametrized,
in the example of the 2D case,
equivalently in terms of the 6 independent numbers that specify $h^{\mu\nu\sigma}$
or in terms of the 6 independent Christoffel symbols. Indeed
one finds that
\begin{equation}
    C_\rho^{\mu\nu}=\frac12\,
    g_{\rho\sigma}\left(g^{\sigma\mu,\nu}+g^{\nu\sigma,\mu}-g^{\mu\nu,\sigma}\right)=
    \frac\ell2\,\eta_{\rho\sigma}\left(h^{\sigma\mu\nu}+h^{\nu\sigma\mu}-h^{\mu\nu\sigma}\right)
\end{equation}
} into the 6 free parameters
of the associated Christoffel symbols $C_\alpha^{\mu \nu}$,
\begin{gather*}
C_{0}^{00}, \quad C_{0}^{01}=C_{0}^{10}, \quad C_{0}^{11}, \\
C_{1}^{00}, \quad C_{1}^{01}=C_{1}^{10}, \quad C_{1}^{11}\ .
\end{gather*}

\section{Some known properties of the $\kappa$-Poincar\'e Hopf algebra
and $\kappa$-Minkowski spacetime}\label{kappaMtotherightDIR}

\subsection{$\kappa$-momentum space}

In this section we describe the construction of the momentum space
motivated by the $\kappa$-Poincar\'e framework ~\cite{lukieIW,majidruegg,kpoinap},
which we shall call here, for short,
the ``$\kappa$-momentum space".
This $\kappa$-momentum space will provide for us an
example of momentum space, of some independent interest,
on which to illustrate in tangible way the efficacy of
the characterization of relative-locality distant observers,
which is our main objective for this manuscript.

In this subsection we follow Ref.~\cite{KowalskiGlikman:2004tz} so
we describe $\kappa$-momentum space as a manifold of the group
$\sf{AN}(3)$ (dubbed also the Borel group), which is, as a manifold,
essentially a half of de Sitter space. The $\sf{AN}(3)$ group is a
subgroup  of the de Sitter $\sf{SO}(4,1)$ group, defined by its Lie
algebra $\sf{an}(3)$ which has the following form:
\begin{equation}\label{j3}
    [{\cal X}^0, {\cal X}^i]=-{i}\ell\, {\cal X}^i\, , \quad [{\cal X}^i,{\cal X}^j]=0\, .
\end{equation}
This algebra is a subalgebra of $\sf{so}(4,1)$ and one can represent
it as an algebra of $5\times5$ real matrices, with the matrices
representing ${\cal X}^i$ being nilpotent.  Knowing the form of the
Lie algebra, one can readily write down a group element. It is
convenient to split it into the product of two elements, one
generated by nilpotent elements ${\cal X}^i$ and the second
generated by the abelian one ${\cal X}^0$,
\begin{equation}
{\sf{AN}}(3)\ni g(p) = \exp(ip_i {\cal X}^i)\,\exp(ip_0 {\cal X}^0)\ .
\label{j4}
\end{equation}
Clearly $p_\mu$ can be thought of as the coordinates on the group manifold.

Since ${\sf{AN}}(3)$ is a subgroup of de Sitter group
$\sf{SO}(4,1)$, $g(p)$ defined by (\ref{j4}) acts naturally on
points of the five dimensional Minkowski space $M^5$. Therefore, if
we take a point ${\cal O}$, the group ${\sf{AN}}(3)$ as a manifold
is just a set of all points of the form $g{\cal O}$. If ${\cal O}$
has coordinates $(0,\ldots,0,1/\ell)$ than the point $g(p){\cal O}$,
with $g(p)$ given by (\ref{j4}) and represented as a $5\times5$
matrix has Minkowski coordinates
 \begin{eqnarray}
 {P_0}(p_0, {p}_i) &=& \frac1\ell\, \sinh
{\ell{p_0}} - \frac{\ell p_{i}^2}{2}\,
e^{-\ell  {p_0}}\, , \nonumber\\
 P_i(p_0, {p}_i) &=&   p_i \, e^{-\ell  {p_0}}\, , \label{j2}\\
 {P_4}(p_0, {p}_i) &=& -\frac1\ell\, \cosh
{\ell{p_0}} +\frac{\ell p_{i}^2}{2}\, e^{-\ell {p_0}}\,
.\nonumber
\end{eqnarray}
One can easily check by direct computation that the coordinates $P_I
= (P_\mu, P_4)$, $ \mu = 0, \ldots, 3$ of these points satisfy the
conditions
\begin{equation}\label{j1}
    P_0^2 -P_1^2-P_2^2-P_3^2-P_4^2=-\frac1{\ell^2}\ ,
\end{equation}
and (assuming that $\ell$ is negative)
\begin{equation}\label{j1a}
P_0+P_4>0\, .
\end{equation}
Thus, as a manifold, the ${\sf{AN}}(3)$ group is an open subset of
the four dimensional de Sitter space (\ref{j1}) defined by the
condition (\ref{j1a}), and the points in this manifold can be
parametrized by coordinates $p_\mu$. It is worth noticing in passing
that the unit element of the group ${\sf{AN}}(3)$, $g(0)$ naturally
corresponds to the zero momentum point $p_\mu=0$ of the momentum
space, whose existence is required for the relative locality
construction \cite{prl}.

Since our momentum space, ${\sf{AN}}(3)$, is defined as a
hyper-surface imbedded in the five dimensional Minkowski space it
possesses a natural induced metric, which can be obtained by
inserting the relations (\ref{j2}) into the five dimensional
Minkowski metric
$$
ds^2=dP_0^2 -dP_1^2-dP_2^2-dP_3^2-dP_4^2\, .
$$
Using (\ref{j1}) one finds that this metric is nothing but the de
Sitter metric in flat coordinates
\begin{equation}\label{j1b}
    ds^2=  dp_0^2-e^{-2\ell p_0}\left(dp_1^2+dp_2^2+dp_3^2\right)\,
    .
\end{equation}
We shall use this form of the metric in the next section, in the derivation
 of the on-shell relation of a particle on the $\kappa$-momentum
space.

Since our momentum space is a group manifold it is natural to assume
that the momentum composition and is defined by the group
multiplication law. If we have two group elements $g(p)$ and $g(q)$
 then their product is a group element
itself so that we can define the momentum composition $\oplus$ as
follows:
\begin{equation}\label{j15}
    g(p)\, g(q) = g(p\oplus q) \ .
\end{equation}
It is worth stressing that since the group multiplication is
associative, the composition $\oplus$ is associative as well.

In the case of the $\sf{AN}(3)$ group elements defined by (\ref{j4})
we find
\begin{equation}\label{j16}
 g(p)\, g(q) =\exp\left(i{\cal X}^i\,(p_i +e^{\ell p_0}\, q_i)\right)\exp\left(i{\cal X}^0\,(p_0 + q_0)\right) \ ,
\end{equation}
so that
\begin{equation}\label{j17}
    (p\oplus q)_i =p_i +e^{\ell p_0}\, q_i\, , \quad (p\oplus q)_0 =p_0 + q_0 \ ,
\end{equation}
which to the leading order in $\ell$ reads
\begin{equation}\label{j17a}
    (p\oplus q)_i =p_i +q_i +\ell\, p_0\, q_i +O(\ell^2)\, , \quad (p\oplus q)_0 =p_0 + q_0 \ .
\end{equation}

We can then introduce $\ominus p$, the ``antipode" of $p$,
using the fact that the inverse of a group element is a group element
itself:
\begin{equation}\label{j18}
    g^{-1}(p)=g(\ominus p)\, , \quad g^{-1}(p)g(p)=1 \Leftrightarrow p\oplus(\ominus p)=0
\end{equation}
and in the case of the $\sf{AN}(3)$ group we find
\begin{equation}\label{j19}
    (\ominus p)_i = -e^{-\ell p_0}\, p_i\, , \quad(\ominus p)_0 = -
    p_0
\end{equation}
and in the leading order we have
\begin{equation}\label{j20}
    (\ominus p)_i = -\left(1-\ell p_0\right)\, p_i+O(\ell^2)\, , \quad(\ominus p)_0 = -
    p_0 \ .
\end{equation}

In closing this subsection let us also observe that when
the momentum space is a Lie group there is a natural way to
construct a free particle action. The idea is to identify the
position space with a linear space dual to the Lie algebra (as a
vector space) and to make use of the canonical pairing between these
dual spaces. Concretely let us define the basis of the vector space
${\cal Y}_\mu$ dual to the Lie algebra $\sf{an}(3)$ as follows:
\begin{equation}\label{j27}
\left<{\cal Y}_\mu, {\cal X}^\nu\right> =\delta^\nu_\mu \ .
\end{equation}
And let us take the space dual to the Lie algebra of $\sf{AN}(3)$ to be
the space of positions so that
\begin{equation}\label{j28}
    x = x^\mu\, {\cal Y}_\mu \ .
\end{equation}
Then the kinetic term of the action of a particle with $\sf{AN}(3)$
momentum space is\footnote{In the mathematical literature the
symplectic form associated with this kinetic term is called Kirillov
symplectic form.}
\begin{equation}\label{j29}
L^{kin}\equiv  -  \left< x, g^{-1} \frac{d}{d\tau}\, g\right> \ .
\end{equation}
Substituting (\ref{j4}), (\ref{j27}), and  (\ref{j28}) into
(\ref{j29}) one easily finds that
\begin{equation}\label{j30}
L^{kin}=x^\mu \dot p_\mu - \ell\, p_i\, x^i\, \dot p_0 \ .
\end{equation}
It is worth noticing that the same procedure can be applied to the
standard case with flat momentum space, when the group associated
with momentum composition is just an abelian group $\sf{R}^4$ (in
our case we get the abelian limit when $\ell\rightarrow\infty$.)

It follows from (\ref{j30}) that positions variables $x^\mu$ have a
nontrivial Poisson bracket. To see this most easily, notice that
with the help of the transformation
\begin{equation}\label{j31}
    x^\mu \rightarrow \bar x^\mu\, , \quad \bar x^0 = x^0
-\ell\, p_i\, x^i\, , \quad\bar x^i = x^i \ ,
\end{equation}
  one can diagonalize the kinetic
Lagrangian (\ref{j30}), $\bar L^{kin} =\bar x^\mu \dot p_\mu$, so
that the Poisson brackets in these new variables read
$$
\{\bar x^\mu, \bar x^\nu\}=0\, , \quad \{\bar x^\mu, p_\nu\}=
\delta^\mu_\nu\, .
$$
Using (\ref{j31}) one easily finds that
\begin{equation}\label{j32}
    \{x^0, x^i\} =-\ell\,  x^i\, , \quad \{x^i, x^j\} = 0 \ ,
\end{equation}
\begin{equation}\label{j33}
    \{x^0, p_0\} =1\, , \quad \{x^i, p_j\} = \delta^i_j\, , \quad \{x^0,
    p_j\}=\ell\, p_j \, .
\end{equation}

\subsection{$\kappa$-momentum space, the $\kappa$-Poincar\'e Hopf algebra
and $\kappa$-Minkowski spacetime}\label{kappaMtotherightSUBDIR}
\def\hx{\hat x}

The characterization of the $\kappa$-momentum space given in the previous subsection
is ideally suited for the purposes of our relative-locality studies,
but it leaves partly implicit the connection
with the $\kappa$-Poincar\'e Hopf algebra and its most popular applications
in the study of spacetime noncommutativity.
In order to expose more clearly this connection
we shall now rederive the characterization
of $\kappa$-momentum space already given in the previous subsection
taking as starting point the role of the $\kappa$-Poincar\'e Hopf algebra
 in the study of
the $\kappa$-Minkowski
noncommutative spacetime\cite{lukieIW,majidruegg,kpoinap}.

Once again the selection of results we mention is due to the
fact that the structure of the relative-locality framework of
Refs.~\cite{prl,grf2nd}, which we adopt,
 essentially requires some ``inspiration"
for an on-shell relation, a law of conservation of momentum
at interactions, and some Poisson brackets.
In this subsection we shall find this ``inspiration"
by revisiting  the most studied formulation
of theories in $\kappa$-Minkowski spacetime, often labeled
as ``bicrossproduct basis"~\cite{majidruegg,kpoinap}
 or ``time-to-the-right basis"~\cite{majidruegg,gacAlessandraFrancesco}.

 Starting from $\kappa$-Minkowski
 noncommutativity~\cite{majidruegg,kpoinap}, which can be thought of
 as the quantization of the Poisson brackets (\ref{j33})
  \begin{equation}
\left[ \hx^{j},\hx^{0} \right] =  i\ell \hx^{j} ~ ,~~~~\left[
\hx^{j},\hx^{k}\right] = 0 ~ , \label{kappadefCOMMUT}
\end{equation}
in this formulation one introduces the Fourier transform $\tilde{\Phi}(k)$
of a given $\kappa$-Minkowski field $\Phi(x)$ using the
time-to-the-right convention
$$\Phi(\hx) = \int d^4x\tilde{\Phi}(k)e^{ik_j\hx^j}e^{ik_0\hx^0}~. $$
This is often equivalently described in terms of the
time-to-the-right Weyl map ${\cal W}_R$ by writing that
$$\Phi(\hx) = {\cal W}_R \left(\int d^4x\tilde{\Phi}(k)e^{ik_\mu x^\mu} \right) \ ,$$
where it is intended that coordinates trivially commute when placed
inside the Weyl map, ${\cal W}_R (x^{j} x^{0}) = {\cal W}_R (x^{0}
x^{j})$, and that taking a function out of the time-to-the-right
Weyl map implies~\cite{gacAlessandraFrancesco} time-to-the-right
ordering, so that for example $\hx^{j} \hx^{0} = {\cal W}_R (x^{j}
x^{0})$, $\hx^{j} \hx^{0} = {\cal W}_R (x^{0} x^{j})$, and
$e^{ik_j\hx^j}e^{ik_0\hx^0} = {\cal W}_R (e^{ik_\mu x^\mu})$.

Then several arguments~\cite{majidruegg,kpoinap}, including the ones
based on the  recently-developed techniques of Noether analysis~\cite{kappanoether,freidkowaNOETHER,nopure},
lead one to find generators of symmetries under
translations, space-rotations and boosts.
For translations one has that
  \begin{equation}
\!P_\mu e^{ik_j\hx^j}e^{ik_0\hx^0} = k_\mu
e^{ik_j\hx^j}e^{ik_0\hx^0}= {\cal W}_R \left( - i \partial_\mu
e^{ik_\nu x^\nu} \right) \label{traslCOMMUT}
\end{equation}
and in general $P_\mu {\cal W}_R \left( f(x) \right) =
{\cal W}_R \left( - i \partial_\mu f(x) \right)$.\\
Similarly one has that the generators of space rotations
are given by
$$R_l e^{ik_j\hx^j}e^{ik_0\hx^0} = \epsilon_{lmn} \hx^m k_n e^{ik_j\hx^j}e^{ik_0\hx^0} $$
and the generators of boosts are given by
$$\! \mathcal{N}_l e^{ik_j\hx^j}e^{ik_0\hx^0} \! = \! \left[ -\hx^{0} k_l \!
 + \! \hx^{l} \left( \frac{1 \! - \! e^{2\ell k_0}}{2\ell}  \!
  + \! \frac{\ell}{2} k_m k_m \right) \right]e^{ik_j\hx^j}e^{ik_0\hx^0} \ .$$

These translations, rotations and boosts are
found to be generators of the $\kappa$-Poincar\'e Hopf algebra,
and their main properties are described, {\it e.g.}, in
Refs.~\cite{majidruegg,kpoinap,gacAlessandraFrancesco}.
In particular one finds a deformed mass Casimir $\mathcal{C}_l$,
obtained from the generators given above
$$\mathcal{C}_l = \left(\frac{2}{\ell}\right)^2 \sinh^2 \left( \frac{\ell}{2} P_0 \right) - e^{-\ell P_0}P_jP_j ~,$$
which can inspire a deformed on-shell relation for relativistic particles.

We also note the following properties of the translation generators
\begin{equation*}
\begin{split}
[P_l, \hx^m] e^{ik_j\hx^j}e^{ik_0\hx^0} &= -i\delta_{l}^{m} e^{ik_j\hx^j}e^{ik_0\hx^0} ,\\
[P_0, \hx^{0}] e^{ik_j\hx^j}e^{ik_0\hx^0} &= -ie^{ik_j\hx^j}e^{ik_0\hx^0} \ ,\\
[P_0, \hx^{l}] e^{ik_j\hx^j}e^{ik_0\hx^0} &= 0 \ ,\\
[P_l, \hx^{0}] e^{ik_j\hx^j}e^{ik_0\hx^0} &= -i\ell k_l
e^{ik_j\hx^j}e^{ik_0\hx^0}
\end{split}
\end{equation*}
which should be compared to (\ref{j33}) of the previous subsection.

And the composition law which we derived in the previous subsection
from the multiplication law on the group $\sf{AN}(3)$, is viewed in
the spacetime-noncommutativity literature as a property
 of products of ``time-to-the-right plane waves",
$$e^{ik_j\hx^j}e^{ik_0\hx^0} e^{ip_j\hx^j}e^{ip_0\hx^0} = e^{ik_j\hx^j} e^{ie^{\ell k_{0}}p_j\hx^j}e^{ik_0\hx^0}e^{ip_0\hx^0}
=e^{i(k_j+e^{\ell k_{0}}p_j)\hx^j}e^{i(k_0+p_0)\hx^0} ~,$$ which
indeed leads us once again to
$$(k \oplus p)_0 = k_{0}+p_{0} \ ,$$
$$(k \oplus p)_j = k_{j} +e^{\ell k_{0}} p_{j} \ .$$
Since this composition law can be derived
within the time-to-the-right formulation
of the $\kappa$-Poincar\'e/$\kappa$-Minkowski framework, which first appeared
in Ref.~\cite{majidruegg} by Majid and Ruegg,
 we shall refer to this composition law as the ``Majid-Ruegg composition law"
 and to the associated affine connection on momentum space as the
``Majid-Ruegg connection".

\section{The example of $\kappa$-Poincar\'e-inspired momentum space
with Majid-Ruegg connection}\label{geometry}
Ref.~\cite{prl}  introduced the idea of using the geodesic distance from the origin to a generic point $p_{\mu}$ in momentum space $\mathcal{P}$ as the mass of a particle.
We shall here argue that according to this proposal
one should view the $\kappa$-Poincar\'e/$\kappa$-Minkowski framework
as a case in which the metric on momentum space is de-Sitter like,
\begin{equation}
g^{\mu\nu}=
\left(
\begin{array}{cccc}
1 & 0 & 0& 0\\
0 & -e^{-2\ell p_{0}} & 0& 0\\
0 & 0 & -e^{-2\ell p_{0}}& 0\\
0 & 0 & 0& -e^{-2\ell p_{0}}\\
\end{array}
\right) \ ,
\label{gdesitter}
\end{equation}
and, as already anticipated, parallel transport is given in terms of
the Majid-Ruegg connection.

The other objective of this section is to establish the torsion
and nonmetricity of this $\kappa$-Poincar\'e-inspired setup.

\subsection{Distance from the origin in a de Sitter momentum space}
In order to calculate the geodesic distance from the origin to a generic point $p_{\mu}=(p_{0},p_{j})$ in momentum space $\mathcal{P}$ we must find
\begin{equation}
D(0,p_{\mu})=\int_{0}^{1} ds \sqrt{g^{\mu\nu}\dot p_{\mu}\dot p_{\nu}} \ ,
\end{equation}
where $p_{\mu}$ is the solution of the geodesic equation
\begin{equation}\label{Geodesic eq.}
\begin{split}
\ddot p_\rho +C_{\rho}^{\mu\nu} \dot p_{\mu}\dot p_{\nu}&=0
\end{split}\ ,
\end{equation}
 $g_{\mu\nu}$ is the metric of $\mathcal{P}$ and $C_{\rho}^{\mu\nu}$ are the Christoffel symbols for the metric $g_{\mu\nu}$.\\

To find an approximate solution consider the metric slightly away from zero, which has the form
\begin{equation}\label{1}
    g^{\mu\nu}=\eta^{\mu\nu} +\ell\, h^{\mu\nu\rho}\, p_\rho + \ldots
\end{equation}
A simple calculation of the Christoffel symbols  to the leading order,
\begin{equation}\label{2}
    C_\rho^{\mu\nu}=\frac12\,
    g_{\rho\sigma}\left(g^{\sigma\mu,\nu}+g^{\nu\sigma,\mu}-g^{\mu\nu,\sigma}\right)=
    \frac\ell2\,\eta_{\rho\sigma}\left(h^{\sigma\mu\nu}+h^{\nu\sigma\mu}-h^{\mu\nu\sigma}\right) \ ,
\end{equation}
shows that the only non vanishing components are:
\begin{equation}\label{Non-zero Gamma}
\begin{split}
C_{0}^{ij}&=-\ell e^{-2\ell p_0} \delta_{ij} \ ,\\
C_{i}^{j0}=C_{i}^{0j}&=-\ell \delta_{ij} \ ,
\end{split}
\end{equation}
so that the  geodesic equation  (\ref{Geodesic eq.}) can be easily solved perturbatively with the boundary conditions
\begin{equation}\label{4}
    p_\mu(0)=0\, , \quad p_\mu(1)=P_\mu \ .
\end{equation}

The solution at leading order is

\begin{equation}\label{geosolution}
    p_\rho(s)=P_\rho\, s+\frac12\,C_\rho{}^{\mu\nu}\, P_\mu\, P_\nu
(s-s^2)
\end{equation}
and
\begin{equation}\label{8}
    \dot p_\rho(s)=P_\rho+\frac12\,C_\rho{}^{\mu\nu}\, P_\mu\, P_\nu
(1-2s) \ .
\end{equation}
To compute the distance one must find
\begin{equation}\label{9}
    \sqrt{g^{\mu\nu}\dot p_\mu(s) \dot p_\nu(s)}=\sqrt{\eta^{\mu\nu}\, P_\mu\,
    P_\nu+C^{\rho\mu\nu}\, P_\rho\, P_\mu\, P_\nu(1-2s) + \ell h^{\mu\nu\rho}\, P_\rho\, P_\mu\,
    P_\nu\, s} \ .
\end{equation}
To do that we use the identity that results from eq. (\ref{2})
\begin{equation}\label{10}
C^{\rho\mu\nu}\, P_\rho\, P_\mu\,
P_\nu=\frac\ell2\,h^{\rho\mu\nu}\, P_\rho\, P_\mu\,
    P_\nu \ .
\end{equation}
So that finally we find
\begin{equation}\label{11}
    \sqrt{g^{\mu\nu}\dot p_\mu(s)\dot p_\nu(s)}=\sqrt{P^2+C^{\rho\mu\nu}\, P_\rho\, P_\mu\, P_\nu} \ .
\end{equation}
Integrating this from 0 to 1 and taking the square we get the final
result
\begin{equation}
    D(0,P_{\mu})=m^2=P^2 +\,C^{\rho\mu\nu}\, P_\rho\, P_\mu\,
    P_\nu \ .
    \label{geodistance}
\end{equation}

Substituting the values of the connections found in eq.\ (\ref{Non-zero Gamma}) we have
\begin{equation} \label{geomass}
    m^2=P_0^2-P_i^2 +\ell P_0 P_i^2 \ ,
\end{equation}
consistently with the leading-order form of the $\kappa$-Poincar\'e inspired
on-shell relation.

\subsection{Momentum space with de Sitter metric and Majid-Ruegg connection: torsion and (non)metricity}

To further investigate the geometrical properties of momentum space we
take the Majid-Ruegg composition law:
\begin{gather*}
(p \oplus q)_0 = p_{0}+q_{0} \ , \\
(k \oplus p)_j = p_{j} +e^{\ell p_{0}} q_{j} \ .
\end{gather*}

Using the Majid-Ruegg composition law, we can define a parallel transport on the momentum space $\mathcal{P}$ as
\begin{equation}
\begin{split}
(p \oplus dq)_\mu = p_\mu + dq_\mu - \Gamma_{\mu}^{\alpha \beta} p_\alpha dq_\beta+ \dots
\end{split}
\end{equation}

In particular for the (leading order) composition law
\begin{equation}
\begin{split}
 (p \oplus q)_0 &= p_0 + q_0 \ ,\\
  (p \oplus q)_i &= p_i + q_i +\ell p_0 q_i \ ,
\end{split}
\end{equation}

we find that the only  non-vanishing components of the connection are:
\begin{equation}
\Gamma_{i}^{0j} =-\ell \delta_{i}^{j} \ .
\end{equation}

Given the components of the connection we can easily find the other geometric properties as torsion, nonmetricity and curvature.

For the torsion, we use~\cite{prl}
\begin{equation}
T_{\mu}^{\alpha \beta}=-\frac{\partial }{\partial p_{\alpha}}\frac{\partial }{\partial q_{\beta}}\left( p\oplus q-q \oplus p\right)_{\mu}=2\Gamma_{\mu}^{[\alpha \beta]}=\Gamma_{\mu}^{\alpha \beta}-\Gamma_{\mu}^{ \beta\alpha}
\end{equation}
to find that at leading order
the only non vanishing components of the torsion tensor are
\begin{equation}
T_{i}^{0j}=-T_{i}^{j0}=\Gamma_{i}^{0j}=-\ell \delta_{i}^{j} \ .
\end{equation}

And for the nonmetricity tensor,
\begin{equation}
N^{\alpha \mu \nu}=\nabla^{\alpha} g^{\mu \nu}= g^{\mu \nu,\alpha}+\Gamma_{\beta}^{\mu \alpha}g^{\beta \nu}+\Gamma_{\beta}^{\nu \alpha}g^{\mu \beta}~,
\end{equation}
the only non-vanishing components to the leading order are
\begin{equation}
\begin{split}
N^{0 i j}&= 2\ell \delta^{ij} \ ,\\
N^{i 0 j}&= \ell \delta^{ij} \ ,\\
N^{i j 0}&= \ell \delta^{ij} \ .\\
\end{split}
\end{equation}

For what concerns the curvature of the connection, determined by~\cite{prl}
\begin{equation}
R^{\alpha \beta \gamma}_{\mu}=2\frac{\partial }{\partial p_{[\alpha}}\frac{\partial }{\partial q_{\beta]}}\frac{\partial }{\partial r_{\gamma}} \biggl((p \oplus q) \oplus r-p \oplus (q \oplus r)\biggr)_{\mu} \ ,
\end{equation}
it is evident that it vanishes by construction in any leading-order analysis
(in a power series in $\ell$ the first contribution to the curvature of the connection
is of order $\ell^2$).
It is worth noticing that in the case of the Majid-Ruegg connection
this curvature vanish exactly (to all orders) as a result of the fact that
the Majid-Ruegg composition law is associative.

\section{Partial anatomy of distant relative-locality observers}\label{setup}

\subsection{A starting point for the description of distant relative-locality
observers}\label{oksofarry}
Let us now return to the preliminary results on translation invariance
reported in Ref.~\cite{prl}, which we already briefly summarized in the
first section, but we shall now analyze in greater detail.
In Ref.~\cite{prl} translation invariance
was explicitly checked only for the
idealized case of the process
we already showed in Figure~\ref{oksofarfig},
with 3 particles of energy-momenta $k_\mu$, $p_\mu$, $q_\mu$
all incoming into the interaction.

Let us note down again here the action $\mathcal{S}^{example}$ which,
according to Ref.~\cite{prl},
could describe the process in  Figure~\ref{oksofarfig}:
\begin{equation}
\begin{split}
\mathcal{S}^{example}= & \int_{-\infty}^{s_{0}}ds\left(x^{\mu}\dot{k}_{\mu}+y^{\mu}\dot{p}_{\mu}
+z^{\mu}\dot{q}_{\mu}+\mathcal{N}_k\mathcal{C}\left[k\right]
+\mathcal{N}_p\mathcal{C}\left[p\right]+\mathcal{N}_q\mathcal{C}\left[q\right]\right) -\xi^{\mu}\mathcal{K}_\mu(s_{0})
\end{split} \ ,
\label{action}
\end{equation}
where ${\cal C}[k]= D^2(k)-m^2$ is the distance of $k_\mu$ from the origin
of momentum space, and the on-shell condition
is ${\cal C}[k]=0$,
while the deformed law of energy-momentum conservation
has been enforced by first introducing a connection-induced composition of
the momenta,
$$\mathcal{K}_\mu(s) \equiv [k(s) \oplus p(s) \oplus q(s)]_\mu \ ,$$
and then adding to the action a boundary term (in this case, at the $s=s_0$ boundary)
with this $\mathcal{K}_\mu$.
The lagrange multipliers enforcing $\mathcal{K}_\mu = 0$ are denoted
by $\xi^{\mu}$ and play the role of ``interaction coordinates"
in the sense of  Ref.~\cite{prl}.
This is a theory on momentum space in the sense that
the ``particle coordinates" $x^{\mu}$, $y^{\mu}$,$z^{\mu}$
are introduced as ``conjugate momenta of the momenta", and for the
action $\mathcal{S}^{example}$ one evidently has that
\begin{gather}
\{x^{\mu} , k_{\nu} \} = \delta^{\mu}_{\nu} \ ,\nonumber\\
\{y^{\mu} , p_{\nu} \} = \delta^{\mu}_{\nu} \ ,\label{canonical}\\
\{z^{\mu} , q_{\nu} \} = \delta^{\mu}_{\nu}\ .\nonumber
\end{gather}

Following again Ref.~\cite{prl} we vary the action $\mathcal{S}^{example}$
keeping the momenta fixed at $s = \pm \infty$
(so that, for the case we are here considering,
one has that $\delta k_\mu \Big|_{s=-\infty} = 0$, $\delta p_\mu \Big|_{s=-\infty} = 0$,
$\delta q_\mu \Big|_{s=-\infty} = 0$)
and we find the equations of motion
\begin{gather*}
\dot k_\mu =0~,~~\dot p_\mu =0~,~~\dot q_\mu =0~,~~\\
\mathcal{C}[k]=0~,~~\mathcal{C}[p]=0~,~~\mathcal{C}[q]=0~,~~\\
\mathcal{K}_\mu=0,\\
\dot x^\mu = \mathcal{N}_k \frac{\delta \mathcal{C}[k]}{\delta k_\mu}~,~~~
\dot y^\mu = \mathcal{N}_p \frac{\delta \mathcal{C}[p]}{\delta p_\mu}~,~~~
\dot z^\mu = \mathcal{N}_q \frac{\delta \mathcal{C}[q]}{\delta q_\mu} \ ,
\end{gather*}
and the boundary conditions at the endpoints of the 3 semi-infinite worldlines
\begin{equation}
x^\mu(s_{0}) = \xi^\nu \frac{\delta \mathcal{K}_\nu}{\delta k_\mu}~,~~~
y^\mu(s_{0}) = \xi^\nu \frac{\delta \mathcal{K}_\nu}{\delta p_\mu}~,~~~
z^\mu(s_{0}) = \xi^\nu \frac{\delta \mathcal{K}_\nu}{\delta q_\mu} \ .
\end{equation}

The relative locality is codified in the fact that
for configurations with $\xi^\mu =0$ the endpoints of the worldlines
must coincide and be located in the origin of the observer
($x^\mu(s_{0}) = y^\mu(s_{0}) = z^\mu(s_{0})=0$),
but for configurations such that $\xi ^\mu \neq 0$
the endpoints of the worldlines do not coincide, since in general
\begin{equation}
\frac{\delta \mathcal{K}_\nu}{\delta k_\mu} \neq \frac{\delta \mathcal{K}_\nu}{\delta p_\mu} \neq \frac{\delta \mathcal{K}_\nu}{\delta q_\mu} \ ,
\end{equation}
so that in the coordinatization of the (in that case, distant)
observer the interaction appears to be nonlocal.

As noticed in Ref.~\cite{prl}, taking as starting point
of the analysis some observer
Alice for whom\footnote{When we compare two observers, Alice and Bob, we
 shall consistently use indices $A$ and $B$ to distinguish between quantities
 determined from one or the other. In particular, here
 we denote with $\xi^\mu_{A}$  the conservation-law
Lagrange multipliers of observer Alice and with $\xi^\mu_{B}$
 the corresponding Lagrange multipliers of observer Bob.}
  $\xi_{A}^\mu \neq 0$, {\it i.e.} an observer distant from
the interaction who sees the interaction as nonlocal,
one can obtain from Alice an observer
Bob for whom $\xi_{B}^\mu =0$
if the transformation from Alice to Bob
for endpoints of coordinates
has the form
\begin{equation}
\begin{split}
x^\mu_{B}(s_{0}) &= x^\mu_{A}(s_{0})
- \xi_{A}^\nu \frac{\delta \mathcal{K}_\nu(s)}{\delta k_\mu(s)}\Big|_{s=s_0} \ ,
\\
y^\mu_{B}(s_{0}) &= y^\mu_{A}(s_{0})
- \xi_{A}^\nu \frac{\delta \mathcal{K}_\nu(s)}{\delta p_\mu(s)}\Big|_{s=s_0} \ ,\\
z^\mu_{B}(s_{0}) &= z^\mu_{A} (s_{0})
- \xi_{A}^\nu \frac{\delta \mathcal{K}_\nu(s)}{\delta q_\mu(s)}\Big|_{s=s_0} \ .
\end{split}
\end{equation}
Such a property for the endpoint is produced of course,
for the choice $b^\nu = \xi_{A}^\nu$,
by the following corresponding prescription for the
 translation transformations:
\begin{equation}
\begin{split}
x^\mu_{B}(s) &= x^\mu_{A}(s) -b^\nu \frac{\delta \mathcal{K}_\nu(s)}{\delta k_\mu(s)} \ ,\\
y^\mu_{B}(s) &= y^\mu_{A}(s) -b^\nu \frac{\delta \mathcal{K}_\nu(s)}{\delta p_\mu(s)} \ ,\\
z^\mu_{B}(s) &= z^\mu_{A}(s) -b^\nu \frac{\delta \mathcal{K}_\nu(s)}{\delta q_\mu(s)}\ ,\\
\label{traslprl}
\xi_{B}^\mu &= \xi_{A}^\mu -b^\mu\ .
\end{split}
\end{equation}
Indeed one finds by direct substitution that these transformations leave
the equations of motion and the boundary conditions unchanged.
And also the action is invariant; indeed
\begin{equation}
\begin{split}
\mathcal{S}_B^{example}&= \int_{-\infty}^{s_{0}}ds\left(x^\mu_{B} \dot{k}_{\mu}+y^\mu_{B} \dot{p}_{\mu}+z^\mu_{B} \dot{q}_{\mu}+\mathcal{N}_k\mathcal{C}\left[k\right]+\mathcal{N}_p\mathcal{C}\left[p\right]+\mathcal{N}_q\mathcal{C}\left[q\right]\right) -\xi^\mu_{B}\mathcal{K}_\mu(s_{0})\\
&=  \int_{-\infty}^{s_{0}}ds\left(\left(x^\mu_{A} -b^\nu \frac{\delta \mathcal{K}_\nu}{\delta k_\mu}\right)\dot{k}_{\mu}+\left(y^\mu_{A} -b^\nu \frac{\delta \mathcal{K}_\nu}{\delta p_\mu}\right) \dot{p}_{\mu}+\left(z^\mu_{A} -b^\nu \frac{\delta \mathcal{K}_\nu}{\delta q_\mu}\right)\dot{q}_{\mu}+\mathcal{N}_k\mathcal{C}\left[k\right]+\mathcal{N}_p\mathcal{C}\left[p\right]+\mathcal{N}_q\mathcal{C}\left[q\right]\right) -\xi^\mu_{B}\mathcal{K}_\mu(s_{0})\\
&=\mathcal{S}_A^{example,bulk}-\int_{-\infty}^{s_{0}}ds\frac{d}{ds}\left(b^\nu \mathcal{K}_\nu\right)-\xi^\mu_{B}\mathcal{K}_\mu(s_{0})\\
&=\mathcal{S}_A^{example,bulk}-(\xi^\mu_{B}+b^\mu)\mathcal{K}_\mu(s_{0})\\
&=\mathcal{S}_A^{example,bulk}-\xi^\mu_{A}\mathcal{K}_\mu(s_{0})\\
&=\mathcal{S}_A^{example}\ ,
\end{split}
\end{equation}
where $\mathcal{S}_A^{example,bulk}$ coincides with $\mathcal{S}_A^{example}$
with the exception of boundary terms.

This also shows that all interactions
are local according to nearby observers
(observers themselves local to the interaction):
if $\xi^\mu_{A} \neq 0$ for observer Alice,
so that in Alice's coordinates the interaction is distant and nonlocal,
one easily finds a observer Bob for whom $\xi^\mu_{B} = 0$,
an observer local to the interaction who witnesses the interaction
as a sharply local interaction in its origin.

For the purposes of the proposal we shall put forward
in the following sections, it is important to notice here
that these observations reported in Ref.~\cite{prl}
actually can be viewed as a prescription for translations generated
by the ``total momentum" $\mathcal{K}_\mu$ (in which however individual momenta
are summed with a nonlinear composition law). In fact,
 in light of (\ref{canonical})
the description of translation transformations given in (\ref{traslprl})
 simply gives
\begin{equation}
\begin{split}
\delta x^\mu_{B}(s) &= x^\mu_{B}(s)  - x^\mu_{A}(s)
= b^\nu \{(k \oplus p \oplus q)_\nu, x^\mu \}
=  b^\nu \{\mathcal{K}_{~\nu}, x^\mu \} =
 - b^\nu \frac{\delta \mathcal{K}_{~\nu}(s)}{\delta k_\mu(s)}\ ,\\
\delta y^\mu_{B}(s) &= y^\mu_{B}(s)  - y^\mu_{A}(s)
= b^\nu \{ (k \oplus p \oplus q)_\nu , y^\mu \}
= b^\nu \{\mathcal{K}_{~\nu} , y^\mu \} =
 -b^\nu \frac{\delta \mathcal{K}_{~\nu}(s)}{\delta p_\mu(s)}\ ,\\
\delta z^\mu_{B}(s) &= z^\mu_{B}(s)  - z^\mu_{A}(s)
=  b^\nu \{ (k \oplus p \oplus q)_\nu , z^\mu \}
=  b^\nu \{\mathcal{K}_{~\nu} , z^\mu \} =
 -b^\nu \frac{\delta \mathcal{K}_{~\nu}(s)}{\delta q_\mu(s)}\ .
\end{split}
\label{traslaprl}
\end{equation}

\subsection{Some properties of our conservation laws}
Our next task is to focus on
another issue which also needs to be fully appreciated in order
to work with relative locality: the issue of ordering momenta in the
nonlinear composition law.

That ordering might be an issue is evident from the fact
that relative-locality momentum spaces can in general
allow~\cite{prl} for interactions characterized
by conservation laws which are possibly noncommutative (torsion)
and/or non-associative (curvature of the connection).
For leading-order analyses, of the type we are here motivating,
only noncommutativity is possible, but that is enough
to introduce quite some novelty with respect to standard absolute-locality
theories.
It should be noticed however that the number of
 truly  different conservation laws is much smaller than one might
naively imagine, as we shall now show for our
illustrative $\kappa$-Poincar\'e-inspired example (a generalization of the argument
shall be provided elsewhere~\cite{spectator}).

Let us first notice that while for arbitrary choices of $k$ and $p$
our composition law is evidently such that $k \oplus p \neq p \oplus k$
(noncommutativity), in the cases of interest when discussing interactions,
cases in which the composition of momenta is used to write a conservation
law, we actually do have
$$k \oplus p =0 \Longleftrightarrow  p \oplus k =0\ .$$
This is easily checked in the case which is of primary interest for us here:
$$0=k_1+p_1+ \ell k_0 p_1 = k_1+p_1+ \ell p_0 k_1\ ,$$
where on the right-hand-side we used in the leading-order correction
the properties $k_0 = - p_0$
and $k_1 = - p_1$ which follow (at zero-th order) from $k \oplus p =0$.

And actually $k \oplus p =0 \Longleftrightarrow  p \oplus k =0$ holds for
any choice~\cite{spectator} of affine connection on momentum space, as shown
by the following chain of properties:
$$k \oplus p =0 \Longrightarrow p = \ominus k \Longrightarrow
 p \oplus k = \ominus k \oplus k =0\ .$$

This observation also simplifies the description of 3-particle interactions.
In fact, since we have established
that $k \oplus p =0 \Longleftrightarrow  p \oplus k =0$
it then evidently follows that\footnote{Note that
from $k \oplus p =0 \Longleftrightarrow  p \oplus k =0$,
which holds for any choice of momentum-space affine connection
(and associated composition law),
it evidently follows
that $(k \oplus p) \oplus q =0 \Longleftrightarrow  q \oplus (k \oplus p) =0$
but unless the composition law is associative
this will not amount to a cyclicity property~\cite{spectator}.
When, as in the case which is here of our primary interest,
the composition law is associative
we then have $q \oplus (k \oplus p) =(q \oplus k) \oplus p = q \oplus k \oplus p $
and a genuine cyclicity property arises.}
$$k \oplus p \oplus q =0 \Longleftrightarrow  q \oplus k \oplus p =0\ .$$
So, while there is no cyclicity property of the rule of composition of generic momenta,
when the rule of composition is used for a conservation law it produces
a conservation law with cyclicity.

\subsection{Boundary terms and conservation of momenta}\label{choices}
So we have seen that the number of truly independent conservation
laws that can be postulated using the deformed composition law ``$\oplus$"
is smaller than one might have naively imagined, because of cyclicity.
For some of the observations we report later on in this manuscript
it is however important to appreciate that different compositions of momenta
that (when set to zero) would produce the same conservation law
still can lead to tangibly different choices of boundary terms
enforcing the conservation laws.

 Let us first illustrate the issue within the specific example
of an interaction with two incoming and one outgoing particle,
with conservation law
$$p \oplus k \oplus (\ominus q) = 0 \ .$$
This conservation law can be enforced
by adding to the action a term of the form $\xi^\mu {\cal K}_\mu$,
with${\cal K}_\mu = [p \oplus k \oplus (\ominus q)]_\mu$
and $\xi^\mu$ are Lagrange multipliers.
But this evidently is not the only choice of constraint term
that enforces the chosen conservation law. For example
let us observe that\footnote{This elementary chain of equivalence
may at first appear striking since
evidently in general $p \oplus k \oplus (\ominus q) \neq (p \oplus k ) - q $,
or, more precisely, if $p \oplus k \oplus (\ominus q) \neq 0$
then $p \oplus k \oplus (\ominus q) \neq (p \oplus k ) - q $.
However the chain of equivalences immediately follows upon observing that
in the special cases where $p \oplus k \oplus (\ominus q) = 0$
(conservation laws) one then has that also $(p \oplus k ) - q =0$.}
$$p \oplus k \oplus (\ominus q) = 0 ~ \Longleftrightarrow ~ p \oplus k = q
~ \Longleftrightarrow (p \oplus k ) - q =0\ ,
$$
and also that
$$p \oplus k \oplus (\ominus q) = 0 ~ \Longleftrightarrow ~ p \oplus k = q
~ \Longleftrightarrow \ominus q \oplus p \oplus k = \ominus q \oplus q=0\ .
$$
So we see that the same conservation law\footnote{It should be noticed
that in Ref.~\cite{gacdsr1}, where nonlinear conservation laws
were analyzed from the perspective of the doubly-special-relativity
research program, a possible role of
such conservation laws of the
type $p^{[in]} - (p^{[out,1]} \oplus p^{[out,2]}) =0$
or
$(p^{[in,1]} \oplus p^{[in,2]}) - (p^{[out,1]} \oplus p^{[out,2]}) =0$
was already motivated on different grounds.} can be enforced by adding a boundary
term of the form $\xi^\mu {\cal K}_\mu$ with ${\cal K}_\mu$
given by any choice among  ${\cal K}_\mu = [p \oplus k \oplus (\ominus q)]_\mu$,
 ${\cal K}_\mu = [(p \oplus k) - q]_\mu$, and
 ${\cal K}_\mu = [(\ominus q) \oplus p \oplus k  ]_\mu$,
However, it is easy to verify (and this will play a role in the analysis reported
in the following section) that these different possible choices of boundary terms
enforcing the same momentum-conservation law actually
produce boundary conditions that are physically different.

In the case of our interest, which is the case of the Majid-Ruegg connection,
we shall be confronted with the observation that
$$(p \oplus k \oplus (\ominus q))_0 = ((p \oplus k) - q)_0
=((\ominus q) \oplus p \oplus k )_0$$
but
$$((p \oplus k) - q)_1  = (p \oplus k \oplus (\ominus q))_1
+ \ell (q_0 - k_0 - q_0)q_1
= ((\ominus q) \oplus p \oplus k )_1 + \ell q_0 (q_1 - k_1 - q_1)~.$$
However, when $((p \oplus k) - q)_\mu =0$ one evidently also has\footnote{We stress
again that
the 3 conservation laws in question are exactly equivalent, equivalent
to all orders in $\ell$. We are however working here to leading order in $\ell$,
and for example the antipode $\ominus$ for the Majid-Ruegg connection
was here determined only to leading order. So the equivalence of the 3 conservation
laws in question is of course verified within our computations only
upon dropping subleading, $O(\ell^2)$, contributions.}.
(neglecting $O(\ell^2)$)
that $\ell (q_0 - k_0 - q_0)q_1 =0 = \ell q_0 (q_1 - k_1 - q_1)$,
so also for the specific case of the Majid-Ruegg connection
 one has this possibility of different boundary terms enforcing the same
 conservations laws, but producing physically-different boundary conditions.

\subsection{A challenge for spacetime-translation invariance in
theories on a relative-locality momentum space}\label{challenge}

We shall now characterize preliminarily the nature of some
consistency conditions that should be enforced in order to produce
a relativistic formulation with relative locality for interacting particles.
As emphasized at the beginning of this section, in the relative-locality
frameworks here of interest essentially what happens is that
the correct notion of translation to distant observers
must act  on the endpoints of worldlines in a way that reflects
the form of the boundary terms used to implement
 the conservation laws.
As also shown at the beginning of this section this notion is never
problematic for semi-infinite worldlines, with a single endpoint.
But we must now highlight a
 challenge which materializes in all instances where two interactions
are causally connected, {\it i.e.} there is a particle ``exchanged" between
the interactions, described by a finite worldline with two endpoints.
In those instances we are going to have
that the conservation laws essentially impose two conditions on the ``exchanged
worldline", for the translation of the two endpoints.
But we must request, for a relativistic description, that the worldline
of distant observer Bob is solution of the same equations of motion
that the initial observer Alice determines, and
 these relativistic demands are not automatically satisfied.

\begin{figure}[hb!]
\begin{center}
\includegraphics[width=0.56 \textwidth]{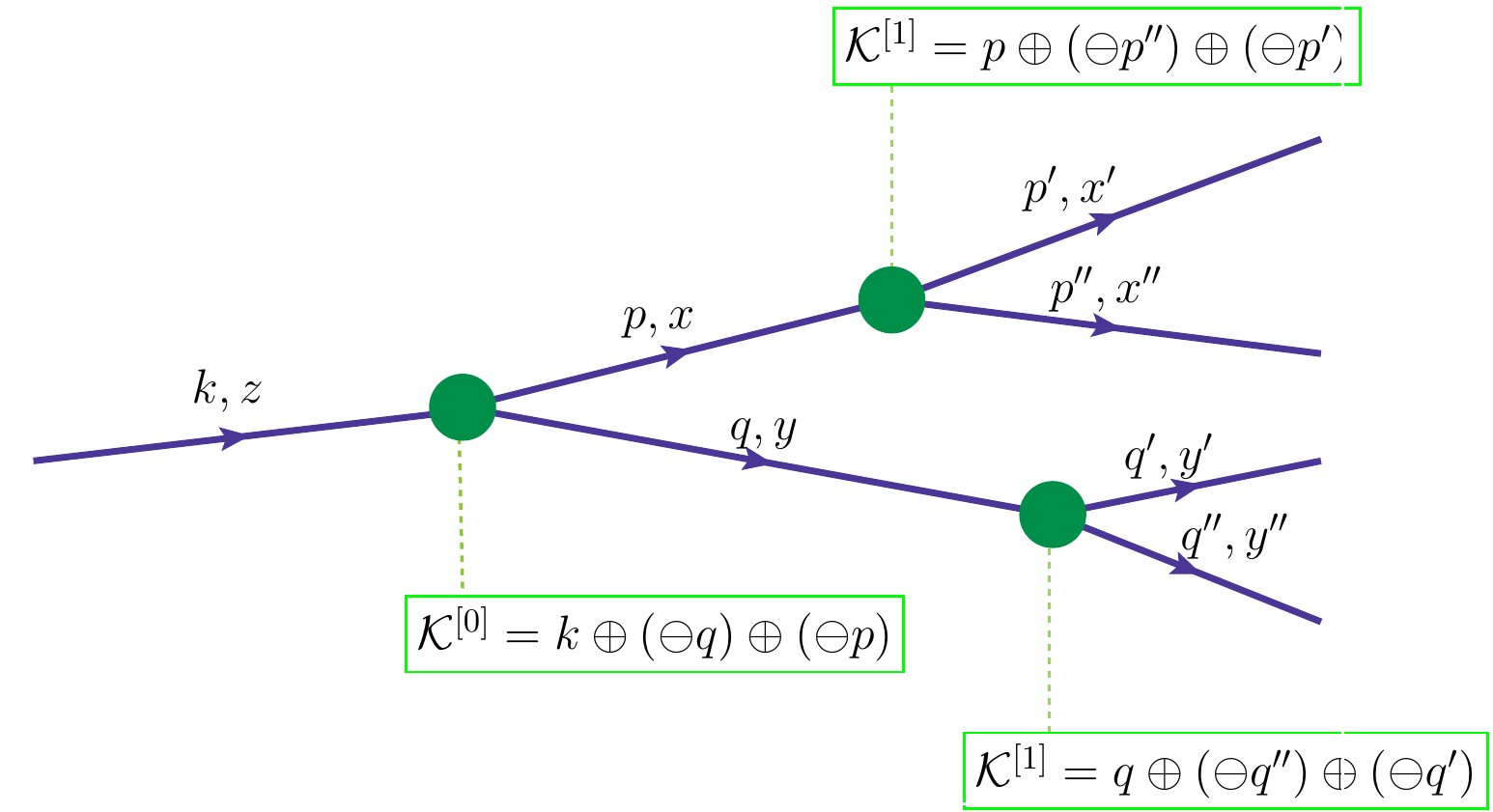}
\caption{The action given and analyzed in this subsection would
be intended for the description of the three causally-connected
interactions shown here. But it appears that such a description
is incompatible with a relativistic description of distant observers }
\label{notoksofar}
\end{center}
\end{figure}

In order to render our concerns more explicit let us consider
a specific example which does not admit the sort
of relativistic description we are here interested in.
For simplicity we consider a case in which
 the on-shell relation is undeformed
 and the symplectic structure is trivial.
 And we consider the situation shown in Figure~\ref{notoksofar},
 in which the two outgoing particles of a first decay
 themselves eventually decay.

A suitable description of the relevant conservation laws is the following:
\begin{equation}
\begin{split}&
0 = \left( k \oplus (\ominus q) \oplus (\ominus p) \right)_\mu = k_\mu -
p_\mu - q_\mu - \ell \delta_\mu^j \left( k_0 q_j - q_0 q_j + k_0 p_j -
p_0 p_j - q_0 p_j \right) \ ,\\&
0 = \left( p \oplus (\ominus p'') \oplus (\ominus p') \right)_\mu = p_
\mu - p'_\mu - p''_\mu - \ell \delta_\mu^j \left( p_0 p''_j - p''_0
p''_j + p_0 p'_j - p'_0 p'_j - p''_0 p'_j \right)\ , \\&
0 = \left( q \oplus (\ominus q'') \oplus (\ominus q') \right)_\mu = q_
\mu - q'_\mu - q''_\mu - \ell \delta_\mu^j \left( q_0 q''_j - q''_0
q''_j + q_0 q'_j - q'_0 q'_j - q''_0 q'_j \right)\ .
\end{split}
\end{equation}
where for definiteness (it is easy to check
that none of the points made in this subsection depend crucially on this choice)
we specified as composition law the one coming from the ``Majid-Ruegg connection".
Evidently the conservation laws concern a first interaction
where a particle of momentum $k$ decays into a particle
of momentum $p$ plus some other particle of momentum $q$,
followed by two more decays,
one where the particle of momentum $p$ decays into
particles of momentum $p'$ and $p"$ and one where
the particle of momentum $q$ decays into
particles of momentum $q'$ and $q"$.

The main observation we here want to convey is that
the following choice of ${\cal K}$'s to be used in writing up
constraints implementing the conservation laws
\begin{equation}\nonumber
\begin{split}
{\cal K}^{[0]} (s_{0})&= k \oplus (\ominus q)\oplus (\ominus p) \ ,\\
{\cal K}^{[1]} (s_{1})&= p \oplus (\ominus p'')\oplus (\ominus p') \ ,\\
{\cal K}^{[2]} (s_{2})&= q \oplus (\ominus q'')\oplus (\ominus q')\ .
\end{split}
\end{equation}
which appears to be a very natural way to implement the conservation
laws as constraints, does not lead to a relativistic description
of distant observers.

To see this let us first write the action which would implement all this:
\begin{equation}
\begin{split}
\mathcal{S}_{A}= & \int_{-\infty}^{0}ds\left(z^{\mu}\dot{k}_{\mu}+\mathcal{N}_k [k^2-m_a^2]\right)+
\int_{s_{0}}^{s_{1}}ds\left({x}^{\mu}\dot{p}_{\mu}+\mathcal{N}_p [p^2-m_b^2]
\right)\\
 & +\int_{s_{0}}^{s_{2}}ds\left(y^{\mu}\dot{q}_{\mu}+\mathcal{N}_q [q^2-m_c^2]
\right)
 +\int_{s_{1}}^{+\infty}ds\left({x'}^{\mu}\dot{p}'_{\mu}+\mathcal{N}_{p'} [p'^2-m_d^2]
 \right)+
 \int_{s_{1}}^{+\infty}ds\left(x''^{\mu}\dot{p}''_{\mu}+\mathcal{N}_{p''} [p''^2-m_e^2]
 \right)\\
  &+ \int_{s_{2}}^{+\infty}ds\left({y'}^{\mu}\dot{q}'_{\mu}+\mathcal{N}_{q'} [q'^2-m_f^2]
 \right)
 + \int_{s_{2}}^{+\infty}ds\left({y''}^{\mu}\dot{q}''_{\mu}+\mathcal{N}_{q''} [q''^2-m_g^2] \right)\\
 & -\xi_{[0]A}^{\mu}\mathcal{K}_{\mu}^{[0]}(s_{0})-\xi_{[1]A}^{\mu}\mathcal{K}_{\mu}^{[1]}(s_{1}) -\xi_{[2]A}^{\mu}\mathcal{K}_{\mu}^{[2]}(s_{2})\ ,
\end{split}
\end{equation}
where we restricted our focus on the undeformed on-shell condition $k^2-m^2=0$
and we allowed for the presence of particles of different mass.
The equations of motion that follow from varying this action evidently
are:
\begin{gather*}
\dot k_\mu =\dot p_\mu =\dot q_\mu =\dot{p'}_\mu =\dot{p''}_\mu =\dot{q'}_\mu =\dot{q''} =0\ ,\\
k^2-m_a^2=p^2-m_b^2=q^2-m_c^2=p'^2-m_d^2=p''^2-m_e^2=q'^2-m_f^2=q''^2-m_g^2=0\ ,\\
\mathcal{K}_\mu^{[0]}=0~,~~~\mathcal{K}_\mu^{[1]}=0~,~~~\mathcal{K}_\mu^{[2]}=0\ ,\\
\dot z^\mu =2 \mathcal{N}_k \delta^{\mu}_{0} k_0-2\mathcal{N}_k \delta^{\mu}_{1} k_1~,~~~\dot x^\mu = 2 \mathcal{N}_p \delta^{\mu}_{0} p_0-2\mathcal{N}_p \delta^{\mu}_{1} p_1~,~~~\dot x'^\mu = 2 \mathcal{N}_{p'} \delta^{\mu}_{0} p'_0-2\mathcal{N}_{p'} \delta^{\mu}_{1} p'_1~,~~~\dot x''^\mu = 2 \mathcal{N}_{p''} \delta^{\mu}_{0} p''_0-2\mathcal{N}_{p''} \delta^{\mu}_{1} p''_1\ ,\\
\dot y^\mu = 2 \mathcal{N}_q \delta^{\mu}_{0} q_0-2\mathcal{N}_q \delta^{\mu}_{1} q_1~,~~~\dot y'^\mu =2 \mathcal{N}_{q'} \delta^{\mu}_{0} q'_0-2\mathcal{N}_{q'} \delta^{\mu}_{1} q'_1~,~~~\dot y''^\mu = 2 \mathcal{N}_{q''} \delta^{\mu}_{0} q''_0-2\mathcal{N}_{q''} \delta^{\mu}_{1} q''_1\ .
\end{gather*}

And for
the boundary conditions at endpoints of worldlines one finds:
\begin{gather*}
z^{\mu}_{A}(s_{0})=  \xi_{[0]A}^{\nu}\frac{\delta{\mathcal{K}}^{[0]} _\nu}{\delta k_{\mu}}= \xi_{[0]A}^{\mu} - \ell \delta^{\mu}_{0}\xi_{[0]A}^{1} (q_{1}+p_{1})\ ,\\
x^{\mu}_{A}(s_{0})=  -\xi_{[0]A}^{\nu}\frac{\delta{\mathcal{K}}^{[0]} _\nu}{\delta p_{\mu}}= \xi_{[0]A}^{\mu} - \ell \delta^{\mu}_{0}\xi_{[0]A}^{1} p_{1}+ \ell \delta^{\mu}_{1}\xi_{[0]A}^{1} (k_{0}-q_{0}-p_{0})
 \ ,\qquad x^{\mu}_{A}(s_{1})=  \xi_{[1]A}^{\nu}\frac{\delta{\mathcal{K}}^{[1]} _\nu}{\delta p_{\mu}}=\xi_{[1]A}^{\mu} - \ell \delta^{\mu}_{0}\xi_{[0]A}^{1} (p_{1}'+p_{1}'')\ ,\\
y^{\mu}_{A}(s_{0})=  -\xi_{[0]A}^{\nu}\frac{\delta{\mathcal{K}}^{[0]} _\nu}{\delta q_{\mu}}=\xi_{[0]A}^{\mu} + \ell \delta^{\mu}_{0}\xi_{[0]A}^{1} (q_{1}+p_{1})+ \ell \delta^{\mu}_{1}\xi_{[0]A}^{1} (k_{0}-q_{0})\ , \qquad y^{\mu}_{A}(s_{2})= \xi_{[2]A}^{\nu}\frac{\delta{\mathcal{K}}^{[2]} _\nu}{\delta q_{\mu}}=\xi_{[2]A}^{\mu} - \ell \delta^{\mu}_{0}\xi_{[0]A}^{1} (q_{1}'+q_{1}'')\ ,\\
x_{A}'^{\mu}(s_{1})=  -\xi_{[1]A}^{\nu}\frac{\delta{\mathcal{K}}^{[1]} _\nu}{\delta p'_{\mu}}=\xi_{[1]A}^{\mu} - \ell \delta^{\mu}_{0}\xi_{[0]A}^{1} p_{1}' + \ell \delta^{\mu}_{1}\xi_{[0]A}^{1} (p_{0}-p_{0}'-p_{0}'')\ ,\\
x_{A}''^{\mu}(s_{1})=  -\xi_{[1]A}^{\nu}\frac{\delta{\mathcal{K}}^{[1]} _\nu}{\delta p''_{\mu}}=\xi_{[1]A}^{\mu} - \ell \delta^{\mu}_{0}\xi_{[0]A}^{1} (p_{1}'+p_{1}'') + \ell \delta^{\mu}_{1}\xi_{[0]A}^{1} (p_{0}-p_{0}'')\ ,\\
y_{A}'^{\mu}(s_{2})=  -\xi_{[2]A}^{\nu}\frac{\delta{\mathcal{K}}^{[2]} _\nu}{\delta q'_{\mu}}=\xi_{[2]A}^{\mu} - \ell \delta^{\mu}_{0}\xi_{[0]A}^{1} q_{1}' + \ell \delta^{\mu}_{1}\xi_{[0]A}^{1} (q_{0}-q_{0}'-q_{0}'')\ ,\\
y_{A}''^{\mu}(s_{2})=  -\xi_{[2]A}^{\nu}\frac{\delta{\mathcal{K}}^{[2]} _\nu}{\delta q''_{\mu}}=\xi_{[2]A}^{\mu} - \ell \delta^{\mu}_{0}\xi_{[0]A}^{1} (q_{1}'+q_{1}'') + \ell \delta^{\mu}_{1}\xi_{[0]A}^{1} (q_{0}-q_{0}'')\ .
\end{gather*}
From this we immediately see that the action $\mathcal{S}_{A}$ does not
admit a relativistic description of distant observers (in relative rest),
at least not in the sense intended in Ref.~\cite{prl}.
And, as announced, the troubles originate from the finite worldlines,
with two endpoints.
For example, according to the observation reported
in Ref.~\cite{prl} (and here summarized in Subsec.~\ref{oksofarry}),
one would like translation transformations such that the
 endpoints of the worldline
of momentum $p$ transform as follows:
\begin{equation}
\begin{split}
x^\mu_{B}(s_{0}) &= x^\mu_{A}(s_{0})
+ b^\nu \frac{\delta \mathcal{K}^{[0]}_\nu}{\delta p_\mu}\ ,\\
x^\mu_{B}(s_{1}) &= x^\mu_{A}(s_{1})
- b^\nu \frac{\delta \mathcal{K}^{[1]}_\nu}{\delta p_\mu}\ .\\
\end{split}
\label{bcright}
\end{equation}
But we also must insist, if the transformation from Alice to Bob
is to be relativistic, that the equations of motion written by Alice and Bob
are the same, so that in particular also for Bob $\dot x^\mu_{B} = 2 \mathcal{N}_p \delta^{\mu}_{0} p_0-2\mathcal{N}_p \delta^{\mu}_{1} p_1$.
However, enforcing both $\dot x_{A}^\mu = 2 \mathcal{N}_p \delta^{\mu}_{0} p_0-2\mathcal{N}_p \delta^{\mu}_{1} p_1$ for Alice
and $\dot x_{B}^\mu = 2 \mathcal{N}_p \delta^{\mu}_{0} p_0-2\mathcal{N}_p \delta^{\mu}_{1} p_1$ for Bob
imposes on our translation transformations that they be rigid translations
of the endpoints, in the sense that for (\ref{bcright}) one should have
\begin{equation}
 \frac{\delta \mathcal{K}^{[0]}_\nu}{\delta p_\mu}
 = - \frac{\delta \mathcal{K}^{[1]}_\nu}{\delta p_\mu}\ .
\label{tricky}
\end{equation}
And it is easy to see that this condition, while automatically verified
at zero-th order is in general not satisfied at $O(\ell)$.

For example for the Majid-Ruegg connection one has that
\begin{equation}
\begin{split}
\mathcal{K}^{[0]}_1 &= [k \oplus (\ominus q)\oplus (\ominus p)]_{1}=k_1 - q_1 - p_1 - \ell (k_0 q_1 - q_0 q_1 + k_0 p_1 - q_0 p_1 - p_0 p_1) \ ,\\
\mathcal{K}^{[1]}_1 &=  [p \oplus (\ominus p'')\oplus (\ominus p')]_{1}
=p_1-p''_1 - p'_1  - \ell (-p''_0 p''_1 - p''_0 p'_1 - p'_0 p'_1 +  p_0 p''_1 +  p_0 p'_1)\ ,
\end{split}
\label{joc1}
\end{equation}
from which it follows that
\begin{equation}
\begin{split}
 \frac{\delta \mathcal{K}^{[0]}_1}{\delta p_0}&=\ell p_1 \ ,\qquad  \frac{\delta \mathcal{K}^{[0]}_1}{\delta p_1}=-1-\ell(k_0-q_{0}-p_0)\ ,\\
\frac{\delta \mathcal{K}^{[1]}_1}{\delta p_0}&=-\ell(p'_1+p''_1)\ , \qquad  \frac{\delta \mathcal{K}^{[1]}_1}{\delta p_1}=1\ .
\end{split}
\label{joc2}
\end{equation}
which indeed confirms that the condition (\ref{tricky}) is satisfied
at zero-th order but violated at $O(\ell)$.


\section{Known relative-locality results for free $\kappa$-Poincar\'e
particles in Hamiltonian description}\label{kbobreview}
The insight gained in the previous section is going to guide us, in the next section,
to a satisfactory relativistic description of interacting particles,
with relative locality, applicable also to cases where particles are ``exchanged",
{\it i.e.} there are finite worldlines.
As a further element of preparation for that task
we find it useful to briefly review the recent results
on the relative locality produced by a ``$\kappa$-Poincar\'e inspired
Hamiltonian" description of free particles.
This is because part of our confidence in the way we shall propose to proceed for the
Lagrangian description of interacting particles
is provided by exposing a consistency with these pre-existing
Hamiltonian free-particle results.

Also for this aside on Hamiltonian description of free particles
on a ``$\kappa$-Minkowski phase space"
we introduce an auxiliary
worldline parameter $s$ and
we denote by $\dot{Q}$ the $s$ derivative of an observable $Q$,
so that $\dot{Q} \equiv \partial Q/\partial s$.\\

On the basis of what was derived in the earlier Section~\ref{kappaMtotherightDIR}
our ``$\kappa$-Minkowski phase-space ansatz" is such that
the  Poisson bracket
for the spacetime coordinates is
 \begin{equation}
\left\{ x^{1},x^{0}\right\} =\ell x^{1},
\label{kappadef}
\end{equation}
spacetime translations are governed by
\begin{gather}
\left\{ x^{0}, p_{0} \right\} =1,\qquad\left\{ x^{1}, p_{0} \right\} =0\ ,\label{timetrasl}
\\ \left\{ x^{0}, p_{1}\right\} =\ell p_{1},\qquad\left\{ x^{1}, p_{1} \right\} =1~.
\label{spacetrasl}
\end{gather}
and the on-shell relation is
\begin{equation}
m^2=p_{0}^{2}-p_{1}^{2}+\ell p_{0} p_{1}^{2}\ .
\label{k-bob dispersion}
\end{equation}
One can then use~\cite{jurekvelISOne,mignemi}
$${\cal H}_p= {\cal N}_p {\cal C}[p] = {\cal N}_p \left(p_{0}^{2}-p_{1}^{2}+\ell p_{0} p_{1}^{2} - m^2 \right)$$
as Hamiltonian
of evolution of the observables on the worldline of a particle
in terms of the worldline parameter $s$.

Hamilton's equations
evidently give the conservation of $p_{0}$ and $p_{1}$ along the worldlines.
And concerning worldlines one finds that
\begin{gather}
\dot{x}^{0}=\left\{ x^{0}, \mathcal{H}_p \right\}
=\frac{\partial\mathcal{H}_p}{\partial p_{0}}\{ x^{0}, p_{0} \}
+\frac{\partial\mathcal{H}_p}{\partial p_{1}}\{ x^{0}, p_{1}\} = {\cal N}_p \left(2p_{0}
-\ell p_{1}^{2} \right)\ ,\nonumber \\
\dot{x}^{1}=\left\{ x^{1}, \mathcal{H}_p\right\}
=\frac{\partial\mathcal{H}_p}{\partial p_{0}}\{ x^{1}, p_{0} \}
+\frac{\partial\mathcal{H}_p}{\partial p_{1}}\{ x^{1}, p_{1}\} = - 2{\cal N}_p \left(p_{1} - \ell p_{0} p_{1}\right)\ ,
~ \nonumber \end{gather}
so that the velocity is\footnote{Note that with our choice of conventions
(signature of the metric)
a particle on shell moving along the positive direction of the $x^1$ axis
has positive $v^1$ and negative $p_1$.}
\begin{equation}
 v = \frac {\dot{x}^1}{\dot{x}^0} = - \frac { p_{1} }{p_{0} } \left( 1 - \ell p_{0} + \frac{1}{2} \ell \frac {p_{1}^{2}} { p_0 }\right) = -  \frac { p_{1} }{\sqrt {p_1^2 + m^2} } + \ell p_{1} \frac { m^2 } {p_1^2 + m^2} \ ,
\end{equation}
where, in light of Eq.~(\ref{k-bob dispersion}),
\begin{equation}
p_0 = \sqrt {p_1^2 + m^2} - \frac{1}{2} \ell p_1^2 \ .
\end{equation}

The worldlines then are
\begin{equation}
x^{1}\left(p_{1},\bar x^{1},\bar x^{0};x^{0}\right)
= \bar x^{1} -  \left(\frac{p_{1}}{\sqrt{p_{1}^{2} + m^{2}}}-\ell p_{1} \frac { m^2 } {p_1^2 + m^2} \right)\left(x^{0} -  \bar x^{0}\right)\ .
\nonumber \end{equation}
In particular, for massless particles these worldlines give
a momentum-independent particle speed:
\begin{equation}
x^{1}\left(p_{1},\bar x^{1},\bar x^{0};x^{0}\right)=\bar x^{1}-\frac{p_{1}}{|p_{1}|}(x^{0} -\bar x^{0})\ .\nonumber\end{equation}

However, as noticed in Ref.~\cite{kappabob},
the fact that worldlines of massless particles
are characterized by ``coordinate velocities" which are momentum independent
does not ensure that
 simultaneously-emitted massless particles of different
momentum are detected simultaneously.
One must factor in the anomalous properties of translations
in $\kappa$-Minkowski, and this is were the relativity of locality
is most vividly exposed.

 To see this it suffices to
consider
a simultaneous emission occurring in the origin of an observer Alice.
This will be described by Alice in terms of two worldlines, a massless particle
 with momentum $p_1^{s}$
and a massless particle
with momentum  $p_1^{h}$, which actually coincide because of the momentum independence
of the coordinate velocity:
\begin{gather}
x^{1}_{[A]p^{s}}(x^{0}_{[A]})=x^{0}_{[A]} ~,~~~
x^{1}_{[A]p^{h}}(x^{0}_{[A]})=x^{0}_{[A]} ~ \label{alicep}
\end{gather}
(where we took both $p_1^{h} <0$ and $p_1^{s}<0$, so that the particles propagate
along the positive direction of the $x^1$ axis).

It is useful to focus on the case of $p_1^{s}$ and  $p_1^{h}$ such
 that $|p_1^{s}| \ll |p_1^{h}|$,
and $|\ell p_1^{s}| \simeq 0$
(the particle with momentum $p_1^{s}$ is soft enough that
it behaves as if $\ell = 0$) while  $|\ell p_1^{h}| \neq 0$, in the sense
 that for the hard particle the effects of $\ell$-deformation  are not negligible.

Then we need to use the fact that
the assignments of coordinates
on points of a worldline adopted by two observers connected
by a generic translation ${\cal T}_{b^{0} , b^{1}}$, with component $b^{0}$
along the $x^{0}$ axis and component $b^{1}$ along the $x^{1}$
axis, is such that
\begin{gather}
x'^{1}=x^{1}+b^{0}\left\{ p_{0} ,x^{1}\right\} +b^{1}\left\{ p_{1},x^{1}\right\} \ ,\nonumber \\
x'^{0}=x^{0}+b^{0}\left\{ p_{0} ,x^{0}\right\} +b^{1}\left\{ p_{1},x^{0}\right\} \ .\nonumber
\end{gather}
Using these we can look~\cite{kappabob} at the two Alice worldlines, given in (\ref{alicep}),
from the perspective
of a second observer, Bob, at rest with respect to Alice at distance $b$ from Alice
(Bob = ${\cal T}_{b , b} \triangleright$ Alice),
local to a detector that the two particles eventually reach.
Of course, in light of the form of the worldlines,
according to Alice's coordinates the two particles reach Bob simultaneously.
But can this distant coincidence of events be trusted?
The two events which according to the coordinates of distant observer Alice are coincident
 are the crossing of Bob's worldline with the worldline of the particle
with momentum $p_1^{s}$ and the
crossing of Bob's worldline with the worldline of the particle
with momentum  $p_1^{h}$.
To clarify the situation we should look at the two worldlines from the perspective
of Bob, the observer who is local to the detection of the particles.

Evidently these Bob worldlines are obtained from Alice worldlines using
the translation transformation codified in (\ref{timetrasl}), (\ref{spacetrasl}).
Acting on a generic Alice
worldline $x^{1}_{[A]}\left(p_{1},\bar x^{1}_{[A]},\bar x^{0}_{[A]};x^{0}_{[A]}\right)$ this gives a Bob
worldline $x^{1}_{[B]}\left(p_{1},\bar x^{1}_{[B]},\bar x^{0}_{[B]};x^{0}_{[B]}\right)$
as follows:
\begin{gather}
x^{1}_{[B]} = x^{1}_{[A]} +b \left\{ p_{0} ,x^{1}_{[A]}\right\} +b \left\{ p_{1},x^{1}_{[A]}\right\}
= x^{1}_{[A]}-b
 \ ,\nonumber \\
x^{0}_{[B]}=x^{0}_{[A]} + b \left\{ p_{0} ,x^{0}_{[A]}\right\} +b \left\{ p_{1},x^{0}_{[A]}\right\}
=x^{0}_{[A]}-b -\ell b p_{1}
  \ .\nonumber \end{gather}
  And specifically for the two worldlines of our interest, given for Alice in
  (\ref{alicep}),
  one then finds
\begin{gather}
x^{1}_{[B]p^{s}}(x^{0}_{[B]})=x^{0}_{[B]}+\ell b p_1^{s} \simeq x^{0}_{[B]}
 \ ,\nonumber \\
x^{1}_{[B]p^{h}}(x^{0}_{[B]})=x^{0}_{[B]}+\ell b p_1^{h}\ .\nonumber
 \end{gather}
The two worldlines, which were coincident according to Alice, are
distinct worldlines for Bob.
And it is established that~\cite{kappabob}
according to Bob, who is at the detector, the two particles reach the detector
at different times: $x^{0}_{[B]} \simeq 0$ for the soft particle and
$x^{0}_{[B]} = -\ell b p^{h}_{1} = \ell b |p^{h}_{1}|$ for the hard particle.
This for the two
massless particles
which, according to the observer Alice who is at the emitter, were emitted simultaneously.
The difference of times of detection at Bob
is governed by the simple formula
\begin{equation}
\Delta t = \ell b |\Delta p_{1}|~. \label{delayk-bob}
\end{equation}

\begin{figure}[h!]
\begin{center}
\includegraphics[scale=0.55]{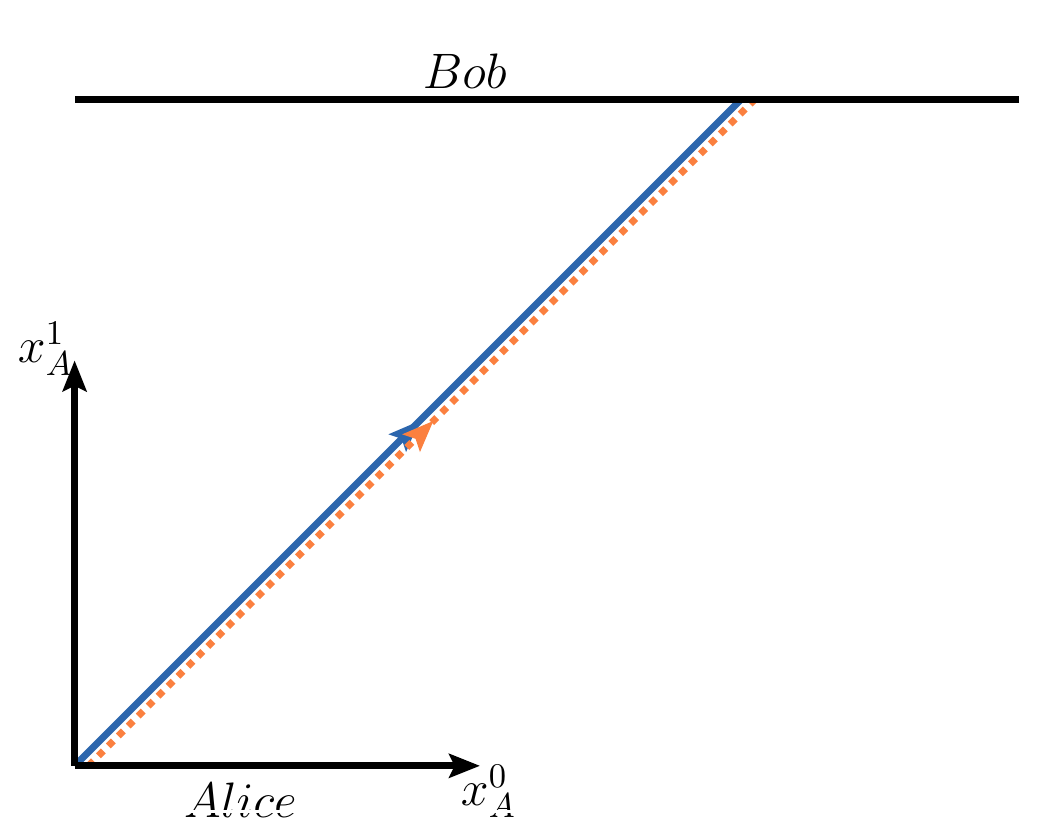}
\includegraphics[scale=0.55]{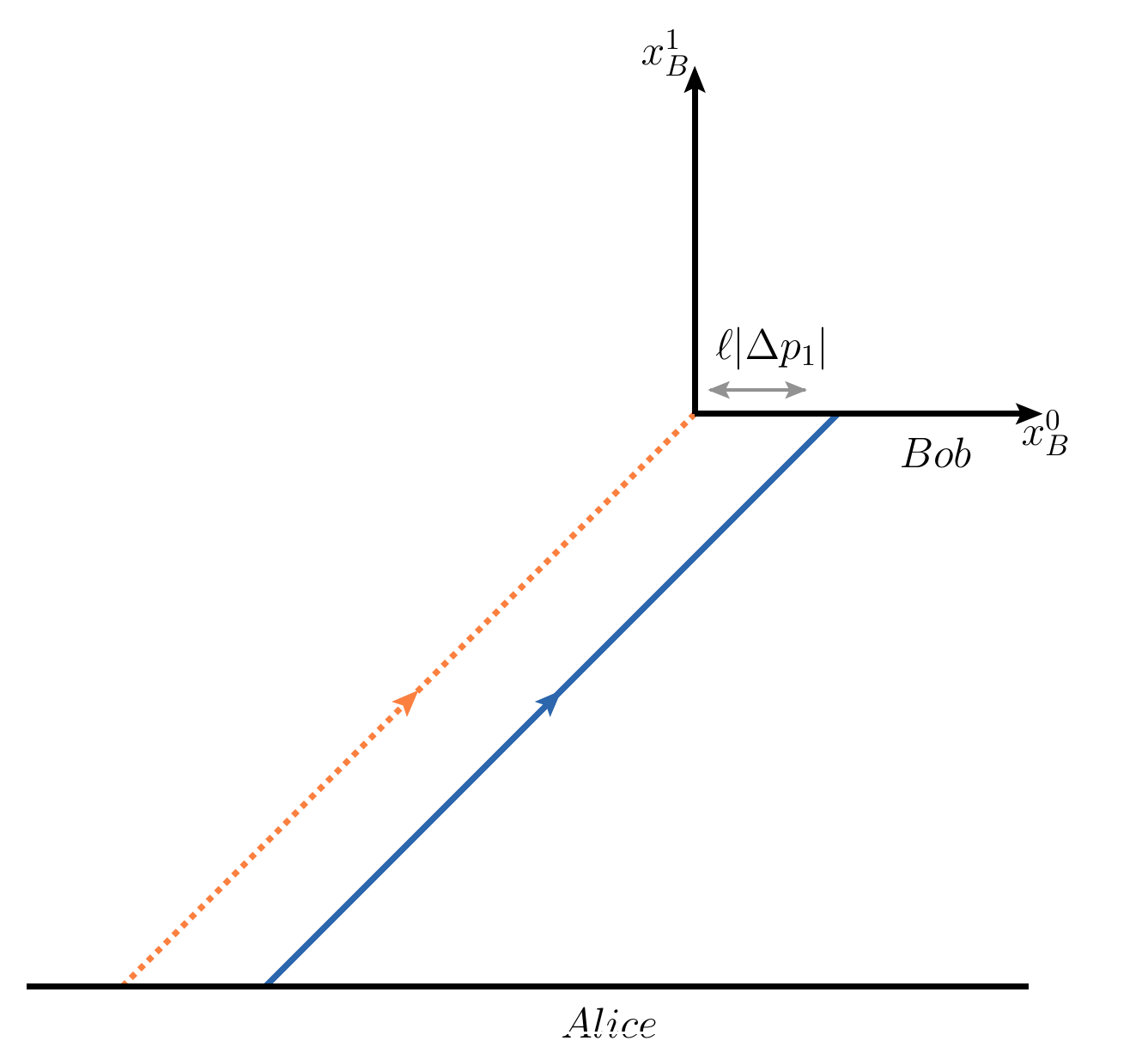}
\caption{Two simultaneously-emitted massless particles
of different momentum in $\kappa$-Minkowski
 are detected at different times.
 The figure shows
 how the simultaneous emission of two such particles and their
 non-simultaneous detection
 is described in the coordinates of observer Alice (left panel),
 who is at the emitter,
 and in the coordinates of observer Bob (right panel),
 who is at the detector.}
\end{center}
\end{figure}

\section{A Lagrangian description of relative locality with interactions}\label{main}

\subsection{$\kappa$-Minkowski symplectic
structure and translations generated by total momentum}\label{secfixin1vertex}
In this section we show that it is possible to have a relativistic
description of pairs of distant observers (in relative rest),
in descriptions of interactions with particle exchanges (finite worldlines)
formulated within the relative-locality framework of Refs.~\cite{prl,grf2nd}.
The main challenge we shall face in this section is
the one characterized in Subsection \ref{challenge}:
relativistic descriptions of a single interaction with relative locality
are rather elementary, but when pairs of interactions a causally connected
the availability
of a relativistic description for distant observers is in no way
assured, and actually before the study we are here reporting
there was no known example where it had been shown to work.

We shall also find reassuring that the Lagrangian description we
obtain for interacting particles, reproduces in an appropriate limit
the known results reviewed in the previous section, concerning
relative locality in a $\kappa$-Poincar\'e-inspired Hamiltonian description
of free particles. In doing so we also provide
an explicit analysis in which the non-trivial geometry
of momentum space is analyzed while adopting a non-standard
symplectic structure.

Indeed, the first point of contact between our Lagrangian description and the
Hamiltonian description reviewed in the previous section is found
in the choice of symplectic structure and on-shell condition
characterizing the ``free part" of the action",
which for the case of 3 particles (of momenta $k_\mu$ incoming and momenta $p_\mu$
and $q_\mu$ outgoing) takes the form:
\begin{equation}
\begin{split}
\mathcal{S}^{bulk}_{\kappa} = & \int_{-\infty}^{s_{0}}ds\left(z^{\mu}\dot{k}_{\mu}-\ell z^{1}k_{1}\dot{k}_{0}+\mathcal{N}_k\mathcal{C}_{\kappa}\left[k\right]\right)+
\int_{s_{0}}^{+\infty}ds\left({x}^{\mu}\dot{p}_{\mu}-\ell x^{1}p_{1}\dot{p}_{0}+\mathcal{N}_p\mathcal{C}_{\kappa}\left[p\right]\right)\\
 & +\int_{s_{0}}^{+\infty}ds\left(y^{\mu}\dot{q}_{\mu}-\ell y^{1}q_{1}\dot{q}_{0}+\mathcal{N}_q\mathcal{C}_{\kappa}\left[q\right]\right)\ ,
 \end{split}
\label{action k}
\end{equation}
where
$${\cal C}_\kappa[k] \equiv k_0^2-k_1^2+\ell k_0 k_1^{2}\ ,$$
so that we implement the on-shell relation of the Hamiltonian $\kappa$-Minkowski
phase-space setup
reviewed in the previous section.
We are adopting $\kappa$-Minkowski Poisson brackets, so that for example
for $x^\mu$
$$\left\{ x^{1},x^{0}\right\} =\ell x^{1}~,$$
and from (\ref{action k})
 one recognizes that our symplectic structure also matches the
 one of the Hamiltonian $\kappa$-Minkowski
phase-space setup
reviewed in the previous section; so that for example for $x^\mu$,$p_\mu$
\begin{gather*}
\left\{ x^{0}, p_{0} \right\} =1,\qquad\left\{ x^{1}, p_{0} \right\} =0\ ,
\\ \left\{ x^{0}, p_{1}\right\} =\ell p_{1},\qquad\left\{ x^{1}, p_{1} \right\} =1~.
\end{gather*}
For what concerns the conservation laws at interactions
we shall adopt the Majid-Ruegg connection.
But, as evident on the basis of the observation we reported
in Section~\ref{setup},
once the conservation laws are specified the construction
of this type of relative-locality theory still leaves open
a choice among possible alternative ways of
implementing  such laws of momentum conservation through some
boundary terms.
We adopt a particular choice
which we favor because it happens
to be immune from
the problem here highlighted in Subsection~\ref{challenge},
which instead is found to affect several alternative
possibilities~\cite{jacktesi,gabrtesi}.
We qualify our choice of momentum-conservation constraints
as the ones that are suitable for a description of translations
in which ``translations are generated by the total momentum", for reasons that
will become clearer in the reminder of this section.
The prescription we adopt will be generalized as we go along, but let
us here start with the case of a single interaction,
whose conservation law is
$$0=k \oplus (\ominus q) \oplus (\ominus p)\ .$$
As already stressed in Subsection~\ref{choices},
such a conservation law could be implemented by several inequivalent
choices of ${\cal K}^{[0]}$
for the constraints on the endpoints of worldlines, including
$${\cal K}^{[0]}=k \oplus (\ominus q) \oplus (\ominus p)\ ,$$
$${\cal K}^{[0]}=(\ominus q) \oplus (\ominus p) \oplus k\ , $$
$${\cal K}^{[0]}=k- (p\oplus q) \ .$$
We find that this latter
option ${\cal K}^{[0]}=k- (p\oplus q) $
admits a consistent relativistic description of distant observers.
Evidence of this will be provided throughout this section.
But let us first notice that this sort of constraints
is very intuitive: they implement the rather standard concept that
the conservation law is such that the total momentum before an interaction
should equal the total momentum after an interaction.
And we shall show that this form of the constraints allows one to
preserve the usual notion that translation transformations
are generated by the total momentum (though of course in our case
the total momentum is obtained in terms of the nonlinear composition law),
even when several interactions are analyzed and particles are exchanged among
some of the interactions.

\begin{figure}[hb!]
\begin{center}
\includegraphics[width=0.44 \textwidth]{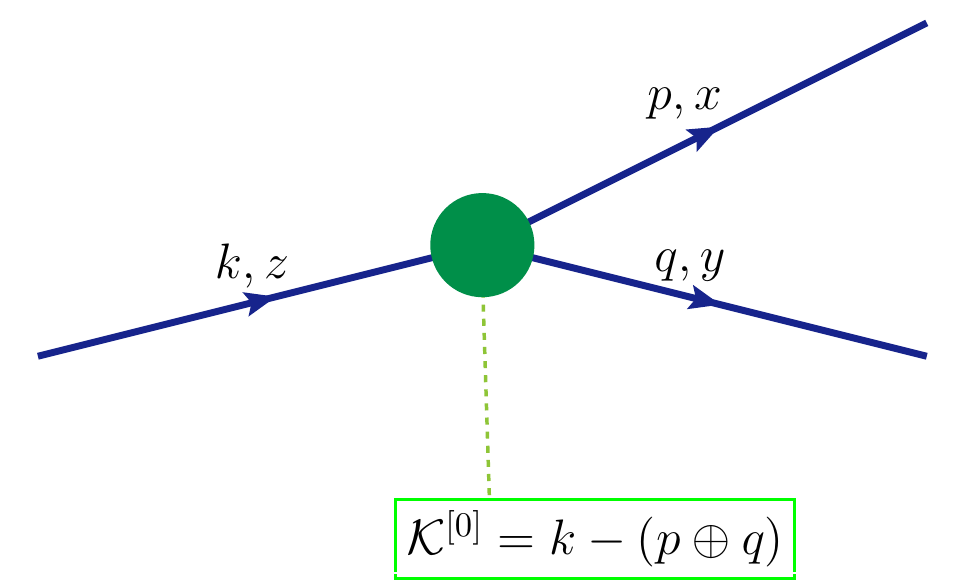}
\caption{The choice of ${\cal K}$ we adopt for the case of a single
interaction with 1 incoming and 2 outgoing particles.}
\label{figfixin1vertex}
\end{center}
\end{figure}

 Essentially our proposal establishes that
there is at least one way (at present we are unable to claim that
it is unique) to address the challenge
we earlier highlighted in  Eq.~(\ref{tricky}). And the conceptual content
of the solution we found for addressing that challenge exemplified
in Eq.~(\ref{tricky}) is,
as shown below, rather simple:
the most basic notion of relativistic translation transformation
is as usual generated by the total momentum acting on worldlines,
but (as also shown in our discussion surrounding Eq.~(\ref{tricky}))
the boundary terms used to enforce the conservation laws require that
endpoints transform under translations in ways governed by
(or at least conditioned by) the boundary terms. We handle the challenge
illustrated by Eq.~(\ref{tricky}) by essentially finding a way to render
these two demands compatible: we enforce the conservation laws through
boundary terms written in such a way that when the worldlines are translated
by the total momentum then the endpoints automatically match
the demands of the boundary terms.

Let us start seeing  how this plays out for a case with a single
 interaction, considering, for the interaction in Figure~\ref{figfixin1vertex},
 the action
 \begin{equation}
\begin{split}
S^{\kappa} =S^\kappa_{bulk}+ S^\kappa_{int} = & \int_{-\infty}^{s_{0}}ds\left(z^{\mu}\dot{k}_{\mu}-\ell z^{1}k_{1}\dot{k}_{0}+\mathcal{N}_k\mathcal{C}_{\kappa}\left[k\right]\right)+
\int_{s_{0}}^{+\infty}ds\left({x}^{\mu}\dot{p}_{\mu}-\ell x^{1}p_{1}\dot{p}_{0}+\mathcal{N}_p\mathcal{C}_{\kappa}\left[p\right]\right)\\
 & +\int_{s_{0}}^{+\infty}ds\left(y^{\mu}\dot{q}_{\mu}-\ell y^{1}q_{1}\dot{q}_{0}+\mathcal{N}_q\mathcal{C}_{\kappa}\left[q\right]\right) -\xi_{[0]}^{\mu}\mathcal{K}_{\mu}^{[0]}(s_{0})\ ,
 \end{split}
\label{action k 3}
\end{equation}
 where indeed for ${\cal K}^{[0]}$  we
 take
 \begin{equation}
{\cal K}^{[0]}_\mu  (s_{0})= k_\mu - (p\oplus q)_\mu  = k_\mu - p_\mu + \ell \delta_\mu^1 q_0 p_1 \ .
\label{K0}
\end{equation}

The equations of motion that follow from our action $S^\kappa$
are of course the same found in the Hamiltonian formulation
of free $\kappa$-Minkowski particles
reviewed in the previous subsection:
\begin{gather}
\dot p_\mu =0~,~~\dot q_\mu =0~,~~\dot k_\mu =0~,~~\nonumber\\
\mathcal{C}_{\kappa}[p]=0~,~~\mathcal{C}_{\kappa}[q]=0~,~~\mathcal{C}_{\kappa}[k]=0\ ,\label{eqmotion3kmomentum}\\
\mathcal{K}_\mu^{[0]}(s_{0})=0\nonumber
\end{gather}
\begin{gather}
\dot x^\mu = \mathcal{N}_p \left(\frac{\delta \mathcal{C}_{\kappa}[p]}{\delta p_\mu} +\ell \delta^\mu_0 \frac{\delta \mathcal{C}_{\kappa}[p]}{\delta p_1} p_1\right)= \delta^{\mu}_{0}\mathcal{N}_p \left( 2 p_{0}-\ell p_{1}^{2}\right)-2\delta^{\mu}_{1}\mathcal{N}_p \left(  p_{1}-\ell p_{0} p_{1}\right)\nonumber\ ,\\
\dot y^\mu = \mathcal{N}_q \left(\frac{\delta \mathcal{C}_{\kappa}[q]}{\delta q_\mu} +\ell \delta^\mu_0 \frac{\delta \mathcal{C}_{\kappa}[q]}{\delta q_1} q_1\right)= \delta^{\mu}_{0}\mathcal{N}_q \left( 2 q_{0}-\ell q_{1}^{2}\right)-2\delta^{\mu}_{1}\mathcal{N}_q \left(  q_{1}-\ell q_{0} q_{1}\right)\label{eqmotion3kspace}\ ,\\
\dot z^\mu = \mathcal{N}_k \left(\frac{\delta \mathcal{C}_{\kappa}[k]}{\delta k_\mu} +\ell \delta^\mu_0 \frac{\delta \mathcal{C}_{\kappa}[k]}{\delta k_1} k_1\right)= \delta^{\mu}_{0}\mathcal{N}_k \left( 2 k_{0}-\ell k_{1}^{2}\right)-2\delta^{\mu}_{1}\mathcal{N}_k \left(  k_{1}-\ell k_{0} k_{1}\right)\ .\nonumber
\end{gather}

And the interaction at $s=s_{0}$ produces the boundary conditions:
\begin{equation}
\begin{split}
x^\mu(s_{0}) &= -\xi_{[0]}^\nu \left(\frac{\delta \mathcal{K}^{[0]}_\nu}{\delta p_\mu}+\ell \delta^\mu_0 \frac{\delta \mathcal{K}^{[0]}_\nu}{\delta p_1}p_1\right)=\xi_{[0]}^\mu +\ell \delta^{\mu}_{0} \xi_{[0]}^1(p_{1}+q_{1})\ ,\\
y^\mu(s_{0}) &= -\xi_{[0]}^\nu \left(\frac{\delta \mathcal{K}^{[0]}_\nu}{\delta q_\mu}+\ell \delta^\mu_0 \frac{\delta \mathcal{K}^{[0]}_\nu}{\delta q_1}q_1\right)=\xi_{[0]}^\mu +\ell \delta^{\mu}_{0} \xi_{[0]}^1 q_{1}+\ell \delta^{\mu}_{0} \xi_{[0]}^1 p_{0}\ ,\\
\label{boundaries3k}
z^\mu(s_{0}) &= \xi_{[0]}^\nu \left(\frac{\delta \mathcal{K}^{[0]}_\nu}{\delta k_\mu}+\ell \delta^\mu_0 \frac{\delta \mathcal{K}^{[0]}_\nu}{\delta k_1}k_1\right)=\xi_{[0]}^\mu +\ell \delta^{\mu}_{0} \xi_{[0]}^1 k_{1}\ .
\end{split}
\end{equation}

The mechanism for relative locality which we already discussed
above is evidently also present here:
the boundary conditions establish that
if the observer is local to the
interaction, {\it i.e.} $\xi^\mu_{[0]} = 0$,
then all endpoints of the semiinfinite worldlines are in the origin
of the observer.
If instead $\xi^\mu_{[0]} \neq 0$ the endpoints of worldlines
do not coincide.

It is also easy to check that
our equations of motion (\ref{eqmotion3kmomentum}), (\ref{eqmotion3kspace})
and boundary conditions (\ref{boundaries3k})
 invariant under deformed translations generated by the total momentum
 acting on coordinates
\begin{equation}
\begin{split}
x_{B}^{0}(s)&=  x_{A}^{0}(s) + b^{\mu} \{ (p\oplus q)_\mu , x^0 \}
=x_{A}^{0}(s)-b^{0}-\ell b^{1}(p_{1}+q_1)\ ,\\
x_{B}^{1}(s)&=x_{A}^{1}(s)+b^{\mu} \{ (p\oplus q)_\mu , x^1 \}={x}_{A}^{1}(s)-b^{1}\ ,\\
y_{B}^{0}(s)&=  y_{A}^{0}(s) + b^{\mu} \{ (p\oplus q)_\mu , y^0 \} =y_{A}^{0}(s)-b^{0}-\ell b^{1}q_1\ ,\\
y_{B}^{1}(s)&=y_{A}^{1}(s) + b^{\mu}\ \{ (p\oplus q)_\mu , y^1 \}
={y}_{A}^{1}(s)-b^{1}-\ell b^{1}p_0\ ,\\
\label{translations3k}
z_{B}^{0}(s)&=  z_{A}^{0}(s)+b^{\mu} \{ k_{\mu} , z^0 \}
= z_{A}^{0}(s)-b^{0}-\ell b^{1}k_{1}\ ,\\
z_{B}^{1}(s)&=z_{A}^{1}(s)+b^{\mu} \{ k_{\mu} , z^1 \} ={z}_{A}^{1}(s)-b^{1}\ .\\
\end{split}
\end{equation}
It should be noticed that we essentially prescribe that a given point
of a given worldline is translated by acting with the total momentum
written in the way that is appropriate for that point of the worldline,
so that, in the specific example here under consideration, all points
with $s<s_0$ are translated by $k_\mu$ whereas all points
with $s > s_0$ are translated by $(p \oplus q)_\mu$.

The invariance of the equations of motion is easily seen by observing
that Eq.~(\ref{eqmotion3kmomentum}) guarantees that $\dot p_\mu =0$, $\dot q_\mu =0$, $\dot k_\mu =0$ and that the translation transformations depend only on momenta.
Considering for example the worldline $x^{\mu}$,
and assuming of course that both observer
Alice and observer Bob adopt the equations of motion
\begin{equation}
\begin{split}
\dot x^\mu_{[A]} &= \mathcal{N}_p \left(\frac{\delta \mathcal{C}_{\kappa}[p]}{\delta p_\mu} +\ell \delta^\mu_0 \frac{\delta \mathcal{C}_{\kappa}[p]}{\delta p_1} p_1\right)\ ,\\
\dot x^\mu_{[B]} &= \mathcal{N}_p \left(\frac{\delta \mathcal{C}_{\kappa}[p]}{\delta p_\mu} +\ell \delta^\mu_0 \frac{\delta \mathcal{C}_{\kappa}[p]}{\delta p_1} p_1\right)\ ,
\end{split}
\end{equation}
one indeed finds that the translation transformations
\begin{equation}
\begin{split}
x_{B}^{0}(s)&=x_{A}^{0}-b^{0}-\ell b^{1}(p_{1}+q_1)\ ,\\
x_{B}^{1}(s)&={x}_{A}^{1}-b^{1}\label{transbob}\\
\end{split}
\end{equation}
are such
that $\dot x^\mu_{[B]}=\dot x^\mu_{[A]}$ (since momenta are conserved).

And the invariance of the boundary conditions is easily seen by
directly checking that
the boundary conditions for Alice are mapped by the translation transformations
into the (identical) boundary conditions for Bob. For example, we have for Alice
\begin{equation}
x^\mu_{[A]}(s_{0}) = -\xi_{[0]A}^\nu\left(\frac{\delta \mathcal{K}^{[0]}_\nu}{\delta p_\mu}+\ell \delta^\mu_0 \frac{\delta \mathcal{K}^{[0]}_\nu}{\delta p_1}p_1\right)=\xi_{[0]A}^\mu +\ell \delta^{\mu}_{0} \xi_{[0]A}^1(p_{1}+q_{1})\ ,
\end{equation}
and the translation transformations
(\ref{transbob}) map this into
\begin{equation}
\begin{split}
x^\mu_{[B]}(s_{0}) &= x^\mu_{[A]}(s_{0})-b^{\mu}-\ell \delta^{\mu}_{0}(p_{1}+q_{1})=-\xi_{[0]A}^\nu\left(\frac{\delta \mathcal{K}^{[0]}_\nu}{\delta p_\mu}+\ell \delta^\mu_0 \frac{\delta \mathcal{K}^{[0]}_\nu}{\delta p_1}p_1\right)-b^{\mu}-\ell \delta^{\mu}_{0}(p_{1}+q_{1})\\
&=\xi_{[0]A}^\mu +\ell \delta^{\mu}_{0} \xi_{[0]A}^1(p_{1}+q_{1})-b^{\mu}-\ell \delta^{\mu}_{0}(p_{1}+q_{1})=-(\xi_{[0]A}-b)^\nu\left(\frac{\delta \mathcal{K}^{[0]}_\nu}{\delta p_\mu}+\ell \delta^\mu_0 \frac{\delta \mathcal{K}^{[0]}_\nu}{\delta p_1}p_1\right)\\
&=-\xi_{[0]B}^\nu\left(\frac{\delta \mathcal{K}^{[0]}_\nu}{\delta p_\mu}+\ell \delta^\mu_0 \frac{\delta \mathcal{K}^{[0]}_\nu}{\delta p_1}p_1\right)\ .
\end{split}
\end{equation}

Besides checking the invariance of the equations of motion and the
boundary conditions, which however already ensure that our translations
are physical symmetries, it is also valuable to apply
the translation transformations (\ref{translations3k})
to the action (\ref{action k 3}), so that we
can find the relation between the action of Alice and
the action of Bob (distant from Alice). We find
\begin{equation*}
\begin{split}
\mathcal{S}_{B}^{\kappa} & =\mathcal{S}_{A}^{\kappa}+\int_{-\infty}^{s_{0}}ds\left(-b^{\mu}\dot{k}_{\mu}\right)+\int_{s_{0}}^{\infty}ds\left(-b^{\mu}\dot{p}_{\mu}-\ell b^{1}q_{1}\dot{p}_{0}\right)\\
 & +\int_{s_{0}}^{\infty}ds\left(-b^{\mu}\dot{q}_{\mu}-\ell b^{1}p_{0}\dot{q}_{1}\right) + \Delta\xi_{\left[0\right]}^{\mu}\mathcal{K}_{\mu}^{[0]}(s_{0})\ .
\end{split}
\end{equation*}
where $\Delta\xi_{[0]}^{\mu}=\xi_{[0]B}^{\mu}-\xi_{[0]A}^{\mu}$ . Substituting $s'=-s$  in the first integral and then relabeling $s'\rightarrow s$ , one then gets
\begin{equation*}
\begin{split}\Delta\mathcal{S}^{\kappa} & =\mathcal{S}_{B}^{\kappa}-\mathcal{S}_{A}^{\kappa}=\int_{s_{0}}^{\infty}ds\left(b^{\mu}\left(\dot{k}_{\mu}-\dot{p}_{\mu}-\dot{q}_{\mu}\right)-\ell b^{1}\left(q_{1}\dot{p}_{0}+p_{0}\dot{q}_{1}\right)\right) +\Delta\xi_{\left[0\right]}^{\mu}\mathcal{K}_{\mu}^{[0]}(s_{0})\ .
\end{split}
\end{equation*}
Then using Eq.~(\ref{K0}) we find
\begin{equation}
\begin{split}\Delta\mathcal{S}^{\kappa} & =\mathcal{S}_{B}^{\kappa}-\mathcal{S}_{A}^{\kappa}=\int_{s_{0}}^{\infty}ds\frac{d}{ds}\left[b^{\mu}{\cal K}_{\mu}^{\left[0\right]}\right] +\Delta\xi_{\left[0\right]}^{\mu}\mathcal{K}_{\mu}^{[0]}(s_{0})\ .
\end{split}
\label{DeltaS}
\end{equation}
The total derivatives contribute to the boundaries in such a way that, for the difference (\ref{DeltaS}) to be null, it must hold
\begin{equation*}
 \left(\Delta\xi_{\left[0\right]}^{\mu}-b^{\mu}\right)\mathcal{K}_{\mu}^{[0]}(s_{0})\ ,
\end{equation*}
from which we see that the $\xi^{\mu}_{[0]}$  translate classically:
\begin{equation}
{\xi^\mu_{[0]}}_B = {\xi^\mu_{[0]}}_A - b^\mu \ .
\end{equation}

And when the observer Alice is distant from the
interaction, {\it i.e.} $\xi^\mu_{[0]A} \neq 0$,
one can always find through such translation
transformations an observer Bob local to the
interaction and for whom the endpoints of worldlines match:
\begin{equation}
\begin{split}
x^\mu_{B}(s_{0}) &= -\xi^\nu_{B} \left(\frac{\delta \mathcal{K}^{[0]}_\nu}{\delta p_\mu}+\ell \delta^\mu_0 \frac{\delta \mathcal{K}^{[0]}_\nu}{\delta p_1}p_1\right)=-(\xi^\nu_{A}-b^\nu) \left(\frac{\delta \mathcal{K}^{[0]}_\nu}{\delta p_\mu}+\ell \delta^\mu_0 \frac{\delta \mathcal{K}^{[0]}_\nu}{\delta p_1}p_1\right)=0\ ,\\
y^\mu_{B}(s_{0}) &= -\xi^\nu_{B} \left(\frac{\delta \mathcal{K}^{[0]}_\nu}{\delta q_\mu}+\ell \delta^\mu_0 \frac{\delta \mathcal{K}^{[0]}_\nu}{\delta q_1}q_1\right)=-(\xi^\nu_{A}-b^\nu) \left(\frac{\delta \mathcal{K}^{[0]}_\nu}{\delta q_\mu}+\ell \delta^\mu_0 \frac{\delta \mathcal{K}^{[0]}_\nu}{\delta q_1}q_1\right)=0\ ,\\
z^\mu_{B}(s_{0}) &= \xi^\nu_{B} \left(\frac{\delta \mathcal{K}^{[0]}_\nu}{\delta k_\mu}+\ell \delta^\mu_0 \frac{\delta \mathcal{K}^{[0]}_\nu}{\delta k_1}k_1\right)=(\xi^\nu_{A}-b^\nu) \left(\frac{\delta \mathcal{K}^{[0]}_\nu}{\delta k_\mu}+\ell \delta^\mu_0 \frac{\delta \mathcal{K}^{[0]}_\nu}{\delta k_1}k_1\right)=0\ .\\
\end{split}
\end{equation}

\subsection{Causally connected interactions and translations
generated by total momentum}\label{mainsubsec}
Our next challenge is to deal with causally-connected interactions.
We show in figure the case we here analyze as illustrative example.

\begin{figure}[hb!]
\begin{center}
\includegraphics[width=0.66 \textwidth]{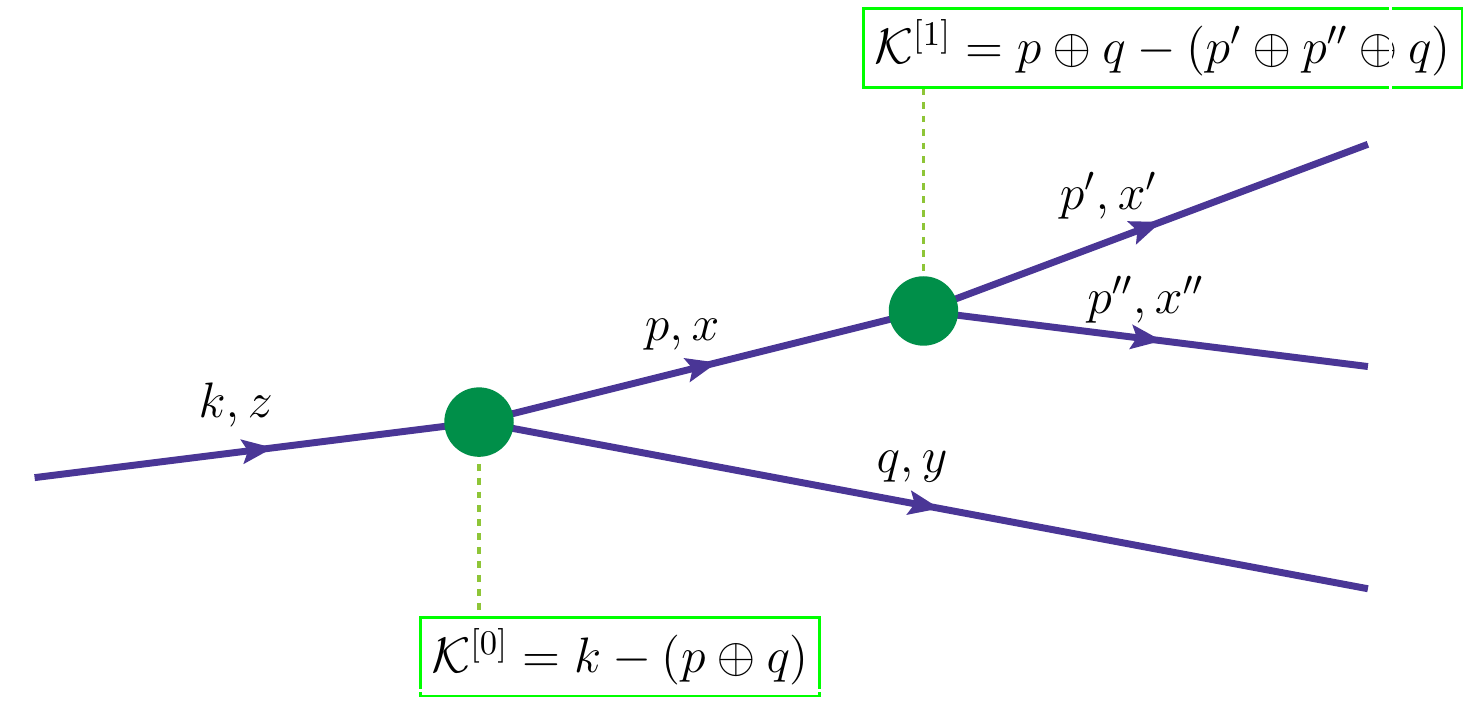}
\caption{The case of causally-connected interactions analyzedin this subsection.}
\end{center}
\end{figure}

Of course, there is no difficulty generalizing
to this case the bulk part of the action:
\begin{equation}
\begin{split}
\mathcal{S}^{\kappa(2)}_{bulk}= & \int_{-\infty}^{s_{0}}ds\left(z^{\mu}\dot{k}_{\mu}-\ell z^{1}k_{1}\dot{k}_{0}+\mathcal{N}_k\mathcal{C}_\kappa\left[k\right]\right)+
\int_{s_{0}}^{s_{1}}ds\left({x}^{\mu}\dot{p}_{\mu}-\ell x^{1}p_{1}\dot{p}_{0}+\mathcal{N}_p\mathcal{C}_\kappa\left[p\right]\right)\\
 &  +\int_{s_{1}}^{+\infty}ds\left({x'}^{\mu}\dot{p'}_{\mu}-\ell{x'}^{1}p'_{1}\dot{p'}_{0}+\mathcal{N}_{p'}\mathcal{C}_\kappa\left[p'\right]\right)+\int_{s_{1}}^{+\infty}ds\left(x''^{\mu}\dot{p''}_{\mu}-\ell x''^{1}p''_{1}\dot{p''}_{0}+\mathcal{N}_{p''}\mathcal{C}_\kappa\left[p''\right]\right)\\
 &+\int_{s_{0}}^{+\infty}ds\left(y^{\mu}\dot{q}_{\mu}-\ell y^{1}q_{1}\dot{q}_{0}+\mathcal{N}_q\mathcal{C}_\kappa\left[q\right]\right)\ .
 \label{actionk3+3free}
\end{split}
\end{equation}

For the description of the interactions we take a case characterized
by the following conservation laws:
$$ k \oplus (\ominus q) \oplus (\ominus p) =0 \ ,$$
$$ p \oplus (\ominus p'')\oplus (\ominus p') =0  \ .$$
And we propose an implementation of these conservation laws
that is compatible with a relativistic description of distant
observers, based on adding to the action constraints with
$${\cal K}^{[0]} = k-(p\oplus q) \ , $$
$${\cal K}^{[1]} = (p\oplus q) - ( p'\oplus p'' \oplus q)  \ ,$$
{\it i.e.}
\begin{gather}
{\cal K}^{[0]}_\mu = k_\mu -(p\oplus q)_\mu  = k_\mu - p_\mu -q_\mu -\ell \delta _\mu ^1 p_0 q_1 \ ,\nonumber\\
{\cal K}^{[1]}_\mu = (p\oplus q)_\mu - ( p'\oplus p'' \oplus q)_\mu = p_\mu  - p_\mu'- p_\mu'' - \ell \delta_\mu^1 \left(-p_0 q_1 + p_0' p_1'' + p_0' q_1 + p_0'' q_1 \right) \ .
\label{K[1]}
\end{gather}

For the constraints we are again implementing our prescription of writing
them in terms of differences between the total momentum before
the interaction and after the interaction.
It may appear that in doing so we included in the constraints some irrelevant
pieces (it is easy to verify that the conservation law ${\cal K}^{[1]}=0$
is actually independent of $q_\mu$, which is the momentum of the particle that
is only a spectator of the interaction occurring at $s=s_1$).
However, as we shall see, those extra pieces,
while irrelevant for the physical content of the conservation laws,
do play a role in the description of translation transformations
and ensure the availability
of a relativistic description of distant observers.

To show this let us then start by writing the full action
for the two-interaction case on which we are presently focusing:
 \begin{equation}
\begin{split}
\mathcal{S}^{\kappa (2)} =\mathcal{S}^{\kappa(2)}_{bulk}+ \mathcal{S}^{\kappa(2)}_{int} = & \int_{-\infty}^{s_{0}}ds\left(z^{\mu}\dot{k}_{\mu}-\ell z^{1}k_{1}\dot{k}_{0}+\mathcal{N}_k\mathcal{C}_\kappa\left[k\right]\right)+
\int_{s_{0}}^{s_{1}}ds\left({x}^{\mu}\dot{p}_{\mu}-\ell x^{1}p_{1}\dot{p}_{0}+\mathcal{N}_p\mathcal{C}_\kappa\left[p\right]\right)\\
 &  +\int_{s_{1}}^{+\infty}ds\left({x'}^{\mu}\dot{p'}_{\mu}-\ell{x'}^{1}p'_{1}\dot{p'}_{0}+\mathcal{N}_{p'}\mathcal{C}_\kappa\left[p'\right]\right)+\int_{s_{1}}^{+\infty}ds\left(x''^{\mu}\dot{p''}_{\mu}-\ell x''^{1}p''_{1}\dot{p''}_{0}+\mathcal{N}_{p''}\mathcal{C}_\kappa\left[p''\right]\right)\\
 &+\int_{s_{0}}^{+\infty}ds\left(y^{\mu}\dot{q}_{\mu}-\ell y^{1}q_{1}\dot{q}_{0}+\mathcal{N}_q\mathcal{C}_\kappa\left[q\right]\right)\\
 & -\xi_{[0]}^{\mu}\mathcal{K}_{\mu}^{[0]}(s_{0})-\xi_{[1]}^{\mu}\mathcal{K}_{\mu}^{[1]}(s_{1})\ ,
 \label{actionk3+3full}
\end{split}
\end{equation}
 where indeed with ${\cal K}^{[0]}$ and ${\cal K}^{[1]}$
we take respectively $k-(p\oplus q)  $ and
 $ (p\oplus q) - ( p'\oplus p'' \oplus q)   $.

It is again straightforward
to derive the equations
 of motion (and constraints)
 that follow from our action $\mathcal{S}^{\kappa (2)}$:
\begin{gather}
\dot p_\mu =0~,~~\dot q_\mu =0~,~~\dot k_\mu =0~,~~\dot p'_\mu =0~,~~\dot p''_\mu =0\ ,\nonumber\\
\mathcal{C}_{\kappa}[p]=0~,~~\mathcal{C}_{\kappa}[q]=0~,~~\mathcal{C}_{\kappa}[k]=0~,~~\mathcal{C}_{\kappa}[p']=0~,~~\mathcal{C}_{\kappa}[p'']=0\label{eqmotion3+3kmomentum}\ ,\\
\mathcal{K}_\mu^{[0]}(s_{0})=0~,~~\mathcal{K}_\mu^{[1]}(s_{1})=0\nonumber\ ,
\end{gather}
\begin{gather}
\dot x^\mu = \mathcal{N}_p \left(\frac{\delta \mathcal{C}_{\kappa}[p]}{\delta p_\mu} +\ell \delta^\mu_0 \frac{\delta \mathcal{C}_{\kappa}[p]}{\delta p_1} p_1\right)= \delta^{\mu}_{0}\mathcal{N}_p \left( 2 p_{0}-\ell p_{1}^{2}\right)-2\delta^{\mu}_{1}\mathcal{N}_p \left(  p_{1}-\ell p_{0} p_{1}\right)\nonumber \ ,\\
\dot y^\mu = \mathcal{N}_q \left(\frac{\delta \mathcal{C}_{\kappa}[q]}{\delta q_\mu} +\ell \delta^\mu_0 \frac{\delta \mathcal{C}_{\kappa}[q]}{\delta q_1} q_1\right)= \delta^{\mu}_{0}\mathcal{N}_q \left( 2 q_{0}-\ell q_{1}^{2}\right)-2\delta^{\mu}_{1}\mathcal{N}_q \left(  q_{1}-\ell q_{0} q_{1}\right)\nonumber \ ,\\
\dot z^\mu = \mathcal{N}_k \left(\frac{\delta \mathcal{C}_{\kappa}[k]}{\delta k_\mu} +\ell \delta^\mu_0 \frac{\delta \mathcal{C}_{\kappa}[k]}{\delta k_1} k_1\right)= \delta^{\mu}_{0}\mathcal{N}_k \left( 2 k_{0}-\ell k_{1}^{2}\right)-2\delta^{\mu}_{1}\mathcal{N}_k \left(  k_{1}-\ell k_{0} k_{1}\right)\nonumber \ ,\\
\dot x'^\mu =\mathcal{N}_{p'} \left(\frac{\delta \mathcal{C}_{\kappa}[p']}{\delta p'_\mu} +\ell \delta^\mu_0 \frac{\delta \mathcal{C}_{\kappa}[p']}{\delta p'_1} p'_1\right)=\delta^{\mu}_{0}\mathcal{N}_{p'} \left( 2 p_{0}'-\ell p_{1}'^{2}\right)-2\delta^{\mu}_{1}\mathcal{N}_{p'} \left(  p_{1}'-\ell p_{0}' p_{1}'\right)\ ,
\label{eqmotion3+3kspace}\\
\dot x''^\mu = \mathcal{N}_p'' \left(\frac{\delta \mathcal{C}_{\kappa}[p'']}{\delta p''_\mu} +\ell \delta^\mu_0 \frac{\delta \mathcal{C}_{\kappa}[k]}{\delta p''_1} p''_1\right)=\delta^{\mu}_{0}\mathcal{N}_{p''} \left( 2 p_{0}''-\ell p_{1}''^{2}\right)-2\delta^{\mu}_{1}\mathcal{N}_{p''} \left(  p_{1}''-\ell p_{0}'' p_{1}''\right)\nonumber\ .
\end{gather}

And also the conditions at the $s=s_{0}$ and $s=s_{1}$ boundaries
produced by the interaction terms are
 of rather standard relative-locality
type:
\begin{gather}
z^\mu(s_{0}) = \xi^\nu_{[0]} \left(\frac{\delta \mathcal{K}^{[0]}_\nu}{\delta k_\mu}+\ell \delta^\mu_0 \frac{\delta \mathcal{K}^{[0]}_\nu}{\delta k_1}k_1\right)=\xi^\mu_{[0]} +\ell \delta^{\mu}_{0} \xi^1_{[0]} k_{1}\nonumber\ ,\\
x^\mu(s_{0}) = -\xi^\nu_{[0]} \left(\frac{\delta \mathcal{K}^{[0]}_\nu}{\delta p_\mu}+\ell \delta^\mu_0 \frac{\delta \mathcal{K}^{[0]}_\nu}{\delta p_1}p_1\right)=\xi^\mu_{[0]} +\ell \delta^{\mu}_{0} \xi^1_{[0]} (p_{1}+q_{1})\ ,
 \qquad x^\mu(s_{1}) = \xi^\nu_{[1]} \left(\frac{\delta \mathcal{K}^{[1]}_\nu}{\delta p_\mu}+\ell \delta^\mu_0 \frac{\delta \mathcal{K}^{[1]}_\nu}{\delta p_1}p_1\right)=\xi^\mu_{[1]} +\ell \delta^{\mu}_{0} \xi^1_{[1]} (p_{1}+q_{1})\nonumber\ ,\\
y^\mu(s_{0})= -\xi^\nu_{[0]} \left(\frac{\delta \mathcal{K}^{[0]}_\nu}{\delta q_\mu}+\ell \delta^\mu_0 \frac{\delta \mathcal{K}^{[0]}_\nu}{\delta q_1}q_1\right)=\xi^\mu_{[0]} +\ell \delta^{\mu}_{0} \xi^1_{[0]} q_{1}+\ell \delta^{\mu}_{1} \xi^1_{[0]} p_{0}
\label{boundaries3+3k}\ ,\\
x'^\mu(s_{1}) = -\xi^\nu_{[1]} \left(\frac{\delta \mathcal{K}^{[1]}_\nu}{\delta p'_\mu}+\ell \delta^\mu_0 \frac{\delta \mathcal{K}^{[1]}_\nu}{\delta p'_1}p'_1\right)=\xi^\mu_{[1]} +\ell \delta^{\mu}_{0} \xi^1_{[1]} (p_{1}'+p_{1}''+q_{1})\nonumber\ ,\\
x''^\mu(s_{1}) = -\xi^\nu_{[1]} \left(\frac{\delta \mathcal{K}^{[1]}_\nu}{\delta p''_\mu}+\ell \delta^\mu_0 \frac{\delta \mathcal{K}^{[1]}_\nu}{\delta p''_1}p''_1\right)=\xi^\mu_{[1]} +\ell \delta^{\mu}_{0} \xi^1_{[1]} (p_{1}''+q_{1})+\ell \delta^{\mu}_{1} \xi^1_{[1]} p_{0}'\nonumber\ .
\end{gather}
 However, thanks to our tailored choice of momentum-conservation
constraints the boundary conditions
at the two endpoints of the finite worldline exchanged by the two interactions
(the finite worldline of particle coordinates $x^\mu(s)$ and momentum $p_\mu$)
match just in the right way to allow implementing
as a relativistic symmetry the following translation
transformations, generated by the total momentum
\begin{equation}
\begin{split}
z_{B}^{0}(s)&=  z_{A}^{0}(s)+b^{\mu} \{ k_{\mu},z^{0}\}
= z_{A}^{0}(s)-b^{0}-\ell b^{1}k_{1}\ ,\\
z_{B}^{1}(s)&=z_{A}^{1}(s)+b^{\mu} \{ k_{\mu},z^{1}\}
={z}_{A}^{1}(s)-b^{1}\ ,\\
x_{B}^{0}(s)&=  x_{A}^{0}(s) + b^\mu
 \{ (p\oplus q)_\mu , x^0\}
=x_{A}^{0}(s) - b^{0}-\ell b^{1}(p_{1}+q_1)\ , \\
x_{B}^{1}(s)&=x_{A}^{1}(s)+b^{\mu}
 \{ (p\oplus q)_\mu , x^0\}
={x}_{A}^{1}(s)-b^{1}\ ,\\
y_{B}^{0}(s)&=  y_{A}^{0}(s)+b^{\mu}  \{ (p\oplus q)_\mu , y^0\}
=y_{A}^{0}(s)-b^{0}-\ell b^{1}q_1\ ,\\
y_{B}^{1}(s)&=y_{A}^{1}(s)+b^{\mu} \{ (p\oplus q)_\mu , y^1\}
={y}_{A}^{1}(s)-b^{1}-\ell b^{1}p_0\ ,\\
{x'}_{B}^{0}(s)&={x'}_{A}^{0}(s)+b^{\mu} \{ (p'\oplus p'' \oplus q)_\mu , x'^0\}
={x'}_{A}^{0}(s)-b^{0}-\ell b^{1}(p'_{1}+p''_{1}+q_1)\ , \\
{x'}_{B}^{1}(s)&={x'}_{A}^{1}(s)+b^{\mu} \{ (p'\oplus p'' \oplus q)_\mu , x'^1\}
={x'}_{A}^{1}(s)-b^{1}\ ,\\
{x''}_{B}^{0}(s)&={x''}_{A}^{0}(s)+b^{\mu}
 \{ (p' \oplus p'' \oplus q)_\mu , x''^0\}
={x''}_{A}^{0}(s)-b^{0}-\ell b^{1}(p''_{1}+q_1)\ , \\
{x''}_{B}^{1}(s)&={x''}_{A}^{1}(s)+b^{\mu}
 \{ (p' \oplus p'' \oplus q)_\mu , x''^1\}
={x''}_{A}^{1}(s)-b^{1}-\ell b^1 p'_0\ .
\label{translations3+3k}
\end{split}
\end{equation}
It is again straightforward to see that these transformations leave the equations of motion (\ref{eqmotion3+3kspace}) unchanged by noticing, as done in the
previous subsection,
 that the only non trivial terms in the deformed translations (\ref{translations3+3k}) depend on momenta and
 the momenta are conserved along the worldlines.

And it is also easy to verify that our translation transformations
leave the boundary conditions unchanged. In order to give an explicit
example let us check the case of $x''^\mu$: substituting the translation calculated
in Eq.~(\ref{translations3+3k})
\begin{gather*}
{x''}_{B}^{0}(s)={x''}_{A}^{0}-b^{0}-\ell b^{1}(p''_{1}+q_1) \ ,\\
{x''}_{B}^{1}(s)={x''}_{A}^{1}-b^{1}-\ell b^1 p'_0 \ ,\\
\xi_B^\mu = \xi_A^\mu -b^\mu \ ,
\end{gather*}
in the boundary conditions (\ref{boundaries3+3k})
\begin{gather*}
 x_B''^0(s_{1}) = - \xi_{[1]B}^\nu \left(\frac{\delta \mathcal{K}^{[1]}_\nu}{\delta p''_0} +\ell  \frac{\delta \mathcal{K}^{[1]}_\nu}{\delta p''_1}p''_1\right) = \xi_{[1]B}^0+\ell \xi_{[1]B}^1( q_1+ p_1'') \ ,\\
x_B''^1(s_{1})= -\xi_{[1]B}^\nu \left(\frac{\delta \mathcal{K}^{[1]}_\nu}{\delta p''_1}\right)= \xi_{[1]B}^1(1+\ell p_0') \ ,
\end{gather*}
we find
\begin{gather*}
x_B''^0(s_{1}) - \xi_{[1]B}^0 - \ell \xi_{[1]B}^1( q_1+ p_1'') = {x''}_{A}^{0}(s_1) - \xi_{[1]A}^0 - \ell \xi_{[1]A}^1( q_1+ p_1'') \ , \\
x_B''^1(s_{1}) -\xi_{[1]B}^1(1+\ell p_0') = x_{A}''^1(s_1) -\xi_{[1]A}^1(1+\ell p_0') \ .
\end{gather*}
which is evidently consistent with our boundary conditions.

Thus we did succeed: even in the case of finite worldlines, causally connecting pairs
of interactions, our prescription for boundary terms
does ensure translational invariance, addressing the challenge highlighted
here in Section~\ref{challenge}. And our relative-locality distant observers
are connected by relativistic transformations generated by the ($\oplus$-deformed)
total momentum.

The invariance of the equations of motion and boundary conditions under our
translations generated by the total momentum is
also manifest in the properties of the action under these translation
 transformations. In fact, it turns out that these translation transformations
 do change the action  $S^{\kappa (2)}$, but
 only by terms
that do not contribute to the equations of motion (once the constraints
are taken into account).
In order to see this explicitly let us start by
noticing that we can split the integral
for the worldline $p$, $x$ in (\ref{actionk3+3full}) in the following way:
\begin{equation*}
 \int_{s_{0}}^{s_{1}}ds\left({x}^{\mu}\dot{p}_{\mu}-\ell{x}^{1}\dot{p}_{0}p_{1}+\mathcal{N}_{p}\mathcal{C}_{\kappa}\left[p\right]\right)= \int_{s_{0}}^{\infty}ds\left({x}^{\mu}\dot{p}_{\mu}-\ell{x}^{1}\dot{p}_{0}p_{1}+\mathcal{N}_{p}\mathcal{C}_{\kappa}\left[p\right]\right)-\int_{s_{1}}^{\infty}ds\left({x}^{\mu}\dot{p}_{\mu}-\ell{x}^{1}\dot{p}_{0}p_{1}
 +\mathcal{N}_{p}\mathcal{C}_{\kappa}\left[p\right]\right)\ .
\end{equation*}
So we can separate in the action (\ref{actionk3+3full}) the contributions relative to the interactions at $s_{0}$ and $s_{1}$
(contributions with boundary at $s_{0}$ and
 contributions with boundary at $s_{1}$).
 The part relative to the vertex $s_{0}$ is the same as the action (\ref{action k 3}) analyzed in the previous section. We consider then only the contributions with boundary
 at $s_{1}$:
\begin{equation*}
 \begin{split}\Delta \mathcal{S}^{\kappa(2)}_{s_{1}} & = -\int_{s_{1}}^{\infty}ds\left(-b^{\mu}\dot{p}_{\mu}-\ell b^{1}q_{1}\dot{p}_{0}\right)+\int_{s_{1}}^{\infty}ds\left(-b^{\mu}\dot{p}_{\mu}'-\ell b^{1}p_{1}''\dot{p}_{0}'-\ell b^{1}q_{1}\dot{p}_{0}'\right)\\
 & +\int_{s_{1}}^{\infty}ds\left(-b^{\mu}\dot{p}_{\mu}''-\ell b^{1}q_{1}\dot{p}_{0}''-\ell b^{1}p_{0}'\dot{p}_{1}''\right)+\Delta\xi_{\left[1\right]}^{\mu}\mathcal{K}_{\mu}^{\left[1\right]}(s_{1})\ .
\end{split}
\end{equation*}
This evidently can be rewritten as
\begin{equation*}
 \begin{split}\Delta\mathcal{S}^{\kappa\left(2\right)}_{s_{1}} & =\int_{s_{1}}^{\infty}ds\left(b^{\mu}\left(\dot{p}_{\mu}-\dot{p}_{\mu}'-\dot{p}_{\mu}''\right)-\ell{b}^{1}\left(-q_{1}\dot{p}_{0}+p_{1}''\dot{p}_{0}'+q_{1}\dot{p}_{0}'+q_{1}\dot{p}_{0}''+p_{0}'\dot{p}_{1}''\right)\right)+\Delta\xi_{\left[1\right]}^{\mu}\mathcal{K}_{\mu}^{\left[1\right]}(s_{1})\ ,
\end{split}
\end{equation*}
which, taking into account Eq.~(\ref{K[1]}), gives
\begin{equation*}
\begin{split}\Delta\mathcal{S}^{\kappa\left(2\right)}_{s_{1}} & = \int_{s_{1}}^{\infty}ds\left(\frac{d}{ds}\left[b^{\mu}{\cal K}_{\mu}^{\left[1\right]}\right]-\ell b^{1}{\cal K}_{0}^{\left[1\right]}\dot{q}_{1}\right) + \Delta\xi_{\left[1\right]}^{\mu}\mathcal{K}_{\mu}^{\left[1\right]}(s_{1})\ .
\end{split}
\end{equation*}
\label{invariance}
The total derivative contributes as before
to the translation of $\xi^\mu_{[1]}$: $\xi^\mu_{[1]B}=\xi^\mu_{[1]A} - b^\mu$.
In addition there is a left over bulk term,
$$\int_{s_{1}}^{\infty}ds\ell b^{1}{\cal K}_{0}^{\left[1\right]}\dot{q}_{1}\ ,$$
but it is evidently inconsequential for what concerns
the equations of motion. In fact, varying this
left-over term one finds
\begin{equation*}
 \int_{s_{1}}^{\infty}ds\ell b^{1}\left(\delta{\cal K}_{0}^{\left[1\right]}\dot{q}_{1}+{\cal K}_{0}^{\left[1\right]}\delta\dot{q}_{1}\right)\ ,
\end{equation*}
{\it i.e.}
\begin{equation*}
 \int_{s_{1}}^{\infty}ds\ell b^{1}\left(\delta{\cal K}_{0}^{\left[1\right]}\dot{q}_{1}-\dot{{\cal K}}_{0}^{\left[1\right]}\delta q_{1}+\frac{d}{ds}{\cal K}_0^{\left[1\right]}\delta q_{1}\right)\ .
\end{equation*}
And the 3 terms in this expression contribute to equations of motion
and boundary conditions only terms which are already fixed to vanish because
of constraints derived from other parts of the action (specifically
$\dot p_\mu =0~,~~\dot q_\mu =0~,~~\dot p'_\mu =0~,~~\dot p''_\mu =0$
and ${\cal K}_0^{\left[1\right]} =0$).

\subsection{Aside on an alternative choice of action}\label{asidevuoto}
In the previous subsection we showed that
causally-connected interactions, with relative locality,
can be formulated consistently with translational invariance,
and therefore
admit a relativistic description of distant observers.
Crucial for our result was noticing that the equations
of motion and the boundary conditions
are invariant under our proposed translation transformations,
generated by the total momentum,
even though those translation transformations did not leave
the action unchanged in the bulk. For completeness in this subsection
we want to show that exactly the same physical proposal of the previous
subsection can be given in terms of a different action,
 with slightly different boundary terms (at endpoints of
worldlines).

Ultimately the difference between the two alternatives we shall then have
amounts to the properties of the two actions under the same
laws of translation transformation: the case in the previous subsection
was such that translation transformations changed the action in the bulk
(but without changing the equations of motion), while the case we
discuss in this subsection will turn out to be such that
translation transformations change the action on the boundary,
but without affecting the boundary conditions.
The two actions give exactly the same physical picture.

\begin{figure}[hb!]
\begin{center}
\includegraphics[width=0.66 \textwidth]{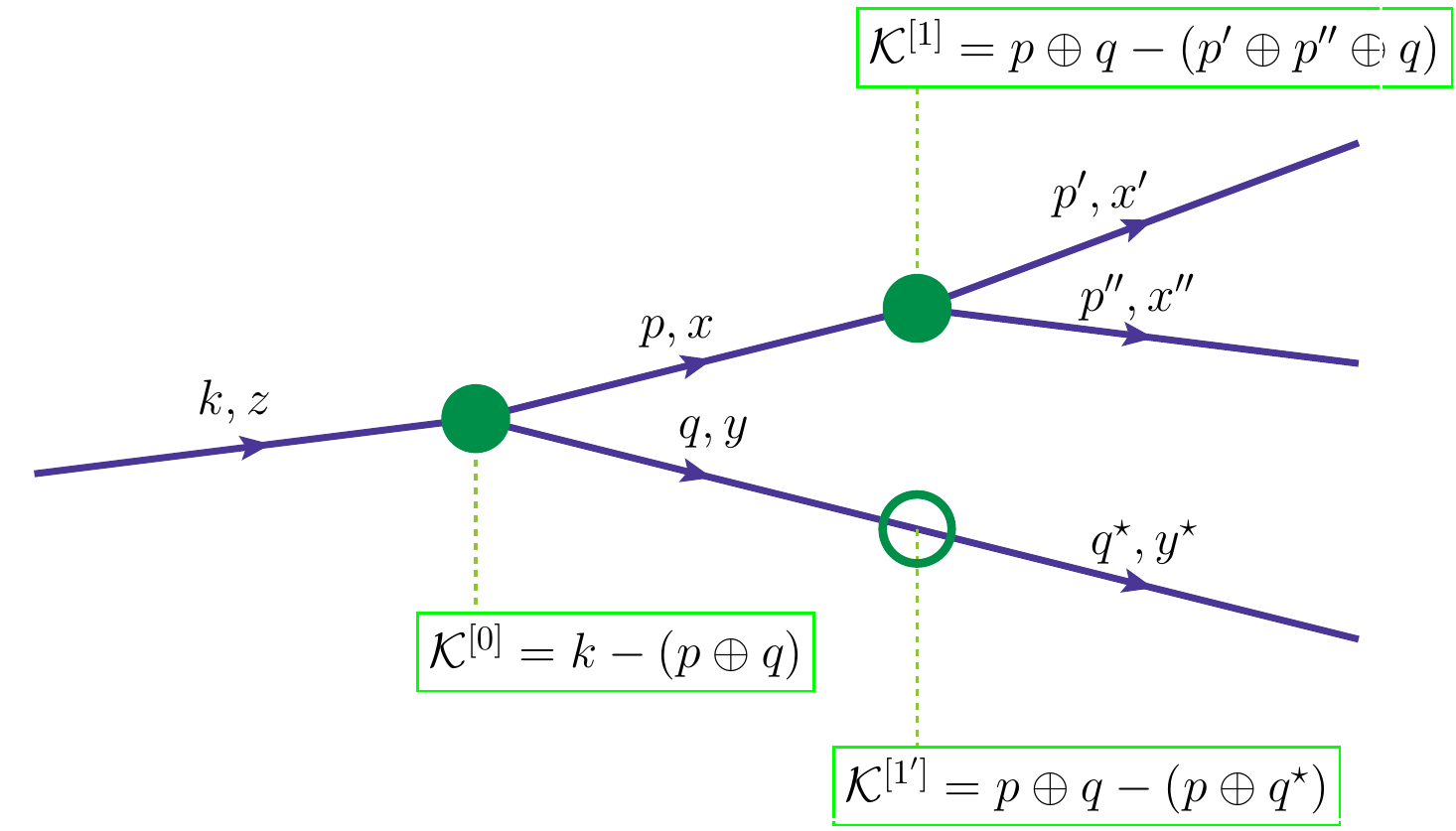}
\caption{Our choices of boundary terms
for a pair of causally-connected interactions,
when using the expedient of a bivalent interaction in combination with
the second trivalent interaction.}\label{figunovuoto}
\end{center}
\end{figure}

We consider exactly the same configuration already analyzed in the
previous subsection, but
(as hinted at in Fig.~\ref{figunovuoto})
in addition to the
constraints given in terms of
$${\cal K}^{[0]} (s_{0})= k-(p\oplus q) \ ,$$
$${\cal K}^{[1]} (s_{1})= (p\oplus q) - ( p'\oplus p'' \oplus q)  \ ,$$
we add, as a technical expedient, another interaction
also at $s=1$, a bivalent interaction (a non-interaction)
characterized by the a constraint given in terms of
$${\cal K}^{[1']} (s_{1})= (p\oplus q) -  (p\oplus q^\star)  \ .$$

The added (fictitious) bivalent interaction
leads to replacing the action $S^{\kappa (2)}$
with
 \begin{equation}
\begin{split}
\mathcal{S}^{\kappa (2')}
= & \int_{-\infty}^{s_{0}}ds\left(z^{\mu}\dot{k}_{\mu}-\ell z^{1}k_{1}\dot{k}_{0}+\mathcal{N}_k\mathcal{C}_\kappa\left[k\right]\right)+
\int_{s_{0}}^{s_{1}}ds\left({x}^{\mu}\dot{p}_{\mu}-\ell x^{1}p_{1}\dot{p}_{0}+\mathcal{N}_p\mathcal{C}_\kappa\left[p\right]\right)\\
 & +\int_{s_{0}}^{s_{1}}ds\left(y^{\mu}\dot{q}_{\mu}-\ell y^{1}q_{1}\dot{q}_{0}+\mathcal{N}_q\mathcal{C}_\kappa\left[q\right]\right)+
 \int_{s_{1}}^{\infty}ds\left(y_{\star}^{\mu}\dot{q}_{\mu}^{\star}-\ell y_{\star}^{1}q_{1}^{\star}\dot{q}_{0}^{\star}+\mathcal{N}_{q^\star}\mathcal{C}_\kappa\left[q^{\star}\right]\right)\\
 & +\int_{s_{1}}^{\infty}ds\left({x'}^{\mu}\dot{p'}_{\mu}-\ell{x'}^{1}p'_{1}\dot{p'}_{0}+\mathcal{N}_{p'}\mathcal{C}_\kappa\left[p'\right]\right)+
 \int_{s_{1}}^{\infty}ds\left(x''^{\mu}\dot{p''}_{\mu}-\ell x''^{1}p''_{1}\dot{p''}_{0}+\mathcal{N}_{p''}\mathcal{C}_\kappa\left[p''\right]\right)\\
 & -\xi_{[0]}^{\mu}\mathcal{K}_{\mu}^{[0]}(s_{0})-\xi_{[1]}^{\mu}\mathcal{K}_{\mu}^{[1]}(s_{1})-\xi_{[1']}^{\mu}\mathcal{K}_{\mu}^{[1']}(s_{1})\ .\label{actionk3+3+2full}
\end{split}
\end{equation}
In Appendix A we show that this action produces exactly the same equations
of motion and boundary conditions as the action considered in the previous
subsection, with only peculiarity that (as suggested by the
drawing in Fig.~\ref{figunovuoto})
the worldline $y^\mu$, $q_\mu$
of the previous subsection gets here fictitiously split into two perfectly-matching
pieces of worldline, a piece labeled again $y^\mu$, $q_\mu$ and a piece
labeled $y_\star^\mu$, $q^\star_\mu$.

The same applies for our description
of translation transformations
generated by the total momentum, which for the action $\mathcal{S}^{\kappa (2')}$
takes the form
\begin{equation}
\begin{split}
z_{B}^{0}(s)&=z_{A}^{0}(s)+b^{\mu}  \{ k_\mu , z^0\}=  z_{A}^{0}(s)-b^{0}-\ell b^{1}k_{1}\ ,\\
z_{B}^{1}(s)&=z_{A}^{1}(s)+b^{\mu}  \{ k_\mu , z^1\}={z}_{A}^{1}(s)-b^{1}\ ,\\
x_{B}^{0}(s)&=x_{A}^{0}(s)+b^{\mu}  \{ (p\oplus q)_\mu , x^0\}= x_{A}^{0}(s)-b^{0}-\ell b^{1}(p_{1}+q_1) \ ,\\
x_{B}^{1}(s)&=x_{A}^{1}(s)+b^{\mu}  \{ (p\oplus q)_\mu , x^1\}={x}_{A}^{1}(s)-b^{1}\ ,\\
y_{B}^{0}(s)&=y_{A}^{0}(s)+b^{\mu}  \{ (p\oplus q)_\mu , y^0\}= y_{A}^{0}(s)-b^{0}-\ell b^{1}q_1\ ,\\
{y_{B}^{1}}(s)&=y_{A}^{1}(s)+b^{\mu}  \{ (p\oplus q)_\mu , y^1\}={y}_{A}^{1}(s)-b^{1}-\ell b^{1}p_0\ ,\\
{y_{B}^{\star 0}}(s)&=y_{A}^{\star 0}(s)+b^{\mu}  \{ (p'\oplus p''\oplus q^{\star})_\mu , y^{\star 0}\}={y_{A}^{\star}}^{0}(s)-b^{0}-\ell b^{1}q^{\star }_1\ ,\\
{y_{B}^{\star 1}}(s)&=y_{A}^{\star1}(s)+b^{\mu}  \{ (p'\oplus p''\oplus q^{\star})_\mu , y^{\star 1}\}={y}_{A}^{1}(s)-b^{1}-\ell b^{1}(p'_0+p''_0)\ ,\\
{x'}_{B}^{0}(s)&=x_{A}'^{0}(s)+b^{\mu}  \{ (p'\oplus p''\oplus q^{\star})_\mu , x'^0\}={x'}_{A}^{0}(s)-b^{0}-\ell b^{1}(p'_{1}+p''_{1}+q^\star_1) \ ,\\
{x'}_{B}^{1}(s)&=x_{A}'^{1}(s)+b^{\mu}  \{ (p'\oplus p''\oplus q^{\star})_\mu , x'^1\}={x'}_{A}^{1}(s)-b^{1}\ ,\\
{x''}_{B}^{0}(s)&=x_{A}''^{0}(s)+b^{\mu}  \{ (p'\oplus p''\oplus q^{\star})_\mu , x''^0\}={x''}_{A}^{0}(s)-b^{0}-\ell b^{1}(p''_{1}+q^\star_1) \ ,\\
{x''}_{B}^{1}(s)&=x_{A}''^{1}(s)+b^{\mu}  \{ (p'\oplus p''\oplus q^{\star})_\mu , x''^1\}={x''}_{A}^{1}(s)-b^{1}-\ell b^1 p'_0\ .
\end{split}
\label{translations3p3p2k}
\end{equation}
These translation transformations are symmetries of
the equations of motion and boundary conditions (given in Appendix A),
 and they essentially are the same translation transformations we discussed
 in the previous subsection, up to splitting again
fictitiously the worldline $y^\mu$, $q_\mu$
into pieces $y^\mu$, $q_\mu$ and $y_\star^\mu$, $q^\star_\mu$.

It is also easy to check that
the action $S^{\kappa (2')}$ does change under this translation
transformations, but only by an amount that can be expressed
in terms of other boundary terms. Indeed repeating the same steps as before, i.e. substituting the relations (\ref{translations3p3p2k}) in the action (\ref{actionk3+3+2full}), and repeating the algebraic manipulations showed previously, we find
 \begin{equation}
\begin{split}
\mathcal{S}^{\kappa (2')}_B = \mathcal{S}^{\kappa (2')}_A+\ell b^1 \mathcal{K}_{0}^{[1]}\mathcal{K}_{1}^{[1']}\ .
\end{split}
\end{equation}
And it is particularly clear that this additional boundary term for
the ``translated action" is irrelevant: it has no implication
on the boundary conditions (it would produce additional boundary conditions
which however are automatically satisfied once the other boundary conditions
are enforced).

\subsection{The case of 3 connected finite worldlines}\label{sec3conn}

At this point we have established that at least in the simplest applications
our prescriptions do provide the desired relativistic picture.
In order to motivate our next consistency check it is useful to
look at available results on relative locality from the following
perspective:\\
$\star$ with the Hamiltonian description of relative locality
for free particles given in
Refs.~\cite{whataboutbob,leeINERTIALlimit,arzkowaRelLoc,kappabob}
one essentially obtains a characterization of relative locality limited
to {\underline{infinite}} worldlines;\\
$\star$ with the Lagrangian description of relative locality
for interacting particles proposed in Ref.~\cite{prl}
the availability of a relativistic description of distant observers
had been checked explicitly only for {\underline{semi-infinite}} worldlines
(a single interaction);\\
$\star$ the results reported so far in this section
 generalize the results for distant observers
of Ref.~\cite{prl} to the case where one
of the worldlines is
{\underline{finite}} (a worldline exchanged between two interactions, establishing
the causal relation between the two interactions).

\begin{figure}[h!]
\begin{center}
\includegraphics[width=0.75 \textwidth]{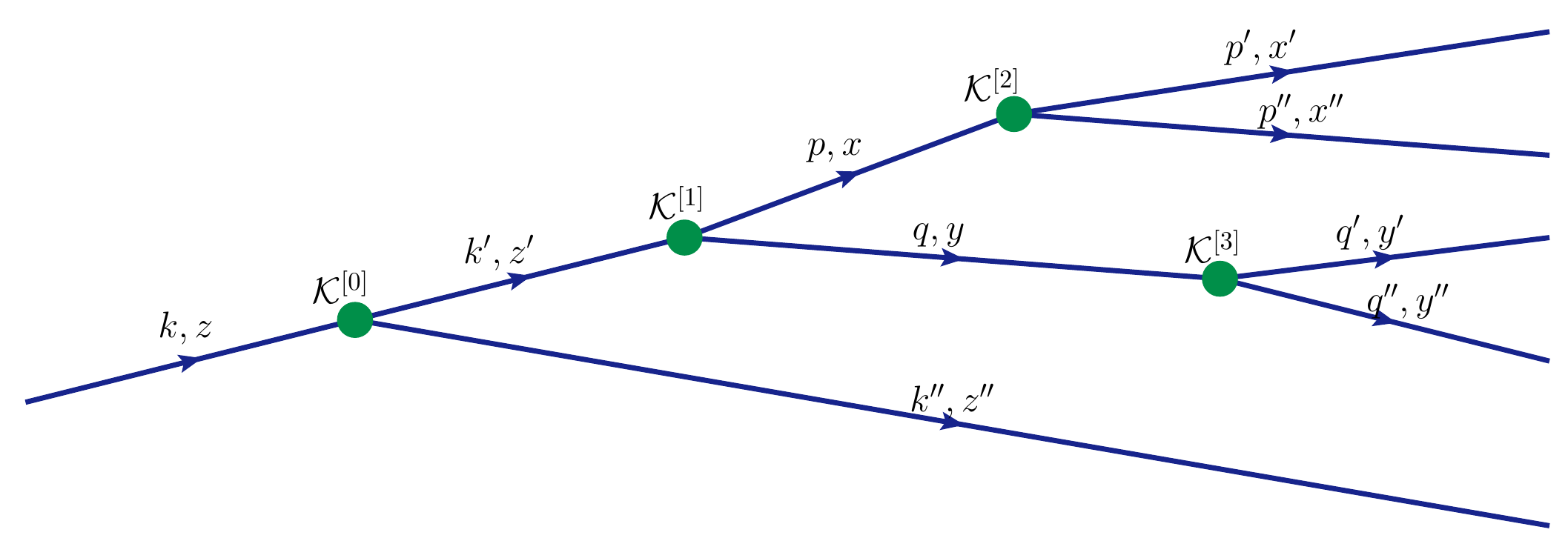}
\caption{The case with 3 connected finite worldlines which we consider
in this subsection.}
\label{figsurrounded}
\end{center}
\end{figure}

In this subsection we provide evidence of the fact that our
prescription is robust also for cases with several finite worldlines.
We actually consider here a case which is very meaningful from this perspective:
the case shown in
Figure~\ref{figsurrounded}, which includes a vertex where 3 finite worldlines meet.

Following the prescription we are advocating
the situation in Figure~\ref{figsurrounded}
 requires handling  boundary terms with
\begin{equation}
\begin{split}
\mathcal{K}^{[0]}&= k - k' \oplus k''\ ,\\
\mathcal{K}^{[1]}&=k' \oplus k'' - p \oplus q \oplus k''\ , \\
\mathcal{K}^{[2]}&=p \oplus q \oplus k''  - p' \oplus p'' \oplus q \oplus k'' \ ,\\
\mathcal{K}^{[3]}&=p' \oplus p'' \oplus q \oplus k'' - p' \oplus p'' \oplus q' \oplus q'' \oplus k''\ .\\
\end{split}
\end{equation}
We therefore describe the chain of interactions in  Figure~\ref{figsurrounded}
through the following action:
\begin{equation}
\begin{split}
\mathcal{S}^{(3conn)}= & \int_{-\infty}^{s_{0}}ds\left(z^{\mu}\dot{k}_{\mu}-\ell z^{1}k_{1}\dot{k}_{0}+\mathcal{N}_{k}\mathcal{C}\left[k\right]\right)
+\int_{s_{0}}^{+\infty}ds\left(z''^{\mu}\dot{k}''_{\mu}-\ell z''^{1}k''_{1}\dot{k}''_{0}+\mathcal{N}_{k''}\mathcal{C}\left[k\right]\right)\\
&+\int_{s_{0}}^{s_{1}}ds\left(z'^{\mu}\dot{k'}_{\mu}-\ell z'^{1}k'_{1}\dot{k}'_{0}+\mathcal{N}_{k'}\mathcal{C}\left[k\right]\right)
+\int_{s_{1}}^{s_{2}}ds\left({x}^{\mu}\dot{p}_{\mu}-\ell x^{1}p_{1}\dot{p}_{0}+\mathcal{N}_{p}\mathcal{C}\left[p\right]\right)\\
& +\int_{s_{2}}^{+\infty}ds\left({x'}^{\mu}\dot{p}'_{\mu}-\ell{x'}^{1}p'_{1}\dot{p}'_{0}+\mathcal{N}_{p'}\mathcal{C}\left[p'\right]\right)
+\int_{s_{2}}^{+\infty}ds\left(x''^{\mu}\dot{p}''_{\mu}-\ell x''^{1}p''_{1}\dot{p}''_{0}+\mathcal{N}_{p''}\mathcal{C}\left[p''\right]\right)\\
&+\int_{s_{1}}^{s_{3}}ds\left(y^{\mu}\dot{q}_{\mu}-\ell y^{1}q_{1}\dot{q}_{0}+\mathcal{N}_{q}\mathcal{C}\left[q\right]\right)
+ \int_{s_{3}}^{+\infty}ds\left(y'^{\mu}\dot{q}'_{\mu}-\ell y'^{1}q'_{1}\dot{q}'_{0}+\mathcal{N}_{q'}\mathcal{C}\left[q'\right]\right)\\
&+\int_{s_{3}}^{+\infty}ds\left( y''^{\mu}\dot{q}''_{\mu}-\ell y''^{1}q''_{1}\dot{q}''_{0}+\mathcal{N}_{q''}\mathcal{C}\left[q''\right]\right)\\
& -\xi_{[0]}^{\mu}\mathcal{K}_{\mu}^{[0]}(s_{0})-\xi_{[1]}^{\mu}\mathcal{K}_{\mu}^{[1]}(s_{1})-\xi_{[2]}^{\mu}\mathcal{K}_{\mu}^{[2]}(s_{2})-\xi_{[3]}^{\mu}\mathcal{K}_{\mu}^{[3]}(s_{3})\ .
\end{split}
\end{equation}

For what concerns equations of motion
and boundary conditions we have
\begin{gather}
\dot p_\mu =0~,~~\dot p'_\mu =0~,~~\dot p''_\mu =0~,~~\dot q_\mu =0~,~~\dot q'_\mu =0~,~~\dot q''_\mu =0~,~~\dot k_\mu =0~,~~\dot k'_\mu =0~,~~\dot k''_\mu =0\nonumber\ ,\\
\mathcal{C}_{\kappa}[p]=0~,~~\mathcal{C}_{\kappa}[p']=0~,~~\mathcal{C}_{\kappa}[p'']
=0~,~~\mathcal{C}_{\kappa}[q]=0~,~~\mathcal{C}_{\kappa}[q']=0~,~~\mathcal{C}_{\kappa}[q'']=0
~,~~\mathcal{C}_{\kappa}[k]=0~,~~\mathcal{C}_{\kappa}[k']=0~,~~\mathcal{C}_{\kappa}[k'']=0\ ,
\nonumber
\end{gather}
\begin{gather}
\dot x^\mu - \mathcal{N}_p \left(\frac{\delta \mathcal{C}_{\kappa}[p]}{\delta p_\mu} +\ell \delta^\mu_0 \frac{\delta \mathcal{C}_{\kappa}[p]}{\delta p_1} p_1\right)=0~,~~~
\dot x'^\mu - \mathcal{N}_{p'} \left(\frac{\delta \mathcal{C}_{\kappa}[p']}{\delta p'_\mu}+\ell \delta^\mu_0 \frac{\delta \mathcal{C}_{\kappa}[p']}{\delta p'_1} p'_1\right)=0~,~~~
\dot x''^\mu - \mathcal{N}_{p''} \left(\frac{\delta \mathcal{C}_{\kappa}[p'']}{\delta p''_\mu} +\ell \delta^\mu_0 \frac{\delta \mathcal{C}_{\kappa}[k]}{\delta p''_1} p''_1\right)=0\nonumber\ ,\\
\dot y^\mu - \mathcal{N}_q \left(\frac{\delta \mathcal{C}_{\kappa}[q]}{\delta q_\mu} +\ell \delta^\mu_0 \frac{\delta \mathcal{C}_{\kappa}[q]}{\delta q_1} q_1\right)=0~,~~~
{\dot y}'^\mu - \mathcal{N}_{q'} \left(\frac{\delta \mathcal{C}_{\kappa}[q']}{\delta q_\mu'} +\ell \delta^\mu_0 \frac{\delta \mathcal{C}_{\kappa}[q']}{\delta q_1'} q_1'\right)=0~,~~~
{\dot y}''^\mu - \mathcal{N}_{q''} \left(\frac{\delta \mathcal{C}_{\kappa}[q'']}{\delta q_\mu''} +\ell \delta^\mu_0 \frac{\delta \mathcal{C}_{\kappa}[q'']}{\delta q_1''} q_1''\right)=0\nonumber\ ,\\
\dot z^\mu - \mathcal{N}_k \left(\frac{\delta \mathcal{C}_{\kappa}[k]}{\delta k_\mu} +\ell \delta^\mu_0 \frac{\delta \mathcal{C}_{\kappa}[k]}{\delta k_1} k_1\right)=0~,~~~
{\dot z}'^\mu - \mathcal{N}_k' \left(\frac{\delta \mathcal{C}_{\kappa}[k']}{\delta k_\mu'} +\ell \delta^\mu_0 \frac{\delta \mathcal{C}_{\kappa}[k']}{\delta k_1} k_1'\right)=0~,~~~
{\dot z}''^\mu - \mathcal{N}_k'' \left(\frac{\delta \mathcal{C}_{\kappa}[k'']}{\delta k_\mu''} +\ell \delta^\mu_0 \frac{\delta \mathcal{C}_{\kappa}[k'']}{\delta k_1''} k_1''\right)=0~,~~~
\nonumber
\end{gather}
\begin{gather}
z^\mu(s_{0}) = \xi^\nu_{[0]} \left(\frac{\delta \mathcal{K}^{[0]}_\nu}{\delta k_\mu}+\ell \delta^\mu_0 \frac{\delta \mathcal{K}^{[0]}_\nu}{\delta k_1}k_1\right)\nonumber\ ,\\
z'^\mu(s_{0}) = -\xi^\nu_{[0]} \left(\frac{\delta \mathcal{K}^{[0]}_\nu}{\delta k'_\mu}+\ell \delta^\mu_0 \frac{\delta \mathcal{K}^{[0]}_\nu}{\delta k'_1}k'_1\right)\nonumber, \qquad
z'^\mu(s_{1}) = \xi^\nu_{[1]} \left(\frac{\delta \mathcal{K}^{[1]}_\nu}{\delta k'_\mu}+\ell \delta^\mu_0 \frac{\delta \mathcal{K}^{[1]}_\nu}{\delta k'_1}k'_1\right)\nonumber\ ,\\
{z''}^\mu(s_{0}) = -\xi^\nu_{[0]} \left(\frac{\delta \mathcal{K}^{[0]}_\nu}{\delta k_\mu''}+\ell \delta^\mu_0 \frac{\delta \mathcal{K}^{[0]}_\nu}{\delta k_1''}k_1''\right)\nonumber\ ,\\
x^\mu(s_{1}) = -\xi^\nu_{[1]} \left(\frac{\delta \mathcal{K}^{[1]}_\nu}{\delta p_\mu}+\ell \delta^\mu_0 \frac{\delta \mathcal{K}^{[1]}_\nu}{\delta p_1}p_1\right)\ , \qquad x^\mu(s_{2}) = \xi^\nu_{[2]} \left(\frac{\delta \mathcal{K}^{[2]}_\nu}{\delta p_\mu}+\ell \delta^\mu_0 \frac{\delta \mathcal{K}^{[2]}_\nu}{\delta p_1}p_1\right)\nonumber\ ,\\
y^\mu(s_{1})= -\xi^\nu_{[1]} \left(\frac{\delta \mathcal{K}^{[1]}_\nu}{\delta q_\mu}+\ell \delta^\mu_0 \frac{\delta \mathcal{K}^{[1]}_\nu}{\delta q_1}q_1\right)\ ,\qquad y^\mu(s_{3})= \xi^\nu_{[3]} \left(\frac{\delta \mathcal{K}^{[3]}_\nu}{\delta q_\mu}+\ell \delta^\mu_0 \frac{\delta \mathcal{K}^{[3]}_\nu}{\delta q_1}q_1\right)\ ,
\nonumber\\
x'^\mu(s_{2}) = -\xi^\nu_{[2]} \left(\frac{\delta \mathcal{K}^{[2]}_\nu}{\delta p'_\mu}+\ell \delta^\mu_0 \frac{\delta \mathcal{K}^{[2]}_\nu}{\delta p'_1}p'_1\right)\nonumber\ ,\\
x''^\mu(s_{2}) = -\xi^\nu_{[2]} \left(\frac{\delta \mathcal{K}^{[2]}_\nu}{\delta p''_\mu}+\ell \delta^\mu_0 \frac{\delta \mathcal{K}^{[2]}_\nu}{\delta p''_1}p''_1\right)\nonumber\ ,\\
y'^{\mu} (s_{3}) = -\xi^\nu_{[3]} \left(\frac{\delta \mathcal{K}^{[3]}_\nu}{\delta q_\mu'}+\ell \delta^\mu_0 \frac{\delta \mathcal{K}^{[3]}_\nu}{\delta q_1'}q_1'\right)\nonumber\ ,\\
y''^{\mu} (s_{3}) = -\xi^\nu_{[3]} \left(\frac{\delta \mathcal{K}^{[3]}_\nu}{\delta q_\mu''}+\ell \delta^\mu_0 \frac{\delta \mathcal{K}^{[3]}_\nu}{\delta q_1''}q_1''\right)\nonumber\ .
\end{gather}

And our notion of translation transformation to
a distant observer is such that
\begin{equation}
\begin{split}
x_{B}^{0}&=x_{A}^{0}+b^{\mu}  \{ (p\oplus q\oplus k'')_\mu , x^0\}= x_{A}^{0}(s)-b^{0}-\ell b^{1}(p_1+q_1+k''_{1})\ ,\\
x_{B}^{1}&=x_{A}^{1}+b^{\mu}  \{ (p\oplus q\oplus k'')_\mu , x^1\}={x}_{A}'^{1}(s)-b^{1} \ ,\\
x_{B}'^{0}&=x_{A}'^{0}+b^{\mu}  \{ (p'\oplus p''\oplus q\oplus k'')_\mu , x'^0\}= x_{A}'^{0}(s)-b^{0}-\ell b^{1}(p'_1+p''_1+q_1+k''_{1})\ ,\\
x_{B}'^{1}&=x_{A}'^{1}+b^{\mu}  \{ (p'\oplus p''\oplus q\oplus k'')_\mu , x'^1\}={x}_{A}'^{1}(s)-b^{1} \ ,\\
x_{B}''^{0}&=x_{A}''^{0}+b^{\mu}  \{ (p'\oplus p''\oplus q\oplus k'')_\mu , x''^0\}= x_{A}''^{0}(s)-b^{0}-\ell b^{1}(p''_1+q_1+k''_{1})\ ,\\
x_{B}''^{1}&=x_{A}''^{1}+b^{\mu}  \{ (p'\oplus p''\oplus q\oplus k'')_\mu , x''^1\}={x}_{A}''^{1}(s)-b^{1} -\ell b^1 p'_0\ ,\\
y_{B}^{0}&=y_{A}^{0}+b^{\mu}  \{ (p\oplus q\oplus k'')_\mu , y^0\}= y_{A}^{0}(s)-b^{0}-\ell b^{1}(q_1+k''_{1})\ ,\\
y_{B}^{1}&=y_{A}^{1}+b^{\mu}  \{ (p\oplus q\oplus k'')_\mu , y^1\}={y}_{A}^{1}(s)-b^{1} -\ell b^{1}p_{0}\ ,\\
y_{B}'^{0}&=y_{A}'^{0}+b^{\mu}  \{ (p'\oplus p''\oplus q'\oplus q''\oplus k'')_\mu , y'^0\}= y_{A}'^{0}(s)-b^{0}-\ell b^{1}(q'_1+q''_1+k''_{1})\ ,\\
y_{B}'^{1}&=y_{A}'^{1}+b^{\mu}  \{ (p'\oplus p''\oplus q'\oplus q''\oplus k'')_\mu , y'^1\}={y}_{A}'^{1}(s)-b^{1} -\ell b^{1}(p'_{0}+p''_{0})\ ,\\
y_{B}''^{0}&=y_{A}''^{0}+b^{\mu}  \{ (p'\oplus p''\oplus q'\oplus q''\oplus k'')_\mu , y''^0\}= y_{A}''^{0}(s)-b^{0}-\ell b^{1}(q''_1+k''_{1})\ ,\\
y_{B}''^{1}&=y_{A}''^{1}+b^{\mu}  \{ (p'\oplus p''\oplus q'\oplus q''\oplus k'')_\mu , y''^1\}={y}_{A}''^{1}(s)-b^{1} -\ell b^{1}(p'_{0}+p''_{0}+q'_0)\ ,\\
z_{B}^{0}&=z_{A}^{0}+b^{\mu}  \{ k_\mu , z^0\}= z_{A}^{0}(s)-b^{0}-\ell b^{1}k_{1}\ ,\\
z_{B}^{1}&=z_{A}^{1}+b^{\mu}  \{ k_\mu , z^1\}={z}_{A}^{1}(s)-b^{1}\ ,\\
z_{B}'^{0}&=z_{A}'^{0}+b^{\mu}  \{ (k'\oplus k'')_\mu , z'^0\}= z_{A}'^{0}(s)-b^{0}-\ell b^{1}(k'_{1}+k''_{1})\ ,\\
z_{B}'^{1}&=z_{A}'^{1}+b^{\mu}  \{ (k'\oplus k'')_\mu , z'^1\}={z}_{A}'^{1}(s)-b^{1}\ ,\\
z_{B}''^{0}&=z_{A}''^{0}+b^{\mu}  \{ (k'\oplus k'')_\mu , z''^0\}= z_{A}''^{0}(s)-b^{0}-\ell b^{1}k''_{1}\ ,\\
z_{B}''^{1}&=z_{A}''^{0}+b^{\mu}  \{ (k'\oplus k'')_\mu , z''^0\}={z}_{A}''^{1}(s)-b^{1}-\ell b^{1}k'_0\ .
\end{split}
\label{trasla3conn}
\end{equation}

It is then easy to check that
also in this case the equations of motion
and boundary conditions are left unchanged
by our notion of translation transformation to
a distant observer.
This is also verifiable by studying the implications
of our translation transformations for the action ${\cal S}^{(3conn)}$,
to which we devote Appendix B.

\section{Implications for the times of arrival of simultaneously-emitted
ultrarelativistic particles}\label{secgrbs}

In the previous section we established the basic notions
and key characterizing results of
our proposal of a first example of prescriptions for boundary terms
ensuring a relativistic description of distant observers
within the relative-locality framework of Ref.~\cite{prl},
with a lagrangian formulation of interacting particles.

In this section we extend the scopes of our analysis
slightly beyond basics, by focusing on a first point
of phenomenological relevance, concerning
observations of distant bursts of massless particles.

In the process we shall also show that there is an appropriate limit
where our more powerful formalism
reproduces the
previous results (here reviewed in Section~\ref{kbobreview}) of the Hamiltonian
description of free $\kappa$-Minkowski particles.

\subsection{Matching Lagrangian and Hamiltonian description of $\kappa$-Minkowski
free particles}\label{kappabobprl}

Let us indeed start this section by
showing that our proposal for translation transformations,
besides fulfilling the demands of
relativistic consistency verified in the previous
section, also has the welcome property of reproducing the
previous results (here reviewed in Section~\ref{kbobreview})
 of the Hamiltonian
description of free $\kappa$-Poincar\'e particles.
Of course this occurs in an appropriate limit of our framework,
since in general our framework describes interacting $\kappa$-Poincar\'e
particles.
A key observation from this perspective is that a particle
is still ``essentially free" when its interactions only involve
exchanges of very small fractions of its momentum.

\begin{figure}[hb!]
\begin{center}
\includegraphics[width=0.72 \textwidth]{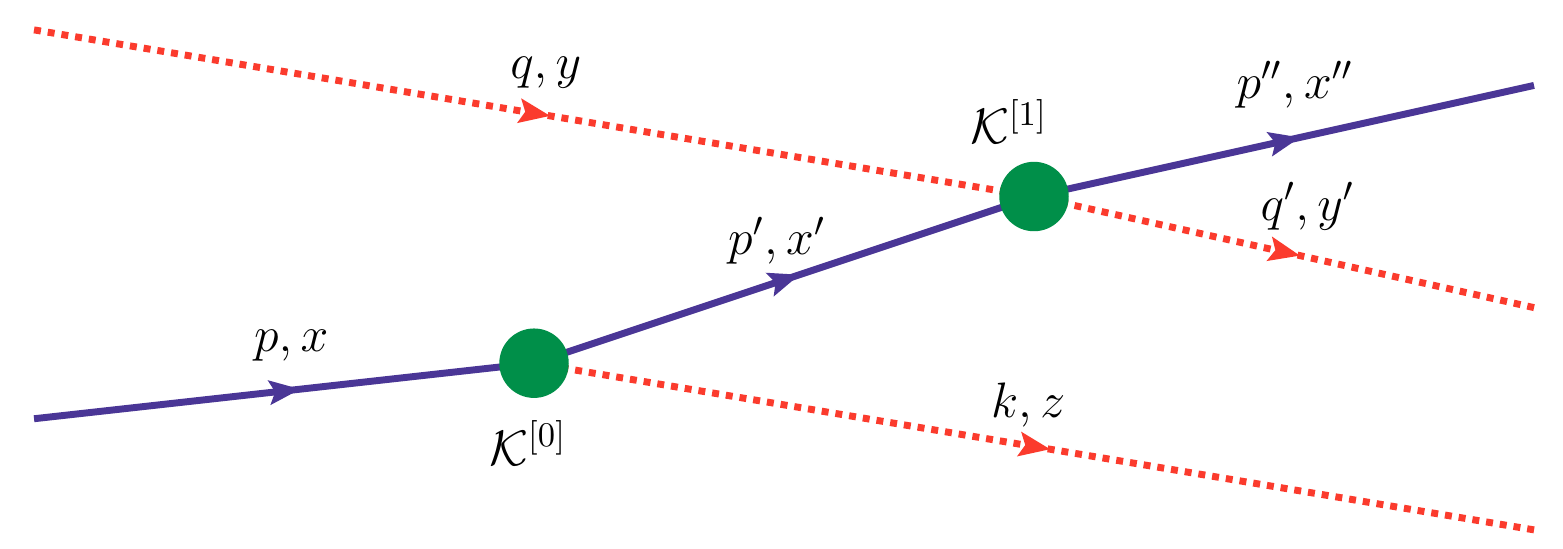}
\caption{Schematics of a pion decaying into a soft and a hard photon, with the hard photon ultimately detected through an interaction
in which it exchanges a small part of its momentum
with a particle in a detector (hard worldlines in solid blue,
soft worldlines in dotted red)}
\label{pic:kbobprl}
\end{center}
\end{figure}

As an illustrative example of a situation where these concepts
apply and the mentioned ``free Hamiltonian limit" is matched,
we consider the situation shown
in Figure~\ref{pic:kbobprl}.

Notice that the situation in Figure~\ref{pic:kbobprl}
is also relevant for the description
of observations of gamma-ray bursts: the incoming blue worldline $p,x$
could be, {\it e.g.}, a highly boosted pion, which decays at the source, producing
a gamma ray ($p',x'$) and a very soft photon ($k,z$);
then the gamma ray propagates freely until its first interaction
at the detector, where it exchanges a small amount of momentum with
a soft particle ($q,y$).
So we can ask if and how the time of detection of the gamma ray
depends on its momentum $p'$; thereby obtaining a prediction
for the large class of studies which is considering possible
energy/time-of-arrival correlations for observations
of gamma-ray bursts
(see, {\it e.g.}, Refs.~\cite{grbgac,magicELLISnew,unoEdue,grb090510}).

An action which is suitable for the relative-locality description
of  the process shown in Figure~\ref{pic:kbobprl} is
 \begin{equation}
\begin{split}
\mathcal{S}^{\kappa (2)} = & \int_{s_{0}}^{+\infty}ds\left(z^{\mu}\dot{k}_{\mu}-\ell z^{1}k_{1}\dot{k}_{0}+\mathcal{N}_k\mathcal{C}_\kappa\left[k\right]\right)+
\int_{-\infty}^{s_{0}}ds\left({x}^{\mu}\dot{p}_{\mu}-\ell x^{1}p_{1}\dot{p}_{0}+\mathcal{N}_p\mathcal{C}_\kappa\left[p\right]\right)\\
 &  +\int_{s_{0}}^{s_{1}}ds\left({x'}^{\mu}\dot{p'}_{\mu}-\ell{x'}^{1}p'_{1}\dot{p}'_0+\mathcal{N}_{p'}\mathcal{C}_\kappa\left[p'\right]\right)+\int_{s_{1}}^{+\infty}ds\left(x''^{\mu}\dot{p''}_{\mu}-\ell x''^{1}p''_{1}\dot{p''}_{0}+\mathcal{N}_{p''}\mathcal{C}_\kappa\left[p''\right]\right)\\
 &+\int_{-\infty}^{s_{1}}ds\left(y^{\mu}\dot{q}_{\mu}-\ell y^{1}q_{1}\dot{q}_{0}+\mathcal{N}_q\mathcal{C}_\kappa\left[q\right]\right) +\int_{s_{1}}^{+\infty}ds\left(y'^{\mu}\dot{q'}_{\mu}-\ell {y}'^{1}{q}_{1}'{\dot{q}}_{0}'+\mathcal{N}_{q'}\mathcal{C}_\kappa\left[{q'}\right]\right)\\
 & -\xi_{[0]}^{\mu}\mathcal{K}_{\mu}^{[0]}(s_{0})
 -\xi_{[1]}^{\mu}\mathcal{K}_{\mu}^{[1]}(s_{1})\ ,
\end{split}
\label{action k-bob}
\end{equation}
where
\begin{gather}\nonumber
{\cal K}^{[0]}_\mu (s_{0})= (q \oplus p)_\mu-(q\oplus p'\oplus k)_\mu = p_\mu - p_\mu' - k_\mu - \ell \delta_\mu^1 ( - q_0 p_1 + q_0 p_1' + q_0 k_1' +p_0'k_1) \ ,\\
{\cal K}^{[1]}_\mu (s_{1})= (q\oplus p'\oplus k)_\mu - (p''\oplus q'\oplus k)_\mu = q_\mu +p_\mu' -p_\mu''-q_\mu'-\ell \delta_\mu^1 (-q_0p_1' - q_0k_1 - p_0'k_1 +p_0''q_1' + p_0''k_1 + q_0'k_1) \ .
\label{constraints}
\end{gather}

And, following the procedure we already used several times,
from this action one obtains easily the
 equations of motion and the constraints,
\begin{gather*}
\dot p_\mu =0~,~~\dot q_\mu =0~,~~\dot k_\mu =0~,~~\dot p'_\mu =0~,~~\dot p''_\mu =0\ ,\\
\mathcal{C}_{\kappa}[p]=0~,~~\mathcal{C}_{\kappa}[q]=0~,~~\mathcal{C}_{\kappa}[k]=0~,~~\mathcal{C}_{\kappa}[p']=0~,~~\mathcal{C}_{\kappa}[p'']=0\ ,\\
\dot z^\mu - \mathcal{N}_k \left(\frac{\delta \mathcal{C}_{\kappa}[k]}{\delta k_\mu} +\ell \delta^\mu_0 \frac{\delta \mathcal{C}_{\kappa}[k]}{\delta k_1} k_1\right)=0~,~~~
\dot y^\mu - \mathcal{N}_q \left(\frac{\delta \mathcal{C}_{\kappa}[q]}{\delta q_\mu} +\ell \delta^\mu_0 \frac{\delta \mathcal{C}_{\kappa}[q]}{\delta q_1} q_1\right)=0~,~~~\\
{\dot y}'^\mu - \mathcal{N}_{q'} \left(\frac{\delta \mathcal{C}_{\kappa}[q']}{\delta q_\mu'} +\ell \delta^\mu_0 \frac{\delta \mathcal{C}_{\kappa}[q']}{\delta q_1'} q_1'\right)=0~,~~~
\dot x^\mu - \mathcal{N}_p \left(\frac{\delta \mathcal{C}_{\kappa}[p]}{\delta p_\mu} +\ell \delta^\mu_0 \frac{\delta \mathcal{C}_{\kappa}[p]}{\delta p_1} p_1\right)=0~,~~~\\
\dot x'^\mu - \mathcal{N}_{p'} \left(\frac{\delta \mathcal{C}_{\kappa}[p']}{\delta p'_\mu} +\ell \delta^\mu_0 \frac{\delta \mathcal{C}_{\kappa}[p']}{\delta p'_1} p'_1\right)=0~,~~~
\dot x''^\mu - \mathcal{N}_{p''} \left(\frac{\delta \mathcal{C}_{\kappa}[p'']}{\delta p''_\mu} +\ell \delta^\mu_0 \frac{\delta \mathcal{C}_{\kappa}[k]}{\delta p''_1} p''_1\right)=0~,~~~
\end{gather*}
and the boundary conditions:
\begin{gather}
z^\mu(s_{0}) = -\xi^\nu_{[0]} \left(\frac{\delta \mathcal{K}^{[0]}_\nu}{\delta k_\mu}+\ell \delta^\mu_0 \frac{\delta \mathcal{K}^{[0]}_\nu}{\delta k_1}k_1\right)\nonumber\ ,\\
x^\mu(s_{0}) = \xi^\nu_{[0]} \left(\frac{\delta \mathcal{K}^{[0]}_\nu}{\delta p_\mu}+\ell \delta^\mu_0 \frac{\delta \mathcal{K}^{[0]}_\nu}{\delta p_1}p_1\right)\nonumber\ ,\\
x'^\mu(s_{0}) = -\xi^\nu_{[0]} \left(\frac{\delta \mathcal{K}^{[0]}_\nu}{\delta p'_\mu}+\ell \delta^\mu_0 \frac{\delta \mathcal{K}^{[0]}_\nu}{\delta p'_1}p'_1\right)\ , \qquad x'^\mu(s_{1}) = \xi^\nu_{[1]} \left(\frac{\delta \mathcal{K}^{[1]}_\nu}{\delta p'_\mu}+\ell \delta^\mu_0 \frac{\delta \mathcal{K}^{[1]}_\nu}{\delta p'_1}p'_1\right)\nonumber\ ,\\
x''^\mu(s_{1}) = -\xi^\nu_{[1]} \left(\frac{\delta \mathcal{K}^{[1]}_\nu}{\delta p''_\mu}+\ell \delta^\mu_0 \frac{\delta \mathcal{K}^{[1]}_\nu}{\delta p''_1}p''_1\right)\nonumber\ ,\\
y^\mu(s_{1})= \xi^\nu_{[1]} \left(\frac{\delta \mathcal{K}^{[1]}_\nu}{\delta q_\mu}+\ell \delta^\mu_0 \frac{\delta \mathcal{K}^{[1]}_\nu}{\delta q_1}q_1\right)\nonumber\ ,\\
y'^{\mu} (s_{1}) = -\xi^\nu_{[1]} \left(\frac{\delta \mathcal{K}^{[1]}_\nu}{\delta q_\mu'}+\ell \delta^\mu_0 \frac{\delta \mathcal{K}^{[1]}_\nu}{\delta q_1'}q_1'\right)\nonumber \ .
\end{gather}

For the first time in this manuscript we are in this section interested
not only in establishing the relativistic properties acquired through our
prescription for the choice of boundary terms, but also on the
predictions of the formalism for what happens to particles.
Evidently here the issue of interest is primarily contained in the
dependence of the time of detection at a given detector
of simultaneously-emitted particles on
the momenta of the particles and on the specific properties
of the interactions involved in the analysis.
We shall analyze this issue arranging the setup in a way that
renders transparent the
comparison with the Hamiltonian treatment of free particles reviewed
in our Section \ref{kbobreview}.
We start by noticing that
for the particle of worldline $x^\mu$, we have
\begin{equation}
x^1 (s) =x^1 (\bar{s})+ v^1(x^0(s)-x^0(\bar{s}))  \label{worldlinek-bob}\ ,
\end{equation}
which in the massless case
(and whenever  $m/{p_1}^2 \ll |\ell p_1|$ takes the simple form
\begin{equation}
 x^1 (s) =x^1 (\bar s)- \frac{p_1}{|p_1|}(x^0 (s) - x^0 (\bar s))\ .
\label{speedmassless}
\end{equation}
In obtaining (\ref{speedmassless}) we used the on-shell relation
\begin{gather*}
p_0=\sqrt{p_1^2 + m^2}-\frac{\ell}{2} p_1^2 \ ,
\end{gather*}
and the fact that for $m/{p_1}^2 \ll |\ell p_1|$
(consistently again with our choice of conventions, which is such
that $v^1>0 \Longrightarrow p^1 <0$)
\begin{gather*}
v^1=\frac{\dot x^1}{\dot x^0}=-\frac{p_1}{p_0}(1-\ell p_0 +\frac{\ell p_1^2}{2 p_0})=-\frac{p_1}{|p_1|} ~.
\end{gather*}

Just as in Sec.~\ref{kbobreview}, we have momentum-independent coordinate speeds for
massless particles, so in particular according to Alice's coordinates
two massless particles of momenta $p_1^s$ and $p_1^h$
simultaneously emitted at Alice (in Alice's spacetime origin)
appear to reach detector Bob simultaneously, apparently establishing
a coincidence of detection events.
But, as stressed already in Sec.~\ref{kbobreview}, the presence of relative locality
evidently requires that
in order to establish
 the dependence of the time of detection
on the momentum of the massless particles we must again transform
the relevant
worldlines to the corresponding description
by an observer Bob local to the detection.
Let us then return to the two-interaction process of Fig.~\ref{pic:kbobprl}
and take as our hard massless particle of momentum $p_1^h$
the particle in that process which we had originally labeled as having
momentum $p'_1$.
For the process of Fig.~\ref{pic:kbobprl}
our description of the transformation from Alice's to Bob's worldlines is
\begin{equation}
\begin{split}
z_{B}^{0}(s)&=  z_{A}^{0}(s)+b^{\mu}  \{ (q\oplus p'\oplus k)_\mu , z^0\} = z_{A}^{0}(s)-b^{0}-\ell b^{1}k_{1} \simeq  z_{A}^{0}(s)-b^{0}\ ,\\
z_{B}^{1}(s)&=z_{A}^{1}(s)+b^{\mu}  \{ (q\oplus p'\oplus k)_\mu , z^1\}
={z}_{A}^{1}(s)-b^{1} -\ell(p_0'+q_0) \simeq {z}_{A}^{1}(s)-b^{1} -\ell b^1 p_0' \ ,\\
x_{B}^{0}(s)&=  x_{A}^{0}(s)+b^{\mu}  \{ (q \oplus p)_\mu , x^0\}  =x_{A}^{0}(s)-b^{0}-\ell b^{1}p_{1} \ ,\\
x_{B}^{1}(s)&=x_{A}^{1}(s)+b^{\mu}  \{ (q \oplus p)_\mu , x^1\}  ={x}_{A}^{1}(s)-b^{1}-\ell q_0  \simeq {x}_{A}^{1}(s)-b^{1} \ ,\\
{x'}_{B}^{0}(s)&=  {x'}_{A}^{0}(s)+b^{\mu}  \{ (q\oplus p'\oplus k)_\mu , x'^0\}
={x'}_{A}^{0}(s)-b^{0}-\ell b^{1}(k_{1}+p'_1) \simeq {x'}_{A}^{0}(s)-b^{0}-\ell b^{1}p'_{1}\ ,\\
{x'}_{B}^{1}(s)&=x_{A}^{1}(s)+b^{\mu}  \{ (q\oplus p'\oplus k)_\mu , x'^1\}
={x'}_{A}^{1}(s)-b^{1}-\ell q_0 \simeq {x'}_{A}^{1}(s)-b^{1}\ , \\
{x''}_{B}^{0}(s)&={x''}_{A}^{0}(s)+b^{\mu} \{ (p''\oplus q'\oplus k)_\mu , x''^0\}
={x''}_{A}^{0}(s)-b^{0}-\ell b^{1}(q_1'+k_1+ p''_{1}) \simeq {x''}_{A}^{0}(s)-b^{0}-\ell b^{1}p''_{1}\ ,\\
{x''}_{B}^{1}(s)&={x''}_{A}^{1}(s)+b^{\mu} \{ (p''\oplus q'\oplus k)_\mu , x''^1\}  ={x''}_{A}^{1}(s)-b^{1}\ ,\\
y_{B}^{0}(s)&=  y_{A}^{0}(s)+b^{\mu}  \{ (q\oplus p'\oplus k)_\mu , y^0\}  =y_{A}^{0}(s)-b^{0}-\ell b^{1}(p'_1+k_1+q_1) \simeq y_{A}^{0}(s)-b^{0}-\ell b^{1}p'_1\ ,\\
y_{B}^{1}(s)&=y_{A}^{1}(s)+b^{\mu}  \{ (q\oplus p'\oplus k)_\mu , y^1\}  ={y}_{A}^{1}(s)-b^{1}\ ,\\
{y'}_{B}^{0}(s)&=  {y'}_{A}^{0}(s)+b^{\mu} \{ (p''\oplus q'\oplus k)_\mu , y'^0\} ={y'_{A}}^{0}(s)-b^{0}-\ell b^{1}(k_1+q'_1) \simeq {y'_{A}}^{0}(s)-b^{0} \ ,\\
{y'}_{B}^{1}(s)&={y'}_{A}^{1}(s)+b^{\mu} \{ (p''\oplus q'\oplus k)_\mu , y'^1\} ={y'_{A}}^{1}(s)-b^{1} - \ell b^1 p_0''\ .
\end{split}
\label{translationsk-bob}
\end{equation}
Using these transformation laws it is easy to recognize that, having dropped
the negligible ``soft terms" from small momenta,
indeed we are obtaining results that are fully consistent
with the ones obtained in the hamiltonian description of free particles.
To see this explicitly let us
consider the situation where, simultaneously to the interaction emitting
the hard particle $x',p'$ in Alice origin, we also have the emission of
a soft photon $x_s, p_s$.\\
And as observer Bob let us take one who
is reached {\underline{in its spacetime origin}}
 by the soft photon emitted by Alice. For the event of detection
 of the hard particle  $x',p'$
 we take one such that it occurs in Bob's {\underline{spatial origin}}.\\
 From a relative-locality perspective the setup we are arranging is
 such that ``Alice is an emitter" (the spatial origin of Alice's coordinate system
 is an ideally compact,
 infinitely small, emitter) and ``Bob is a detector"
(the spatial origin of Bob's coordinate system
 is an ideally compact,
 infinitely small, detector). The two worldlines we focus on, a soft and
 a hard worldline,
 both originate from Alice's spacetime origin (they are both emitted by Alice,
 in the spatial origin of Alice's frame of reference, and both
 at time $t_{Alice}=0$)
 and both end up being detected by Bob, but, while by construction the soft
 particle reaches Bob's spacetime origin, the time at which the hard particle
 reaches Bob spatial origin is to be determined through our analysis.

Reasoning  as usual
at first order in $\ell$, it is easy to verify that Bob describes
the ``interaction coordinate" ${\xi_B^{[1]}}^\mu$ of the interaction at $s=s_1$
as coincident with the $s=s_1$ endpoints of the
worldlines $x',p'$; $x'',p''$; $q,y$; $q',y'$:
\begin{equation}
{\xi_B^{[1]}}^\mu = {x_B'}^\mu(s_{1}) = {x_B''}^\mu(s_{1}) = y_B^\mu(s_{1}) = {y_B'}^\mu(s_{1}) \ .
 \label{delaycoordPRE}
\end{equation}
We take into account that there are no relative-locality effects
in the description given by Bob whenever an interaction occurs ``in the vicinity of Bob":
our leading-order analysis assumes the observatories have sensitivity
sufficient to expose manifestation of relativity of locality
of order $\ell p_h L$ (where $L$ is the distance from the interaction-event
to the origin
of the observer and $p_h$ is a ``suitably high" momentum),
with $L$ set in this case by the distance Alice-Bob, so even a hard-particle
interaction which is at a distance $d$ from the origin of Bob
will be treated as absolutely local by Bob if $d \ll L$.\\
According to this both ``detection events" are absolutely local
for observer Bob: of course this is true for
the  event of detection of the soft photon $x_s, p_s$
(which we did not even specify since its softness ensures us of its
absolute locality) and it is also true for the interaction-event
of ``detection near Bob"
of the hard particle $x',p'$. Ultimately this allows us
 to handle the time component of the coordinate fourvector (\ref{delaycoordPRE})
  as the actual delay that Bob measures
between the two detection times:
\begin{equation}
 \Delta t = {\xi_B^{[1]}}^0 = {x_B'}^0(s_{1}) = {x_B''}^0(s_{1}) = y_B^0(s_{1}) = {y_B'}^0(s_{1}) \label{delaycoord}\ .
\end{equation}

\begin{figure}[h!]
\begin{center}
\includegraphics[width=0.44 \textwidth]{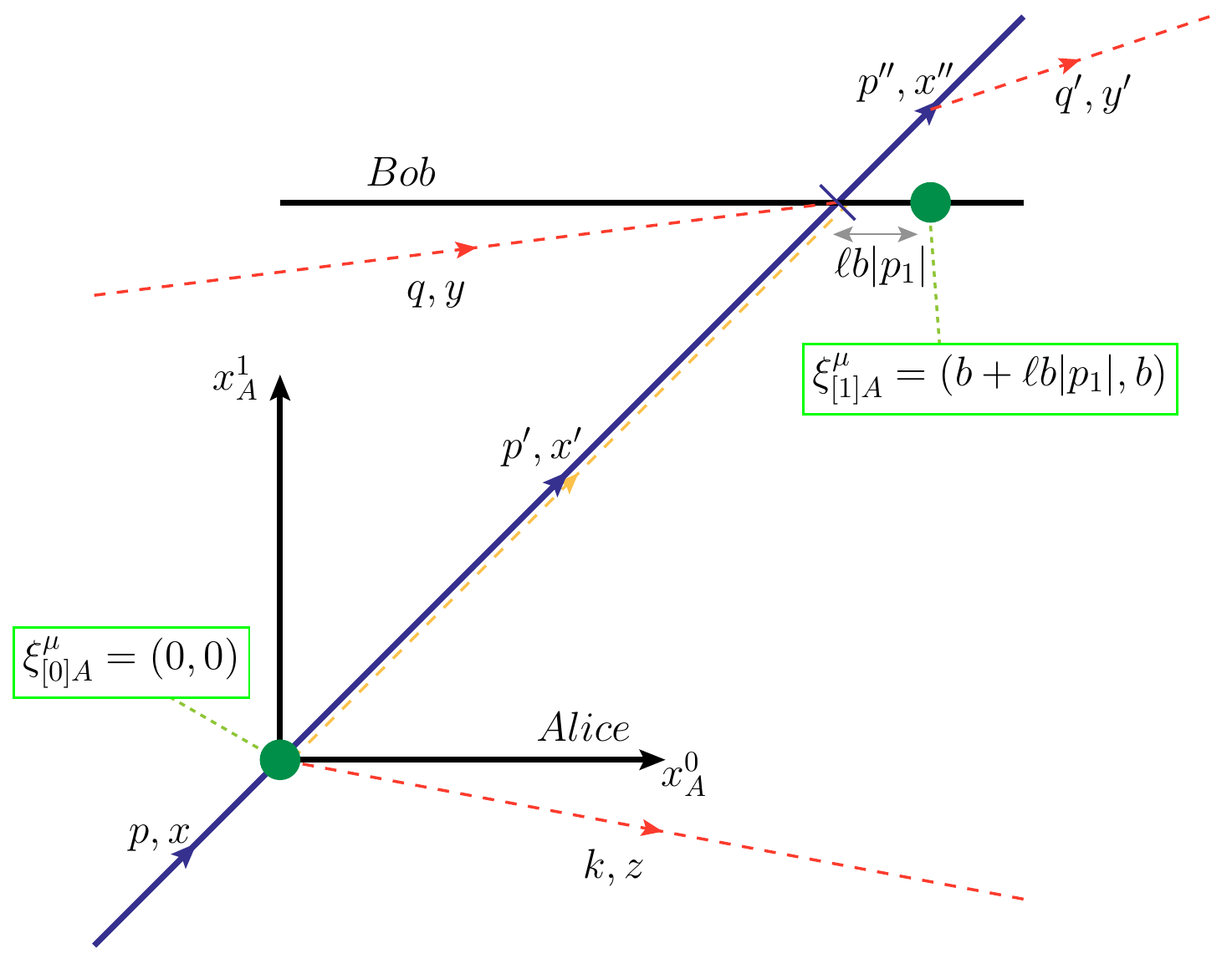}
\includegraphics[width=0.44 \textwidth]{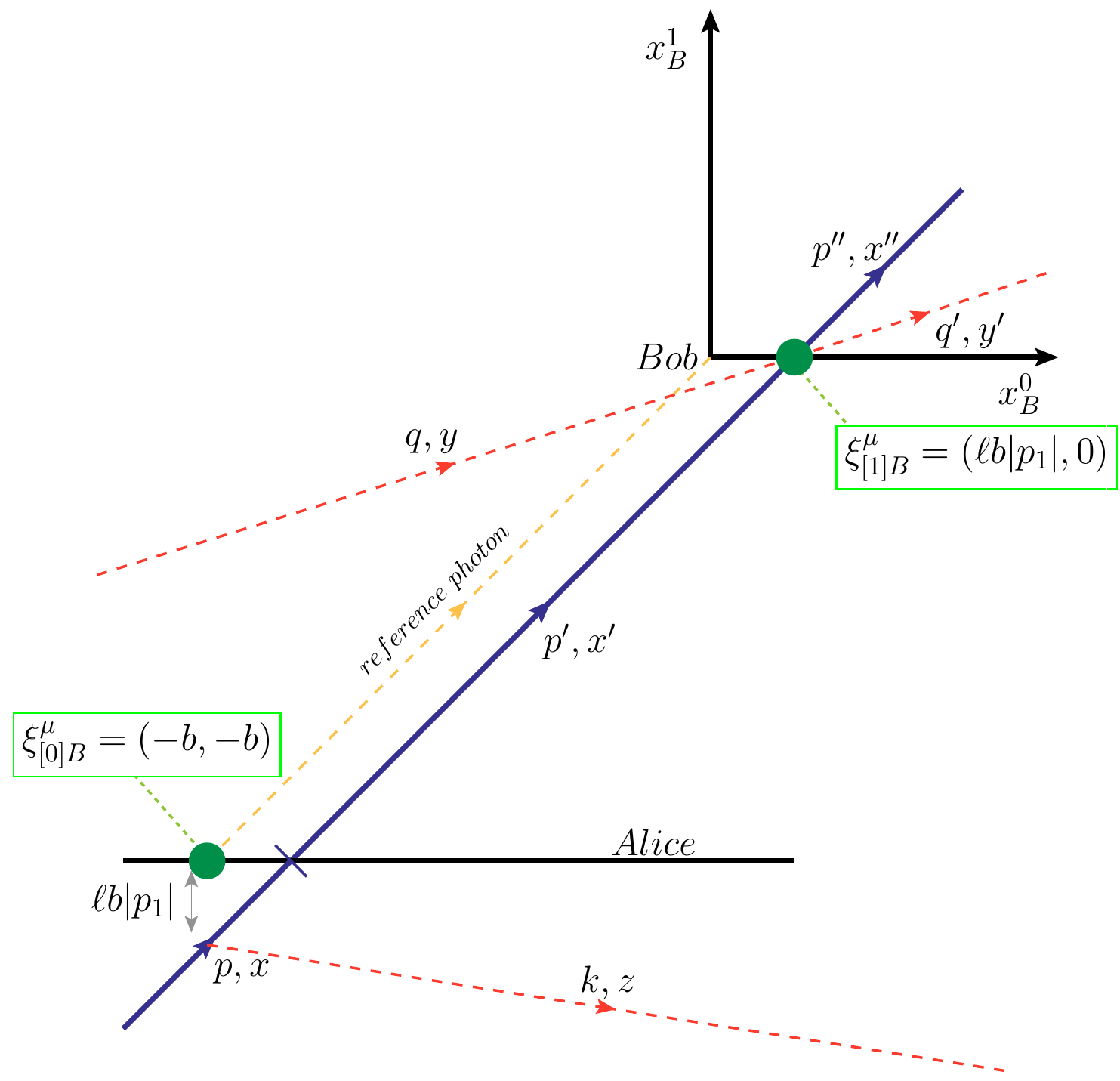}
\caption{Schematic description of the time delay derived in this subsection.
The various agents in the analysis are shown both as described by Alice
(left panel) and as described by Bob (right panel). These are spacetime graphs
(in a 2D spacetime) showing the actual worldlines of particles. In addition to the
two hard interactions we are considering
(qualitatively described already in Figure~\ref{pic:kbobprl}),
we also show (as the orange-dotted worldline) a soft photon going from Alice's
origin to Bob's origin. As shown in the figure we have arranged the calculations in
this section so that all emissions and detections occur in the spatial origin
of either Alice or Bob (but because of the relative locality according to Alice
the hard detections at/near Bob would be nonlocal interactions and according to Bob
the hard emissions at Alice would be nonlocal processes). We also show,
as the bulky green dots,
the formal positions of the interaction points, as coded in the
formal ``interaction coordinates" $\xi^\mu$.}
\label{alicebobkappabobprl}
\end{center}
\end{figure}

From the equations (\ref{translationsk-bob}) relative to the worldline $x',p'$, it follows that
\begin{equation}
 {x'_A}^1(s_{1}) = {x'_B}^1(s_{1}) + b^1 = b^1\ ,
\end{equation}
from which, considering the worldlines (\ref{speedmassless}),
 it follows that
  (assuming indeed $m/(p'_1)^2 \ll |\ell p'_1|$)
 Alice ``sees'' the $s=s_{1}$ endpoint of the worldline $x',p'$ at the coordinates
\begin{equation}
{x'}^\mu_A(s_{1}) = {x'}^\mu_B(s_{1}) + b^\mu = b^\mu = (b,b)\ .
\end{equation}
And then, from the equations (\ref{translationsk-bob}) and (\ref{delaycoord}), it follows that Bob measures the delay
\begin{equation}
\Delta t = {x_B'}^0(s_{1}) = {x'}_{A}^{0}(s_{1})-b^{0}-\ell b^{1}p'_{1} = \ell b |p'_{1}| \ ,
\label{delay}
\end{equation}
in agreement with the result (\ref{delayk-bob}) found in the Hamiltonian description. These findings are summarized in Figure \ref{alicebobkappabobprl}.

Since we are dealing with a momentum-space with torsion, {\it i.e. the
momentum composition law is noncommutative},
it is interesting to check whether this result establishing agreement with
the Hamiltonian description of free particles also holds for
other choices of ordering of momenta in the conservation laws
(and accordingly in the boundary conditions).

 An interesting alternative for the conservation laws and boundary conditions
of the process in Figure~\ref{pic:kbobprl}
is the following
\begin{gather}\nonumber
{\cal K}^{[0]}_\mu (s_{0})= (p \oplus q)_\mu-(k\oplus  p' \oplus q)_\mu = p_\mu - p_\mu' - k_\mu - \ell \delta_\mu^1 ( - p_0 q_1 + k_0 p_1' + k_0 q_1 +p_0' q_1) \ ,\\
{\cal K}^{[1]}_\mu (s_{1})= (k\oplus  p' \oplus q)_\mu - (k \oplus  p'' \oplus q')_\mu = p_\mu' +q_\mu -p_\mu''-q_\mu'-\ell \delta_\mu^1 (-k_0p_1' - k_0 q_1 - p_0' q_1 +k_0 p_1'' + k_0 q_1' + p_0'' q_1' ) \ .
\end{gather}
Going from the previous version of the boundary conditions to this one
does change several things in the analysis, but it easy
to see that it does not change anything about the ``free particle" $x',p'$.
With these conservation laws and boundary conditions the relationships
between  Alice's worldline $x'$ and Bob's worldline $x'$
are codified in
\begin{equation}
\begin{split}
{x'}_{B}^{0}(s)&=  {x'}_{A}^{0}(s)+b^{\mu} \{ (k\oplus p'\oplus q)_\mu , x'^0\}
={x'}_{A}^{0}(s)-b^{0}-\ell b^{1}(q_{1}+p'_1) \simeq {x'}_{A}^{0}(s)-b^{0}-\ell b^{1}p'_{1}\ ,\\
{x'}_{B}^{1}(s)&=x_{A}^{1}(s)+b^{\mu} \{ (k\oplus p'\oplus q)_\mu , x'^1\} =x_{A}^{1}(s)-b^{1}-\ell b^{1}k_{0}  \simeq {x'}_{A}^{1}(s)-b^{1} \ .
\end{split}
\end{equation}
And using the equation of motion (\ref{speedmassless})
one easily checks that then the relevant particle reaches
Bob's spatial origin at the time
\begin{equation}
\Delta t = {x_B'}^0(s_{1}) = {x'}_{A}^{0}(s_{1})-b^{0}-\ell b^{1}p'_{1} = \ell b |p'_{1}| \ ,
\end{equation}
in perfect agreement with the result of Eq.~(\ref{delay}), which had been
obtained with the other choice of ordering of momenta in the conservation laws.

So we find evidence of the fact that the properties of ``free particles"
(particles exchanging only small fractions of their momentum)
are insensitive to the ordering chosen for the law of composition
of momenta.

And for what concerns bursts of simultaneously emitted massless particles,
such as in a gamma-ray-burst, this derivation predicts differences in times
of arrival governed by the formula
$$\Delta t_{arrival} = \ell L |\Delta p_1|\ ,$$
where $L$ is the distance from source to detector (the corresponding translation
from observer at the source to observer at the detector has $b^\mu=(L,L)$)
and $|\Delta p_1|$ is the difference of momentum among the two massless
particles whose arrival times differ by $\Delta t_{arrival}$.

The derivation in this subsection establishes this result for cases
where the interaction at the source
emitting the particle of interest  only involves one hard particle
in the in state and one hard particle in the out state (all other particles
involved in the interactions being soft).

\subsection{More on  observations of distant
 bursts of massless particles}\label{secsuperpion}

In the previous subsection, in showing
that our proposal (in an appropriate limit) matches the predictions
of previous Hamiltonian descriptions of relative locality for free
particles, we also showed that, at least for certain types of emission
and detection interactions,
our proposal predicts time-of-detection
delays $\Delta t_{arrival} = \ell L |\Delta p_1|$
between simultaneously-emitted massless
particles with momentum difference $|\Delta p_1|$.
This is very interesting because, as established in several studies reported
over the last decade, such an effect is testable, even if $\ell$ is as small
as the Planck length\footnote{We introduced $\ell$ as a momentum-space property,
with dimensions of inverse momentum. When we mention the possibility
of $\ell$ of order
the Planck length we are essentially using jargon, a compact way to describe cases
where $\ell^{-1}$ is of order the Planck scale.}
(or even one or two orders of magnitude smaller
than the Planck length~\cite{grbgac,magicELLISnew,unoEdue,grb090510}).

We derived this time-delay result assuming certain types of emission
and detection interactions.
But evidently the structure of our formalism
is such that it would not be surprising to find that the
times of  detection depended
on the actual emission and detection interactions involved.
In this subsection we intend to establish that this is indeed the case,
and that the torsion of momentum space plays a crucial role in the
relevant analysis.

It suffices to modify the analysis of the previous subsection
in rather minor way for us to show that the times of detection
of simultaneously emitted particles depend not only
on the momenta of the particles but also on the actual nature
of the emitting interaction.
We find that in order for this to occur there must be
at least 3 hard particles in total, among in and out particles
of the emission interaction.\\
As an example of this we consider here explicitly the case
of a ultraenergetic particle at rest decaying into two particles, both hard,
one of which is the particle detected at our observatory.

As shown in figure we arrange the analysis in exactly the same way as
in the previous section, with a tri-valent vertex for the emission
interaction and a four-valent vertex for the detection. And the kinematics at the
four-valent vertex is left unchanged, involving a soft particle in the in state
and a soft particle in the out state. We only change the kinematics of
the emission vertex, now assuming that all particles involved are hard.

\begin{figure}[h!]
\begin{center}
\includegraphics[width=0.66 \textwidth]{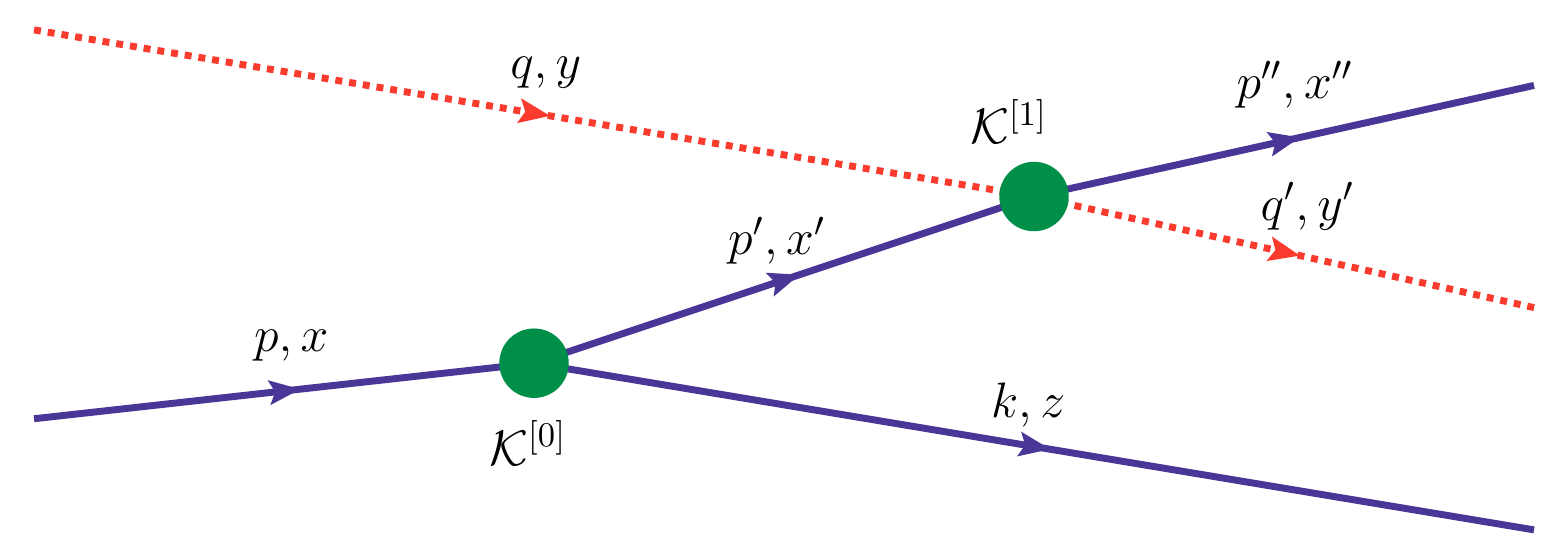}
\caption{Schematic description of a case where a hard ultrarelativistic particle
originating from a hard emission interaction
(one hard particle in, two hard particles out) is detected
in a soft interaction (only one hard particle in and only one hard particle
out). Solid-blue lines are for hard particles, dashed-red lines are for soft particles.}\label{figsuperpion}
\end{center}
\end{figure}

And we shall again consider two possible choices of conservation-enforcing
boundary conditions, suitable for exploring the role of the noncommutativity
of the law of composition of momenta. The same
two possible choices of conservation-enforcing
boundary conditions already considered in the previous subsection.

Let us start again by analyzing as first possibility
\begin{gather}\nonumber
{\cal K}^{[0]}_\mu (s_{0})= (q \oplus p)_\mu-(q\oplus p'\oplus k)_\mu = p_\mu - p_\mu' - k_\mu - \ell \delta_\mu^1 ( - q_0 p_1 + q_0 p_1' + q_0 k_1' +p_0'k_1)\ , \\
{\cal K}^{[1]}_\mu (s_{1})= (q\oplus p'\oplus k)_\mu - (p''\oplus q'\oplus k)_\mu = q_\mu +p_\mu' -p_\mu''-q_\mu'-\ell \delta_\mu^1 (-q_0p_1' - q_0k_1 - p_0'k_1 +p_0''q_1' + p_0''k_1 + q_0'k_1) \ .
\end{gather}
The worldlines seen by observer/detector Bob, distant from the emission,
that follow from this choice of boundary terms
have been already given in Eq.~(\ref{speedmassless}).
The main difference between the situation in the previous
subsection and the situation we are now analyzing is that
the ``primary", the  particle incoming to the emission interaction,
is at rest, with $p_{1}=0$, which also implies that the two outgoing particles
of the emission interaction must both be hard.
 For the worldlines involved in the emission interaction this leads to
\begin{equation}
\begin{split}
x_{B}^{0}(s)&=x_{A}^{0}(s)+b^{\mu}  \{ (q \oplus p)_\mu , x^0\} =x_{A}^{0}(s)-b^{0}-\ell b^{1}p_{1} =x_{A}^{0}(s)-b^{0}\ ,\\
x_{B}^{1}(s)&=x_{A}^{1}(s)+b^{\mu}  \{ (q \oplus p)_\mu , x^1\}={x}_{A}^{1}(s)-b^{1}-\ell q_0\simeq {x}_{A}^{1}(s)-b^{1}\ ,\\
{x'}_{B}^{0}(s)&={x'}_{A}^{0}(s)+b^{\mu}  \{ (q\oplus p'\oplus k)_\mu , x'^0\}={x'}_{A}^{0}(s)-b^{0}-\ell b^{1}(k_{1}+p'_1)={x'}_{A}^{0}(s)-b^{0}\ ,\\
{x'}_{B}^{1}(s)&=x_{A}^{1}(s)+b^{\mu}  \{ (q\oplus p'\oplus k)_\mu , x'^1\} ={x'}_{A}^{1}(s)-b^{1}-\ell q_0\simeq{x'}_{A}^{1}(s)-b^{1}\ ,\\
z_{B}^{0}(s)&=z_{A}^{0}(s)+b^{\mu}  \{ (q\oplus p'\oplus k)_\mu , z^0\}= z_{A}^{0}(s)-b^{0}-\ell b^{1}k_{1} \ ,\\
z_{B}^{1}(s)&=z_{A}^{1}(s)+b^{\mu}  \{ (q\oplus p'\oplus k)_\mu , z^1\}={z}_{A}^{1}(s)-b^{1} -\ell(p_0'+q_0) \simeq {z}_{A}^{1}(s)-b^{1} -\ell b^1 p_0' \ .
\end{split}
\label{translationsnodelay}
\end{equation}

And from this one easily sees that the particle $p',x'$, the particle
then detected at Bob, translates classically, without any deformation term.
{\underline{So this time we have that no momentum dependence of the times
of arrival is predicted}}
$$t_{detection}= x_{B}'^{0}(s_{1})=x_{A}'^{0}(s_{1})-b^{0}=0~.$$

Next we show that in this case with the emission interaction involving only hard
particles the noncommutativity of the composition law,
which had turned out to be uninfluential in the previous subsection,
 does play a highly non-trivial role.\\
To see this
let us consider, as in the previous subsection,
the following alternative choice of ${\cal K}$'s for the boundary terms
\begin{gather}\nonumber
{\cal K}^{[0]}_\mu = (p \oplus q)_\mu-(k\oplus  p' \oplus q)_\mu = p_\mu - p_\mu' - k_\mu - \ell \delta_\mu^1 ( - p_0 q_1 + k_0 p_1' + k_0 q_1 +p_0' q_1)\ , \\
{\cal K}^{[1]}_\mu = (k\oplus  p' \oplus q)_\mu - (k \oplus  p'' \oplus q')_\mu = p_\mu' +q_\mu -p_\mu''-q_\mu'-\ell \delta_\mu^1 (-k_0p_1' - k_0 q_1 - p_0' q_1 +k_0 p_1'' + k_0 q_1' + p_0'' q_1' ) \ .
\end{gather}
Focusing again on the worldline $x' , p'$ detected at Bob
we now find
\begin{equation}
\begin{split}
{x'}_{B}^{0}(s)&={x'}_{A}^{0}(s)+b^{\mu}  \{ (k\oplus  p' \oplus q)_\mu , x'^0\}={x'}_{A}^{0}(s)-b^{0}-\ell b^{1}(q_{1}+p'_1) \simeq {x'}_{A}^{0}(s)-b^{0}-\ell b^{1}p'_{1}\ ,\\
{x'}_{B}^{1}(s)&=x_{A}^{1}(s)+b^{\mu}  \{ (k\oplus  p' \oplus q)_\mu , x'^1\} ={x'}_{A}^{1}(s)-b^{1}-\ell b^{1}k_{0} \ .
\end{split}
\end{equation}
And from the equation of motion (\ref{speedmassless})
one now deduces that
$${x'}_{B}^{1}(s)={x'}_{B}^{0}(s)-\ell b^{1}(k_{0}-p'_{1}) \ ,$$
which in turn implies that the time of detection at Bob
of the particle with worldline $x' , p'$
is
\begin{equation}
 t_{detection} = {x_B'}^0(s_{1})  =-\ell b^{1}(p'_{1}-k_{0})=2\ell b^{1}|p'_{1}| \ .
\end{equation}
{\underline{The dependence of the time of detection on the momentum of
the particle being detected is back!}}\\
{\underline{And this dependence is twice as strong as the dependence
on momentum found in the previous subsection!}}

\subsection{Nonmetricity, torsion and time delays}
The results we obtained in this subsection are rather striking
and deserve to be summarized and discussed in relation
with previous related results.

For what concerns
times of detection of simultaneously emitted massless particles of  momentum $p'_1$,
emitted from a source at a distance $L$ from the detector we analyzed
3 situations:\\
{\bf (case A)} the emission interaction involves only one hard incoming particle
and one hard outgoing particle, all other particles in the emission
interaction being soft: \\
the times of arrival have a dependence
on momentum governed by
$$t_{detection} = \ell L |p'_{1}|$$
and this result is independent of the position occupied by the
momentum $p'_\mu$ in our noncommutative composition law\\
{\bf (case B)} the emission interaction is the decay of a ultra-high-energy
 particle at rest, involves a total of 3 hard particles,
and the momentum $p'_\mu$ appears in the composition of momenta
to the {\underline{left}} of a hard particle: \\
the times of arrival have no dependence
on momentum
$$t_{detection} = 0$$
{\bf (case C)} the emission interaction is the decay of a ultra-high-energy
 particle at rest, involves a total of 3 hard particles,
and the momentum $p'_\mu$ appears in the composition of momenta
to the {\underline{right}} of a hard particle: \\
the times of arrival have the following dependence
on momentum
$$t_{detection} = 2 \ell L |p'_{1}|$$
(twice as large as in the case A).

We obtained this results working with our $\kappa$-Poincar\'e-inspired
momentum space, with nonmetricity and torsion.\\
The fact that it would be possible with such a momentum space to
have that simultaneously emitted massless particles are not detected
simultaneously at the same detector could be expected on the
basis of the analysis reported in Ref.~\cite{leelaurentGRB},
whose main message was that indeed nonmetricity should result
in the possibility of having  simultaneously emitted massless particles
that are not detected simultaneously.\\
The nature and scopes of our study were such that we could for the first
time investigate how the presence of both torsion and nonmetricity could
affect these time-of-detection delays. And we evidently found that torsion
can have striking effects, effects capable of changing the predicted
new effect at order 1, and therefore effects that are as much within
reach of ongoing and forthcoming experiments as the pure-nonmetricity (no torsion)
effects.

It should be stressed that what we found might even underestimate
the significance of the
effects of torsion on time delays (at least the effects on time delays
of torsion, when also nonmetricity is present). This is because we contemplated
a total of only 3 cases for what concerns the kinematics and the conservation
laws that are relevant for such an analysis. But even within the confines
of our preliminary
investigation we found a type of dependence of the time delays,
not only on momenta of observed particles but also on interactions that
produced them,
which had never been encountered before in the literature and would
therefore provide a very distinguishing feature of the model of momentum
space we here adopted as illustrative example.

\section{An alternative choice of symplectic structure}\label{itsgeometry}
We showed in the previous section that the class of theories
we are studying can have striking manifestation, whose magnitude is extremely
small but within the reach of our observatories, at least when they observe
particles at rather high energy from
very distant astros ({\it e.g.} multiGeV photons from gamma-ray bursts
at redshift greater than 1, as discussed in Ref.~\cite{grbgac,unoEdue}
and references therein).\\
It is interesting to ask which of the novel feature of the framework
we analyzed should be deemed responsible for these striking and testable
novel effects. Specifically: is this due exclusively
to the geometry of momentum space, codified in the momentum-space
 metric and connection?
or is there also a role played by the choice we made above of a
a ``$\kappa$-Minkowski inspired" symplectic structure?

In this section we provide evidence of the fact that the choice of symplectic structure
is completely irrelevant. The physical content of these relative-locality theories
is fully codified in the geometry of momentum space.
We argue this by adopting here the same geometry of momentum space as in the previous
sections, the ``$\kappa$-momentum space", but changing the symplectic structure.
We find that the predictions indeed do not change.

We have verified this for all the applications of ``$\kappa$-momentum space"
discussed in the previous sections. But let us report here explicitly only
one case, the particularly noteworthy case studied in the previous
 Subsection.\ref{secsuperpion}.

 We then describe the process in Fig.~\ref{figsuperpion} by the following action
 \begin{equation}
\begin{split}
\mathcal{S}^{\kappa (2)} = & \int_{s_{0}}^{+\infty}ds\left(z^{\mu}\dot{k}_{\mu} + \mathcal{N}_k\mathcal{C}_\kappa\left[k\right]\right)+
\int_{-\infty}^{s_{0}}ds\left({x}^{\mu}\dot{p}_{\mu} + \mathcal{N}_p\mathcal{C}_\kappa\left[p\right]\right)\\
 &  +\int_{s_{0}}^{s_{1}}ds\left({x'}^{\mu}\dot{p'}_{\mu} + \mathcal{N}_{p'}\mathcal{C}_\kappa\left[p'\right]\right) + \int_{s_{1}}^{+\infty}ds\left(x''^{\mu}\dot{p''}_{\mu} + \mathcal{N}_{p''}\mathcal{C}_\kappa\left[p''\right]\right)\\
 &+\int_{-\infty}^{s_{1}}ds\left(y^{\mu}\dot{q}_{\mu}+\mathcal{N}_q\mathcal{C}_\kappa\left[q\right]\right) +\int_{s_{1}}^{+\infty}ds\left(y'^{\mu}\dot{q'}_{\mu} + \mathcal{N}_{q'}\mathcal{C}_\kappa\left[{q'}\right]\right)\\
 & -\xi_{[0]}^{\mu}\mathcal{K}_{\mu}^{[0]}(s_{0})
 -\xi_{[1]}^{\mu}\mathcal{K}_{\mu}^{[1]}(s_{1}) \ ,
\end{split}
\label{action k-bob COM}
\end{equation}
which we take to be identical to the corresponding one written
in  Subsection.\ref{secsuperpion}, with the exception of the evident
change from the ``$\kappa$-Minkowski symplectic structure" of
Subsection.\ref{secsuperpion} to the adoption here of a trivial
symplectic structure:
\begin{gather}
\left\{ x^{1},x^{0}\right\} = 0 \ , \nonumber \\
\left\{ x^{0} , p_{0} \right\} =1,\qquad\left\{ x^{1} , p_{0} \right\} =0\ ,
\\ \left\{ x^{0} , p_{1}\right\} =0,\qquad\left\{ x^{1} , p_{1}\right\} =1~.
\end{gather}
The metric and connection on momentum space are still the ones
of the ``$\kappa$-momentum space", and we consider again
the same boundary terms also considered in Subsection.\ref{secsuperpion}:
\begin{gather}\nonumber
{\cal K}^{[0]}_\mu (s_{0})= (q \oplus p)_\mu-(q\oplus p'\oplus k)_\mu = p_\mu - p_\mu' - k_\mu - \ell \delta_\mu^1 ( - q_0 p_1 + q_0 p_1' + q_0 k_1' +p_0'k_1) \ ,\\
{\cal K}^{[1]}_\mu (s_{1})= (q\oplus p'\oplus k)_\mu - (p''\oplus q'\oplus k)_\mu = q_\mu +p_\mu' -p_\mu''-q_\mu'-\ell \delta_\mu^1 (-q_0p_1' - q_0k_1 - p_0'k_1 +p_0''q_1' + p_0''k_1 + q_0'k_1) \ .
\label{constraints COM}
\end{gather}

The change of symplectic structure does lead to some changes
in the
 equations of motion
\begin{gather*}
\dot p_\mu =0~,~~\dot q_\mu =0~,~~\dot k_\mu =0~,~~\dot p'_\mu =0~,~~\dot p''_\mu =0\\
\mathcal{C}_{\kappa}[p]=0~,~~\mathcal{C}_{\kappa}[q]=0~,~~\mathcal{C}_{\kappa}[k]=0~,~~\mathcal{C}_{\kappa}[p']=0~,~~\mathcal{C}_{\kappa}[p'']=0\\
\dot z^\mu - \mathcal{N}_k \frac{\delta \mathcal{C}_{\kappa}[k]}{\delta k_\mu} =0~,~~~
\dot y^\mu - \mathcal{N}_q \frac{\delta \mathcal{C}_{\kappa}[q]}{\delta q_\mu} =0~,~~~\\
{\dot y}'^\mu - \mathcal{N}_{q'} \frac{\delta \mathcal{C}_{\kappa}[q']}{\delta q_\mu'} =0~,~~~
\dot x^\mu - \mathcal{N}_p \frac{\delta \mathcal{C}_{\kappa}[p]}{\delta p_\mu} =0~,~~~\\
\dot x'^\mu - \mathcal{N}_{p'} \frac{\delta \mathcal{C}_{\kappa}[p']}{\delta p'_\mu} =0~,~~~
\dot x''^\mu - \mathcal{N}_{p''} \frac{\delta \mathcal{C}_{\kappa}[p'']}{\delta p''_\mu} =0~,~~~
\end{gather*}
and in the boundary conditions:
\begin{gather}
z^\mu(s_{0}) = -\xi^\nu_{[0]} \frac{\delta \mathcal{K}^{[0]}_\nu}{\delta k_\mu} \nonumber\\
x^\mu(s_{0}) = \xi^\nu_{[0]} \frac{\delta \mathcal{K}^{[0]}_\nu}{\delta p_\mu} \nonumber\\
x'^\mu(s_{0}) = -\xi^\nu_{[0]} \frac{\delta \mathcal{K}^{[0]}_\nu}{\delta p'_\mu} \qquad x'^\mu(s_{1}) = \xi^\nu_{[1]} \frac{\delta \mathcal{K}^{[1]}_\nu}{\delta p'_\mu} \nonumber\\
x''^\mu(s_{1}) = -\xi^\nu_{[1]} \frac{\delta \mathcal{K}^{[1]}_\nu}{\delta p''_\mu}\nonumber\\
y^\mu(s_{1})= \xi^\nu_{[1]} \frac{\delta \mathcal{K}^{[1]}_\nu}{\delta q_\mu}\nonumber\\
y'^{\mu} (s_{1}) = -\xi^\nu_{[1]} \frac{\delta \mathcal{K}^{[1]}_\nu}{\delta q_\mu'} \ , \nonumber\\
\end{gather}
but in spite of this the predictions remain unchanged.

The changes in the equations of motion appear to take
a rather tangible form at intermediate stages of analysis
For example
for the particle of worldline $x^\mu$, we have
\begin{equation}
\dot {x}^\mu = {\cal N}_p \frac {\delta {\cal C}_\kappa [p]}{\delta p_\mu} = \delta^\mu_0 {\cal N}_p \left( 2p_0+\ell p_1^2\right) -2 \delta^\mu_1 {\cal N}_p (p_1 -\ell p_0 p_1)\ ,
\end{equation}
from which it follows that
in the massless case
\begin{gather}
 x^1 (s) =x^1 (\bar s)- \left( \frac{p_1}{|p_1|} -\ell p_1 \right)(x^0(s)-x^0(\bar s)) \ ,
\label{eq. motion COM}
\end{gather}
and this is quite different from the corresponding formula
obtained in Subsection.\ref{secsuperpion}.
But there are other aspects of the analysis which are
affected by the change of symplectic structure,
and ultimately the predictions of time of detection
remain unchanged.

To see this let us focus again on the case
in which the ``primary", the  particle incoming to the emission interaction,
is at rest, with $p_{1}=0$, which also implies that the two outgoing particles
of the emission interaction must both be hard.
 For the worldlines involved in the emission interaction this leads to
\begin{equation}
\begin{split}
x_{B}^{0}(s)&=x_{A}^{0}(s)+b^{\mu}  \{ (q \oplus p)_\mu , x^0\} =x_{A}^{0}(s)-b^{0} \ , \\
x_{B}^{1}(s)&=x_{A}^{1}(s)+b^{\mu}  \{ (q \oplus p)_\mu , x^1\}={x}_{A}^{1}(s)-b^{1}-\ell q_0\simeq {x}_{A}^{1}(s)-b^{1} \ , \\
{x'}_{B}^{0}(s)&={x'}_{A}^{0}(s)+b^{\mu}  \{ (q\oplus p'\oplus k)_\mu , x'^0\}={x'}_{A}^{0}(s)-b^{0}-\ell b^{1}k_{1} \ , \\
{x'}_{B}^{1}(s)&=x_{A}^{1}(s)+b^{\mu}  \{ (q\oplus p'\oplus k)_\mu , x'^1\} ={x'}_{A}^{1}(s)-b^{1}-\ell q_0\simeq{x'}_{A}^{1}(s)-b^{1} \ , \\
z_{B}^{0}(s)&=z_{A}^{0}(s)+b^{\mu}  \{ (q\oplus p'\oplus k)_\mu , z^0\}= z_{A}^{0}(s)-b^{0} \ , \\
z_{B}^{1}(s)&=z_{A}^{1}(s)+b^{\mu}  \{ (q\oplus p'\oplus k)_\mu , z^1\}={z}_{A}^{1}(s)-b^{1} -\ell(p_0'+q_0) \simeq {z}_{A}^{1}(s)-b^{1} -\ell b^1 p_0' \ .\\
\end{split}
\label{translationsnodelay COM}
\end{equation}

To find the time of detection of the particle $x',p'$ as seen by Bob, who is at Alice coordinates $b^\mu = (b,b)$, we exploit the fact that,
assuming that the particle $x',p'$ crosses Bob spatial origin at $s=s_1$,
\begin{equation}
{x'}^1_A(s_1) = {x'}^1_B(s_1) + b^1 = b^1 = b\ ,
\end{equation}
from which, using the equations of motion (\ref{eq. motion COM}), rewritten for the particle $x',p'$ as seen by Alice, it follows that
\begin{equation}
 {x'}^0_A(s_1) = {x'}^1_A(s_1) \left( 1 - \ell p'_1 \right) =  b - \ell b p'_1 \ .
\end{equation}
Using again the equations of motion (\ref{eq. motion COM}),
rewritten for the particle $x',p'$ as seen by Bob, together with Eq. (\ref{translationsnodelay COM}), we finally find
\begin{equation}
t_{detection}= x_{B}'^{0}(s_{1})=x_{A}'^{0}(s_{1})-b^{0}- \ell b^1k_1 = -\ell b \left( k_1 + p_1' \right)=0~.
\end{equation}

So we have no momentum dependence of the times
of detection exactly for the choice of boundary terms
that also in Subsect.~\ref{secsuperpion} produced
the same momentum independence of times of detection.

Let us consider now, as in Subsect.~\ref{secsuperpion},
the following alternative choice of ${\cal K}$'s for the boundary terms
\begin{gather}\nonumber
{\cal K}^{[0]}_\mu = (p \oplus q)_\mu-(k\oplus  p' \oplus q)_\mu = p_\mu - p_\mu' - k_\mu - \ell \delta_\mu^1 ( - p_0 q_1 + k_0 p_1' + k_0 q_1 +p_0' q_1) \ ,\\
{\cal K}^{[1]}_\mu = (k\oplus  p' \oplus q)_\mu - (k \oplus  p'' \oplus q')_\mu = p_\mu' +q_\mu -p_\mu''-q_\mu'-\ell \delta_\mu^1 (-k_0p_1' - k_0 q_1 - p_0' q_1 +k_0 p_1'' + k_0 q_1' + p_0'' q_1' ) \ .
\end{gather}
Focusing again on the worldline $x' , p'$ detected at Bob
we now find
\begin{equation}
\begin{split}
{x'}_{B}^{0}(s)&=  {x'}_{A}^{0}(s)+b^{\mu}  \{ (k\oplus  p' \oplus q)_\mu , x'^0\} = {x'}_{A}^{0}(s)-b^{0}-\ell b^{1}q_{1} \simeq {x'}_{A}^{0}(s)-b^{0} \ ,\\
{x'}_{B}^{1}(s)&=x_{A}^{1}(s)+b^{\mu}  \{ (k\oplus  p' \oplus q)_\mu , x'^1\} ={x'}_{A}^{1}(s)-b^{1}-\ell b^{1}k_{0} \ .\\
\end{split}
\end{equation}
And from the equation of motion (\ref{eq. motion COM}) for the particle $x',p'$ as seen by Bob, one now deduces that
$${x'}_{B}^{1}(s)={x'}_{B}^{0}(s) -\ell b^{1}(k_{0}-p'_{1}) \ ,$$
which in turn implies that the time of detection at Bob
of the particle with worldline $x' , p'$
is
\begin{equation}
 t_{detection} = {x_B'}^0(s_{1})  =\ell b^{1}(k_{0} - p'_{1})=2\ell b^{1}|p'_{1}| \ .
\end{equation}
And this once again shows that, in spite of the change of symplectic structure
the results of Subsect.~\ref{secsuperpion} are exactly reproduced:
we got dependence on momentum of the times of detection given
by $2\ell b^{1}|p'_{1}|$ for exactly the same choice of boundary terms that also in
Subsect.~\ref{secsuperpion} produced
dependence on momentum of the times of detection given by $2\ell b^{1}|p'_{1}|$.

\section{Summary and outlook}~\label{closingsec}
The young idea of relative locality of course as a long way to go
before taking full shape, but it is encouraging that
already some aspects of it which might have been naively perceived
as unsurmountable challenges actually turned out, upon closer inspection,
to be manifestations of the internal strength of the logical
structure of the relative-locality proposal.
A striking example of this is the analysis reported in Ref.~\cite{soccerball},
which addressed some potential challenges for the description
of macroscopic bodies. And we feel that we reported in this manuscript
a similar accomplishment, by showing that the complication
of the nonlinear boundary terms used for enforcing momentum-conservation laws
does admit, in spite of its first-look appearance,
a fully satisfactory relativistic
description of distant observers.

Besides its use for the development of new theories, we should stress
in this closing remarks also how  awareness of a possible
relativity of locality can empower certain analyses: the ``$\kappa$-Poincar\'e
phase spaces" which were used extensively in our manuscript were invented
long before the recent conceptualization of relative locality
(see, {\it e.g.}, Ref.~\cite{lukiePHASESPACE,lukiegacPHASESPACE,jurekEARLYphasespace}).
 But the analysis of such $\kappa$-Poincar\'e
phase spaces had remained confined for many years
at the level of theories of only free particles.
No such interacting particle theory had been found, and we now
understand why: without the awareness of the possibility of a relativity
of locality one could not have introduced the description of interactions
that actually works, which we instead here produced with rather little effort.
We expect that several such instances may be discovered in future studies,
other instances in which an already well-known theoretical framework
is shown to host relativity of locality, and is then much better understood
once the awareness of the relativity of locality is exploited.

The fact we here settled several issues relevant for translational invariance
may be an important step forward
for relative locality, but several other such steps need to be taken.
In particular, the description of relative locality for distantly boosted
observers is at present reasonably well understood only for simple
models of relative locality for free
particles~\cite{whataboutbob,leeINERTIALlimit,arzkowaRelLoc,kappabob}.
The description of distantly boosted observers within the relative-locality
framework for interacting particles
of Refs.~\cite{prl,grf2nd}, which we here adopted, will probably require
facing challenges of similar magnitude to (but different nature from)
the ones we studied here.

Of course, of primary importance for the relative-locality framework
are its phenomenological consequences, and specifically its ability
to predict effects that (while surely minute) are within the reach
of the sensitivities of ongoing or foreseeable experimental studies.
That such opportunities would be found was already clear
on the basis of some of the observations reported in Ref.~\cite{prl},
and  the results of Ref.~\cite{leelaurentGRB}
connecting nonmetricity to time delays
(of the type here considered in Sec.~\ref{secgrbs})
with encouraging quantitative estimates already gave additional tangibility
to these expectations. In a sort of much-welcome corollary to
our main work on translational invariance for cases with causally connected
interactions, we here exposed, in Sec.~\ref{secgrbs}, first
evidence of some striking (minute but ``observably large")
manifestations of the torsion of momentum space,
in a case  where nonmetricity
and torsion are both present (our ``$\kappa$-momentum space").
It is perhaps not surprising that the most striking phenomenological
consequences of the geometry of momentum space would be found when
both nonmetricity and torsion are present. This is likely going to be
the ``closest target" for the phenomenology of relative-locality momentum
spaces, and we feel that it may therefore deserve priority
in future investigations of relative locality.

\section*{NOTE ADDED}
As we were in the final stages of preparation of this manuscript,
we became aware of the study reported in Ref.~\cite{flagiuRL},
which takes as starting point a characterization of
the ``$\kappa$-momentum space",
just like here in
Sections~\ref{kappaMtotherightDIR} and~\ref{geometry}
we got our analysis started by a
characterization of
the ``$\kappa$-momentum space".
There are some differences in style and focus, but
the characterizations of ``$\kappa$-momentum space" given here
and in Ref.~\cite{flagiuRL} are fully consistent with one another.
The issues for the relativistic description of distant observers
within the framework of Refs.~\cite{prl,grf2nd},
which are the main objective of the study we reported in this manuscript,
 (main results in
 Sections~\ref{setup}~\ref{main}~\ref{secgrbs}),
were not considered in Ref.~\cite{flagiuRL}. Instead Ref.~\cite{flagiuRL}
investigates certain issues relevant for the implementation
of $\kappa$-Poincar\'e boosts in a relative-locality setting of
the type advocated in Refs.~\cite{prl,grf2nd}.
We expect that the interplay between the relativistic description of distant observers
we provided here and the properties of $\kappa$-Poincar\'e boosts
highlighted in Ref.~\cite{flagiuRL} could make for an entertaining
future project.

\section*{ACKNOWLEDGEMENTS}
First of all we are grateful for encouragement and numerous valuable discussions
to Laurent Freidel and Lee Smolin.
Also very valuable were conversations with Niccolo' Loret and Antonino Marciano'.
And we are grateful for the interest shown by participants
to the XXIX Max Born symposium, where two of us (GAC and JKG) presented
parts of the results here contained in Sections~\ref{geometry}~\ref{setup}~\ref{main}.
MA is supported by EU Marie Curie
Actions under a Marie Curie Intra-European Fellowship.
The work of JKG was supported in part by a
Polish Ministry of Science and Higher Education grant 182/N-QGC/2008/0.

\appendix

\section{Equations of motion from the action of Subsection~\ref{asidevuoto}}
In this appendix
we write down explicitly the equations of motion, constraints and
boundary conditions that are relevant for the discussion
given in Subsec.~\ref{asidevuoto}.

These are the
equations of motion  and
boundary conditions
that follow from the action
 \begin{equation}
\begin{split}
\mathcal{S}^{\kappa (2')}
= & \int_{-\infty}^{s_{0}}ds\left(z^{\mu}\dot{k}_{\mu}-\ell z^{1}k_{1}\dot{k}_{0}+\mathcal{N}_k\mathcal{C}_\kappa\left[k\right]\right)+
\int_{s_{0}}^{s_{1}}ds\left({x}^{\mu}\dot{p}_{\mu}-\ell x^{1}p_{1}\dot{p}_{0}+\mathcal{N}_p\mathcal{C}_\kappa\left[p\right]\right)\\
 & +\int_{s_{0}}^{s_{1}}ds\left(y^{\mu}\dot{q}_{\mu}-\ell y^{1}q_{1}\dot{q}_{0}+\mathcal{N}_q\mathcal{C}_\kappa\left[q\right]\right)+
 \int_{s_{1}}^{\infty}ds\left(y_{\star}^{\mu}\dot{q}_{\mu}^{\star}-\ell y_{\star}^{1}q_{1}^{\star}\dot{q}_{0}^{\star}+\mathcal{N}_{q^\star}\mathcal{C}_\kappa\left[q^{\star}\right]\right)\\
 & +\int_{s_{1}}^{\infty}ds\left({x'}^{\mu}\dot{p'}_{\mu}-\ell{x'}^{1}p'_{1}\dot{p'}_{0}+\mathcal{N}_{p'}\mathcal{C}_\kappa\left[p'\right]\right)+
 \int_{s_{1}}^{\infty}ds\left(x''^{\mu}\dot{p''}_{\mu}-\ell x''^{1}p''_{1}\dot{p''}_{0}+\mathcal{N}_{p''}\mathcal{C}_\kappa\left[p''\right]\right)\\
 & -\xi_{[0]}^{\mu}\mathcal{K}_{\mu}^{[0]}(s_{0})-\xi_{[1]}^{\mu}\mathcal{K}_{\mu}^{[1]}(s_{1})-\xi_{[1']}^{\mu}\mathcal{K}_{\mu}^{[1']}(s_{1}) ~,
\end{split}
\end{equation}
and one easily finds
\begin{gather}
\dot p_\mu =0~,~~\dot q_\mu =0~,~~\dot k_\mu =0~,~~\dot p'_\mu =0~,~~\dot p''_\mu =0~,~~\dot q^\star_\mu =0~,~~\nonumber\\
\mathcal{C}_{\kappa}[p]=0~,~~\mathcal{C}_{\kappa}[q]=0~,~~\mathcal{C}_{\kappa}[k]=0~,~~\mathcal{C}_{\kappa}[p']=0~,~~\mathcal{C}_{\kappa}[p'']=0~,~~\mathcal{C}_{\kappa}[q^\star]=0~,~~\\
\mathcal{K}_\mu^{[0]}=0~,~~\mathcal{K}_\mu^{[1]}=0~,~~\mathcal{K}_\mu^{[1']}=0 \ . \nonumber
\label{eqmotion3+3+2kmomentum}
\end{gather}
\begin{gather}
\dot x^\mu = \mathcal{N}_p \left(\frac{\delta \mathcal{C}_{\kappa}[p]}{\delta p_\mu} +\ell \delta^\mu_0 \frac{\delta \mathcal{C}_{\kappa}[p]}{\delta p_1} p_1\right)= \delta^{\mu}_{0}\mathcal{N}_p \left( 2 p_{0}-\ell p_{1}^{2}\right)-2\delta^{\mu}_{1}\mathcal{N}_p \left(  p_{1}-\ell p_{0} p_{1}\right) \ ,\nonumber\\
\dot y^\mu = \mathcal{N}_q \left(\frac{\delta \mathcal{C}_{\kappa}[q]}{\delta q_\mu} +\ell \delta^\mu_0 \frac{\delta \mathcal{C}_{\kappa}[q]}{\delta q_1} q_1\right)= \delta^{\mu}_{0}\mathcal{N}_q \left( 2 q_{0}-\ell q_{1}^{2}\right)-2\delta^{\mu}_{1}\mathcal{N}_q \left(  q_{1}-\ell q_{0} q_{1}\right)\label{eqmotion3+3+2kspace} \ ,\\
\dot z^\mu = \mathcal{N}_k \left(\frac{\delta \mathcal{C}_{\kappa}[k]}{\delta k_\mu} +\ell \delta^\mu_0 \frac{\delta \mathcal{C}_{\kappa}[k]}{\delta k_1} k_1\right)= \delta^{\mu}_{0}\mathcal{N}_k \left( 2 k_{0}-\ell k_{1}^{2}\right)-2\delta^{\mu}_{1}\mathcal{N}_k \left(  k_{1}-\ell k_{0} k_{1}\right) \ , \nonumber\\
\dot x'^\mu =\mathcal{N}_{p'} \left(\frac{\delta \mathcal{C}_{\kappa}[p']}{\delta p'_\mu} +\ell \delta^\mu_0 \frac{\delta \mathcal{C}_{\kappa}[p']}{\delta p'_1} p'_1\right)=\delta^{\mu}_{0}\mathcal{N}_{p'} \left( 2 p_{0}'-\ell p_{1}'^{2}\right)-2\delta^{\mu}_{1}\mathcal{N}_{p'} \left(  p_{1}'-\ell p_{0}' p_{1}'\right) \ ,\nonumber\\
\dot x''^\mu = \mathcal{N}_p'' \left(\frac{\delta \mathcal{C}_{\kappa}[p'']}{\delta p''_\mu} +\ell \delta^\mu_0 \frac{\delta \mathcal{C}_{\kappa}[k]}{\delta p''_1} p''_1\right)=\delta^{\mu}_{0}\mathcal{N}_{p''} \left( 2 p_{0}''-\ell p_{1}''^{2}\right)-2\delta^{\mu}_{1}\mathcal{N}_{p''} \left(  p_{1}''-\ell p_{0}'' p_{1}''\right) \ ,\nonumber\\
\dot y^{\star\mu} = \mathcal{N}_{q^\star} \left(\frac{\delta \mathcal{C}_{\kappa}[q^\star]}{\delta q^\star_\mu} +\ell \delta^\mu_0 \frac{\delta \mathcal{C}_{\kappa}[q^\star]}{\delta q^\star_1} q^\star_1\right)=\delta^{\mu}_{0}\mathcal{N}_{q^\star} \left( 2 q^\star_{0}-\ell q_{1}^{\star2}\right)-2\delta^{\mu}_{1}\mathcal{N}_{q^\star} \left(  q^\star_{1}-\ell q^\star_{0} q^\star_{1}\right) \ .\nonumber
\end{gather}
It is easy to recognize that
these are the same equations of motion that one also obtains from
the action $\mathcal{S}^{\kappa (2)}$ (of Subsec.~\ref{mainsubsec}),
up to splitting fictitiously
(as suggested by the
drawing in Fig.~\ref{figunovuoto})
the worldline $y^\mu$, $q_\mu$
  into two perfectly-matching
pieces of worldline, a piece labeled again $y^\mu$, $q_\mu$ and a piece
labeled $y_\star^\mu$, $q^\star_\mu$.

Similarly one has that from $\mathcal{S}^{\kappa (2')}$
it follows that
the conditions at the $s=s_{0}$ and $s=s_{1}$ boundaries
 are
\begin{gather}
z^\mu(s_{0}) = \xi^\nu_{[0]} \left(\frac{\delta \mathcal{K}^{[0]}_\nu}{\delta k_\mu}+\ell \delta^\mu_0 \frac{\delta \mathcal{K}^{[0]}_\nu}{\delta k_1}k_1\right)=\xi^\mu_{[0]} +\ell \delta^{\mu}_{0} \xi^1_{[0]} k_{1} \ ,\nonumber\\
x^\mu(s_{0}) = -\xi^\nu_{[0]} \left(\frac{\delta \mathcal{K}^{[0]}_\nu}{\delta p_\mu}+\ell \delta^\mu_0 \frac{\delta \mathcal{K}^{[0]}_\nu}{\delta p_1}p_1\right)=\xi^\mu_{[0]} +\ell \delta^{\mu}_{0} \xi^1_{[0]} (p_{1}+q_{1})  \ , \nonumber \\
x^\mu(s_{1}) = \xi^\nu_{[1]} \left(\frac{\delta \mathcal{K}^{[1]}_\nu}{\delta p_\mu}+\ell \delta^\mu_0 \frac{\delta \mathcal{K}^{[1]}_\nu}{\delta p_1}p_1\right)=\xi^\mu_{[1]} +\ell \delta^{\mu}_{0} \xi^1_{[1]} (p_{1}+q_{1}) \ , \nonumber\\
\!\!\!\!\!\!\!\!\!\!\!\!\!\! \!\!\!\!\!\!\!\! \!\!\!\! y^\mu(s_{0})
= -\xi^\nu_{[0]} \left(\frac{\delta \mathcal{K}^{[0]}_\nu}{\delta q_\mu}
+\ell \delta^\mu_0 \frac{\delta \mathcal{K}^{[0]}_\nu}{\delta q_1}q_1\right)=\xi^\mu_{[0]} +\ell \delta^{\mu}_{0} \xi^1_{[0]} q_{1}+\ell \delta^{\mu}_{1} \xi^1_{[0]} p_{0}
 \ , \nonumber \\ y^\mu(s_{1})= \xi^\nu_{[1']} \left(\frac{\delta \mathcal{K}^{[1']}_\nu}{\delta q_\mu}+\ell \delta^\mu_0 \frac{\delta \mathcal{K}^{[1']}_\nu}{\delta q_1}q_1\right)=\xi^\mu_{[1']} +\ell \delta^{\mu}_{0} \xi^1_{[1']} q_{1}+\ell \delta^{\mu}_{1} \xi^1_{[1']} p_{0}
 \ , \nonumber\\
x'^\mu(s_{1}) = -\xi^\nu_{[1]} \left(\frac{\delta \mathcal{K}^{[1]}_\nu}{\delta p'_\mu}+\ell \delta^\mu_0 \frac{\delta \mathcal{K}^{[1]}_\nu}{\delta p'_1}p'_1\right)=\xi^\mu_{[1]} +\ell \delta^{\mu}_{0} \xi^1_{[1]} (p_{1}'+p_{1}''+q_{1})\label{boundaries3+3+2k}  \ , \\
x''^\mu(s_{1}) = -\xi^\nu_{[1]} \left(\frac{\delta \mathcal{K}^{[1]}_\nu}{\delta p''_\mu}+\ell \delta^\mu_0 \frac{\delta \mathcal{K}^{[1]}_\nu}{\delta p''_1}p''_1\right)=\xi^\mu_{[1]} +\ell \delta^{\mu}_{0} \xi^1_{[1]} (p_{1}''+q_{1})+\ell \delta^{\mu}_{1} \xi^1_{[1]} p_{0}' \ , \nonumber\\
y^{\star\mu} (s_{1}) = -\xi^\nu_{[1']} \left(\frac{\delta \mathcal{K}^{[1']}_\nu}{\delta q^\star_\mu}+\ell \delta^\mu_0 \frac{\delta \mathcal{K}^{[1']}_\nu}{\delta q^\star_1}q^\star_1\right)=\xi^\mu_{[1']} +\ell \delta^{\mu}_{0} \xi^1_{[1']} q_{1}^{\star}+\ell \delta^{\mu}_{1} \xi^1_{[1']} p_{0} \ ,
\nonumber
\end{gather}
and these essentially reproduce the boundary conditions obtained
from  $\mathcal{S}^{\kappa (2)}$ in Subsec.~\ref{mainsubsec},
up to boundary conditions that ensure the perfect matching
of the two ``pieces of worldline" $y^\mu$, $q_\mu$
and $y_\star^\mu$, $q^\star_\mu$ into a single worldline.

\section{Translation transformations of the action
in Subsection~\ref{sec3conn}}
We here study
the effect of our translation transformations, given in (\ref{trasla3conn}),
on the action ${\cal S}^{(3conn)}$ discussed in Subsection~\ref{sec3conn}.

The action ${\cal S}^{(3conn)}$
includes quite a few terms, and we find it to be convenient
to organize them in a way that helps one to keep track of all contributions.
A useful expedient is to rewrite
the integrations concerning finite worldlines,
 of the kind $\int_{s_0}^{s_1}$, splitting them
 into two semi-infinite integrals:
\begin{equation}
 \int_{s_0}^{s_1} = \int_{s_0}^\infty -\int_{s_1}^\infty\ .
\end{equation}
We can then rewrite our action as a sum of semi-infinite integrals
and in particular we can opt for the following form
\begin{equation}
{\cal S}^{(3conn)} = {\cal S}^{(3conn)[s_0]} + {\cal S}^{(3conn)[s_1]}
+ {\cal S}^{(3conn)[s_2]}+ {\cal S}^{(3conn)[s_3]} \ ,
 \end{equation}
where in ${\cal S}^{[{\bar s}]}$ we include all semi-infinite integrals
with a boundary at $s ={\bar s}$.

Using manipulations we already discussed in Subsec.~\ref{invariance}
one easily finds that the action for observer Bob
can be written as
\begin{equation}
\begin{split}
\mathcal{S}^{(3conn)}_B= &\mathcal{S}_A^{(3conn)[s_0]}+\mathcal{S}_A^{(3conn)[s_1]}   -\int_{s_{2}}^{+\infty}ds\left({x}^{\mu}\dot{p}_{\mu}-\ell x^{1}p_{1}\dot{p}_{0}+\mathcal{N}_{p}\mathcal{C}\left[p\right]\right)
 +\int_{s_{2}}^{+\infty}ds\left({x'}^{\mu}\dot{p}'_{\mu}-\ell{x'}^{1}p'_{1}\dot{p}'_{0}+\mathcal{N}_{p'}\mathcal{C}\left[p'\right]\right)\\
&+\int_{s_{2}}^{+\infty}ds\left(x''^{\mu}\dot{p}''_{\mu}-\ell x''^{1}p''_{1}\dot{p}''_{0}+\mathcal{N}_{p''}\mathcal{C}\left[p''\right]\right)\\
&-\int_{s_{3}}^{+\infty}ds\left(y^{\mu}\dot{q}_{\mu}-\ell y^{1}q_{1}\dot{q}_{0}+\mathcal{N}_{q}\mathcal{C}\left[q\right]\right)
+ \int_{s_{3}}^{+\infty}ds\left(y'^{\mu}\dot{q}'_{\mu}-\ell y'^{1}q'_{1}\dot{q}'_{0}+\mathcal{N}_{q'}\mathcal{C}\left[q'\right]\right)\\
&+\int_{s_{3}}^{+\infty}ds\left( y''^{\mu}\dot{q}''_{\mu}-\ell y''^{1}q''_{1}\dot{q}''_{0}+\mathcal{N}_{q''}\mathcal{C}\left[q''\right]\right)\\
& -(\Delta\xi_{[0]}^{\mu}+b^\mu)\mathcal{K}_{\mu}^{[0]}(s_{0})-(\Delta\xi_{[1]}^{\mu}+b^\mu)\mathcal{K}_{\mu}^{[1]}(s_{1})-\xi_{[2]B}^{\mu}\mathcal{K}_{\mu}^{[2]}(s_{2})-\xi_{[3]B}^{\mu}\mathcal{K}_{\mu}^{[3]}(s_{3})\\
&-\int_{s_{1}}^{+\infty}ds\left( \ell b^1 \mathcal{K}_{0}^{[1]} \dot k''_1\right) \ ,
\end{split}
\end{equation}
where $\mathcal{S}_A ^{(3conn)[s_0]}$ and $\mathcal{S}_A^{(3conn)[s_1]}$
are described from Alice's
perspective, and we already showed above that the
term $\int_{s_{1}}^{+\infty}ds\left( \ell b^1 \mathcal{K}_{0}^{[1]} \dot k''_1\right)$
can be dropped since it
does not contribute to the equations of motion and the boundary
conditions.

Next we can focus on the contributions from integrals with
a boundary at $s=s_{2}$, finding that
\begin{equation}
\begin{split}
\mathcal{S}_B^{(3conn)[s_2]}= &-\int_{s_{2}}^{+\infty}ds\left({x}^{\mu}\dot{p}_{\mu}
-\ell x^{1}p_{1}\dot{p}_{0}+\mathcal{N}_{p}\mathcal{C}\left[p\right]\right)
 +\int_{s_{2}}^{+\infty}ds\left({x'}^{\mu}\dot{p}'_{\mu}-\ell{x'}^{1}p'_{1}\dot{p}'_{0}
 +\mathcal{N}_{p'}\mathcal{C}\left[p'\right]\right)\\
&+\int_{s_{2}}^{+\infty}ds\left(x''^{\mu}\dot{p}''_{\mu}-\ell x''^{1}p''_{1}\dot{p}''_{0}+\mathcal{N}_{p''}\mathcal{C}\left[p''\right]\right)\\
&-\xi_{[2]B}^{\mu}\mathcal{K}_{\mu}^{[2]}(s_{2}) \ , \\
\end{split}
\end{equation}
after applying our notion of  translation transformation,
can be written as
\begin{equation}
\begin{split}
\mathcal{S}_B^{(3conn)[s_2]}=&\mathcal{S}_A^{(3conn)[s_2]}
-\int_{s_{2}}^{+\infty}ds\left(-{b}^{\mu}\dot{p}_{\mu}-\ell b^1(q_{1}+k''_{1})\dot{p}_{0}\right)
 +\int_{s_{2}}^{+\infty}ds\left(-{b}^{\mu}\dot{p}'_{\mu}-\ell b^1(p''_1+q_1+k''_1)\dot{p}'_{0}\right)\\
&+\int_{s_{2}}^{+\infty}ds\left(-b^{\mu}\dot{p}''_{\mu}-\ell b^1(q_{1}+k''_1)\dot{p}''_{0}-\ell b^1 p'_0 \dot {p}''_1\right)
-\Delta\xi_{[2]}^{\mu}\mathcal{K}_{\mu}^{[2]}(s_{2})\\
=&\mathcal{S}_A^{(3conn)[s_2]}+\int_{s_{2}}^{+\infty}ds \frac{d}{ds}(b^\mu \mathcal{K}_{\mu}^{[2]})-\Delta\xi_{[2]}^{\mu}\mathcal{K}_{\mu}^{[2]}(s_{2})\\
&-\int_{s_{2}}^{+\infty}ds\left( \ell b^1 \mathcal{K}_{0}^{[2]} (\dot q_1+\dot k''_1)\right) \ .
\end{split}
\end{equation}
The last term is again of the type that does not contribute to equations of motion and boundary conditions
(once the conservation laws are enforced) and can therefore be dropped,
while for the contribution
$$\int_{s_{2}}^{+\infty}ds \frac{d}{ds}(b^\mu \mathcal{K}_{\mu}^{[2]})-\Delta\xi_{[2]}^{\mu}\mathcal{K}_{\mu}^{[2]}(s_{2}) \ , $$
we find that, since $\Delta\xi_{[2]}^{\mu}=\xi_{[2]B}^{\mu}-\xi_{[2]A}^{\mu}=-b^{\mu}$, it only produces a boundary term at $s=+\infty$:
$$\int_{s_{2}}^{+\infty}ds \frac{d}{ds}(b^\mu \mathcal{K}_{\mu}^{[2]})-\Delta\xi_{[2]}^{\mu}\mathcal{K}_{\mu}^{[2]} = b^\mu \mathcal{K}_{\mu}^{[2]}(\infty)-(\Delta\xi_{[2]}^{\mu} +b^{\mu})\mathcal{K}_{\mu}^{[2]}(s_{2})=b^\mu \mathcal{K}_{\mu}^{[2]} (\infty) \ , $$
and the boundary term at infinity
does not contribute to the physics
since  momenta are not varied at $\pm \infty$.

Similarly for contributions to the action by semi-infinite integrals
with a boundary at $s=s_{3}$ we have
\begin{equation}
\begin{split}
\mathcal{S}_B^{(3conn)[3]}=
&-\int_{s_{3}}^{+\infty}ds\left(y^{\mu}\dot{q}_{\mu}-\ell y^{1}q_{1}\dot{q}_{0}+\mathcal{N}_{q}\mathcal{C}\left[q\right]\right)
+ \int_{s_{3}}^{+\infty}ds\left(y'^{\mu}\dot{q}'_{\mu}-\ell y'^{1}q'_{1}\dot{q}'_{0}+\mathcal{N}_{q'}\mathcal{C}\left[q'\right]\right)\\
&+\int_{s_{3}}^{+\infty}ds\left( y''^{\mu}\dot{q}''_{\mu}-\ell y''^{1}q''_{1}\dot{q}''_{0}+\mathcal{N}_{q''}\mathcal{C}\left[q''\right]\right)\\
& -\xi_{[3]B}^{\mu}\mathcal{K}_{\mu}^{[3]}(s_{3}) \ ,
\end{split}
\end{equation}
which,  using again properties
of our translation transformations, can be written as
\begin{equation}
\begin{split}
\mathcal{S}_B^{(3conn)[3]}=&\mathcal{S}_A^{(3conn)[3]}
-\int_{s_{3}}^{\infty}ds\left(-b^{\mu}\dot{q}_{\mu}-\ell b^{1}k''_{1}\dot{q}_{0}-\ell b^1 (p'_0+p''_0)\dot q_1\right)
+ \int_{s_{3}}^{\infty}ds\left(-b^{\mu}\dot{q}'_{\mu}-\ell b^{1}(q''_{1}+k''_1)\dot{q}'_{0}-\ell b^1(p'_0+p''_0)\dot {q}'_1\right)\\
&+\int_{s_{3}}^{\infty}ds\left( -b^{\mu}\dot{q}''_{\mu}-\ell b^{1}k''_{1}\dot{q}''_{0} -\ell b^1 (p'_0+p''_0+q'_0)\dot {q}''_1\right)-\Delta\xi_{[3]}^{\mu}\mathcal{K}_{\mu}^{[3]}(s_{3})\\
=&\mathcal{S}_A^{(3conn)[3]}+\int_{s_{3}}^{\infty}ds \frac{d}{ds}(b^\mu \mathcal{K}_{\mu}^{[3]})-\Delta\xi_{[3]}^{\mu}\mathcal{K}_{\mu}^{[3]}(s_{3})\\
&-\int_{s_{3}}^{\infty}ds\left( \ell b^1 \mathcal{K}_{0}^{[3]} \dot {k}''_1+\ell b^1 \mathcal{K}_{1}^{[3]}( \dot {p}'_0+ \dot {p}''_0)\right) \ .
\end{split}
\end{equation}
And again we notice that
$\int_{s_{3}}^{\infty}ds\left( \ell b^1 \mathcal{K}_{0}^{[3]} \dot {k}''_1+\ell b^1 \mathcal{K}_{1}^{[3]}( \dot {p}'_0+ \dot {p}''_0)\right)$
can be dropped since it does not contribute to equations of motion
and boundary terms,
and as before we have  that, since  $\xi_{[3]B}^{\mu}=\xi_{[3]A}^{\mu}-b^{\mu}$,
$$\int_{s_{3}}^{\infty}ds \frac{d}{ds}(b^\mu \mathcal{K}_{\mu}^{[3]})-\Delta\xi_{[3]}^{\mu}\mathcal{K}_{\mu}^{[3]}(s_{3}) = b^\mu \mathcal{K}_{\mu}^{[3]} (\infty)-(\Delta\xi_{[3]}^{\mu} +b^{\mu})\mathcal{K}_{\mu}^{[3]}(s_{3})=b^\mu \mathcal{K}_{\mu}^{[3]}(\infty) \ , $$
{\it i.e.} once again the only left-over piece
is a boundary term at infinity that
can be safely dropped.

Combining these observations we conclude that both Alice and Bob
can use the same action principle to characterize the equations
of motion and boundary conditions for the chain of interactions we  analyzed
in Subsec.~\ref{sec3conn}.

\end{document}